\documentclass[aip,pop,reprint,10pt,superscriptaddress,showpacs]{revtex4-2}
\usepackage[left=0.90cm,right=0.90cm,top=0.92cm,bottom=0.92cm]{geometry}
\usepackage{amsfonts} 
\usepackage{amsmath}
\usepackage{amssymb}
\usepackage{graphicx}
\usepackage{subfigure}
\usepackage{color}
\usepackage{soul}
\usepackage{cancel}
\usepackage[none]{hyphenat}
\usepackage{ulem}
\begin{document}
\title{Dust-ion-acoustic damped solitary waves and shocks in  laboratory and Saturn's E-ring magnetized nonthermal dusty plasmas with anisotropic ion pressure and dust-charge fluctuation}
\author{Num Prasad  Acharya}
\affiliation{Central Department of Physics, Tribhuvan University, Kirtipur, Kathmandu 44613, Nepal}
\affiliation{Department of Physics, Mahendra Multiple Campus, Tribhuvan University, Ghorahi 22415, Dang, Nepal}
\author{Suresh Basnet}
\email{sbplasma1986@gmail.com}
\affiliation{Central Department of Physics, Tribhuvan University, Kirtipur, Kathmandu 44613, Nepal}
\author{Amar P. Misra}
\email{apmisra@visva-bharati.ac.in}
\affiliation{Department of Mathematics, Siksha Bhavana, Visva-Bharati University, Santiniketan-731 235, India}
\author{Raju Khanal}
\affiliation{Central Department of Physics, Tribhuvan University, Kirtipur, Kathmandu 44613, Nepal}
\begin{abstract}
We study the oblique propagation of weakly nonlinear dust-ion-acoustic (DIA) solitary waves (SWs) and shocks in collisional magnetized nonthermal dusty plasmas that are relevant in laboratory and space (Saturn's E-ring) environments. We consider plasmas to be composed of $q$-nonextensive hot electrons, thermal positive ions, and immobile negatively charged dust grains immersed in a static magnetic field and take into account the effects of ion creation (source), and ion loss (sink), ion-neutral and ion-dust collisions, anisotropic ion pressure and dust-charge fluctuations on the evolution of small-amplitude SWs and shocks. The ion-neutral collision enhancement equilibrium dust-charge number is self-consistently determined using Newton's Raphson method. We found that in laboratory dusty plasmas with adiabatic dust-charge variation [i.e., when the dust charging frequency ($\nu_{\rm{ch}}$) is much higher than the dust-plasma oscillation frequency ($\omega_{\rm{pd}}$)], the DIA solitary waves (DIASWs) get damped by the effects of the ion-dust and ion-neutral collisions, whereas the ion creation and ion loss leads to the amplification of solitary waves, and they appear as only compressive types with positive potential. On the other hand, in Saturn's E-ring plasmas, where the collisional and ion creation or ion loss effects are insignificant, the non-adiabaticity of dust-charge variation can give rise to the evolution of either damped DIASWs or DIA shocks, depending on the smallness of the ratios  $\nu_{\rm{ch}}/\omega_{\rm{pd}}$ or $\omega_{\rm{pd}}/\nu_{\rm{ch}}$, respectively. Furthermore, two critical values of the nonextensive parameter $q$ exist, below (or above) which, the DIASWs and shocks can appear as rarefactive (or compressive) types. The characteristics of DIASWs and shocks are also analyzed numerically for parameters relevant to the laboratory and Saturn's E-ring plasmas.    
\end{abstract}
\maketitle
\section{INTRODUCTION} \label{sec-intro}
The presence of micron to nanometer-sized negatively or positively charged particulates, called dust particulates, cannot be avoided in laboratory, space, and astrophysical plasma environments, and plasmas with such additional species (charged grains) exhibit significantly distinct behaviors than usual electron-ion plasmas \citep{shotorban2011nonstationary,merlino2006dusty}. In the contexts of space and astrophysical environments, dusty plasmas get encountered in cometary tails, asteroid zones, planetary rings, magnetosphere, the lower part of the Earth's ionosphere, the lunar surface, and many more \citep{whipple1985electrostatics,horanyi1986dynamics,sheehan1990device}. On the other hand, dusty plasmas are also observed in laboratory environments such as tokamak plasmas, glow discharge plasmas, and low-temperature plasma processing (fabrication of semiconductor devices via dry etching) \citep{chen1995industrial,merlino2006dusty}. Typically, charged dust particles are more massive than ions (about $10^{14}$ times the mass of a hydrogen ion, but it depends on the size and types of dust particles) and hence electrons, and the magnitude of dust-charge number varies from $100$ to $20,000$ depending on the plasma system \citep{merlino2006dusty,merlino2009dust}. As the dust particles are submerged (either naturally occurred  or externally inserted) in a plasma system, they can acquire negative or positive charges by collecting energetic electrons and positive ions or by releasing electrons from their surface via different processes, e.g., thermionic emission, photo-electron emission due to ultra-violet radiation, secondary electron emission by energetic ions and electrons or sometimes absorption of excessive flux of positive ions \citep{barkan1994charging,sickafoose2000photoelectric}. When the thermal flux of electrons residing on the surface of dust grains dominates over the flux of positive ions and electron emission, the dust particles become negatively charged. On the contrary, the dust particles can acquire a positive charge when the flux of positive ions to the dust surface and electron emission dominates over the flux of plasma electrons \citep{walch1995charging}. Whatever the charge acquired by the dust grains in plasmas, the existence of plasma wave spectra is modified by the stationary charged dust grains \citep{tribeche2009effect}. 
\par 
In dusty plasmas, the dust-charge fluctuation is a dynamic process and a complex mechanism, which arises due to inherent random fluctuations of background ion and electron currents flowing into the dust-grain surface \citep{asgari2011stochastic}. The discrete nature of the charge carriers also causes the charge fluctuations \citep{cui1994fluctuations} affecting various  processes and structures in dusty plasmas, such as plasma sheath, ion-flow at the sheath boundary, Dust-Mach number, wave phase velocity of Dust-acoustic (DA) and dust-ion-acoustic (DIA) waves, DA and DIA solitons and shocks, etc. \citep{misra2004electrostatic,baluku2008solitons,ghosh2000nonlinear,tribeche2001current,vranjevs2002ion}. 
The dust-charge fluctuation in unmagnetized dusty plasmas produces a dissipation on the dynamics of DIA  waves (DIAWs) in giving rise to the generation of DIA shock waves in space and laboratory plasmas \citep{mamun2002role,duha2009dust}. The propagation of DIAWs in plasmas with positive and negative dust-charge variations has been studied by Mowafy \textit{et al.} in weakly inhomogeneous, unmagnetized, and collisionless dusty plasmas. They reported that the positively charged dust grains modify the characteristics and the polarity of solitary waves \citep{mowafy2008effect}. Experimental and theoretical studies show that charged dust grains in a plasma alter the space and time scales of plasma oscillations and produce new wave eigenmodes, namely the DA or DIA mode \citep{rao1990dust}, dust-lattice mode \citep{melandso1996lattice}, dust Bernstein-Greene-Kruskal modes \citep{tribeche2000effects} etc. The modification of DIAWs \citep{shukla1992dust}  has received considerable interest, and it is confirmed experimentally in low-temperature dusty plasma devices \citep{barkan1995laboratory,merlino1998laboratory}.  Shukla and Silin \citep{shukla1992dust}theoretically predicted low-frequency DIAWs in collisionless unmagnetized dusty plasmas (e.g., in the F-ring of Saturn). They showed that in the charge neutrality at equilibrium state, i.e., $en_{\mathrm{e0}} = en_{\mathrm{i0}} + q_{\mathrm{d0}}n_{\mathrm{d0}}$, if the strong inequality, $n_{\mathrm{i0}} \gg n_{\mathrm{e0}}$ holds, a dusty plasma containing negatively charged dust grains can support the existence of DIAWs. Here, $e$ is the elementary charge, $q_{\rm{d0}}$ is the charge of background dust grains, and $n_{j0}$ is the equilibrium number density of $j$-th species particles. Barkan \textit{et al.} experimentally investigated the characteristics of DA waves in dusty plasmas \citep{barkan1995laboratory}. Later, Nakamura \textit{et al.} also experimentally studied the linear and nonlinear features of DIAWs in a laboratory dusty plasma \citep{nakamura1999observation}. In addition to the dust charge variation, the DA and DIA wave characteristics are also significantly altered by the effects of ion creation (source) and ion loss (sink) in plasma flow  \citep{d1997ionization,ghosh2005effects,tamang2018solitary,mehdipoor2022characteristics}. While the source term causes to excite the DA and DIA waves, the collisions of ions with charged dust grains and neutral atoms reduce the growth rate and amplitude of such waves \citep{d1997ionization,ghosh2005effects,mehdipoor2022characteristics}.
In another investigation,  Ghosh et al. studied the effects of ionization on the nonlinear propagation of DIAWs, and they found that the amplitude of DIAWs exponentially grows over time while ion-neutral and ion-dust drag force reduces the wave's growth rate \cite{ghosh2005effects}.
\par
Electrostatic waves exhibit distinct behaviors in magnetized and unmagnetized plasmas, and they have been extensively studied in both environments \citep{melzer2021physics,ghosh2001small}. In weakly collisional or collisionless plasmas, the plasma anisotropy takes place due to a strong magnetic field. In this scenario, plasma particles exhibit different behaviors along and  perpendicular to the external magnetic field \citep{choi2007dust}. In the past few years, the effects of anisotropic ion pressure on DIA solitary waves (SWs) in magnetized dusty plasmas  have been extensively studied. It was found that the perpendicular and parallel components of the anisotropic ion pressure have significant effects on the characteristics of DIASWs \citep{choi2007dust,adnan2014effect,almas2023oblique}. In space and astrophysical plasmas like Earth's magnetosphere and planetary rings, the presence of charged dust grains and their charge variation can significantly influence the formation of both compressive and rarefactive solitary waves and shocks. The linear and nonlinear theories of DIAWs have been developed by several authors to explore the fundamental properties of dust-plasma interactions in planetary ring structures \citep{ghosh2001smal,mamun1998solitary, mamun1998nonlinear,khan2007linear,el2004dust,yang2012collision,el2015linear,misra2015complex}. The ring system of the giant planet Saturn is divided into several rings: E, G, F, A, B, C, and D from the inner side to the outer side edge in which A, B, and C are the main rings of Saturn and are massive and more turbid, while D, E, F, and G are the faint rings and are of diffused types. As visible components, micron, and sub-micron-sized particles are found in E, F, and G rings of Saturn \citep{spilker1997passage,morfill1983saturn}. El-Labany \textit{et al.} \citep{el2009effects} studied the nonlinear theory of DIASWs in magnetized dusty plasmas with two-temperature electrons in the contexts of Saturn's E-ring.  
Using a fluid model, Shohaib \textit{et al.} studied the linear and nonlinear properties of DIAWs in dusty plasmas with nonextensive electrons relevant to Saturn's B ring. They reported the existence of both compressive and rarefactive solitons and analyzed their properties by the effects of the nonextensive parameter
\citep{shohaib2022interaction}. In another investigation, Ali \textit{et al.} reported the existence of dust-acoustic solitary waves (compressive and rarefactive) and periodic structures in Saturn's inner magnetosphere dusty plasmas containing two groups of superthermal electrons\citep{ali2023attributes}.
\par
Boltzmann–Gibbs (B-G) statistics has been widely considered in applying to macroscopic ergodic equilibrium systems. Numerous earlier studies reported that the B-G statistics for the particle distribution may not be sufficient to explain nonequilibrium systems involving long-range interactions, such as gravitating and plasma systems \citep{vergou2009non,lavagno2011nonextensive}. In these systems, the particle distribution considerably deviates from the B-G statistics and follows the nonextensive statistics proposed by Tsallis \citep{Tsallis1998nonextensive}, which is a generalized form of B-G statistics and frequently encountered in space, astrophysical, and laboratory plasmas \citep{yoon2019thermodynamic,liu2009nonextensive}. Tsallis modeled the nonextensive entropy by assuming a composition law for two independent thermodynamic systems $R$ with entropy $S_{\text{q}}(R)$ and $T$ with entropy $S_{\text{q}}(T)$. The composite entropy of the resultant system is obtained by using the concept of pseudo-additivity property of nonextensive entropy, which is written as\citep{Tsallis1998nonextensive} 
\begin{equation}
S_{\mathrm{q}}\left(R + T\right) = S_{\mathrm{q}}\left(R\right) + S_{\mathrm{q}}\left(T\right) + \left(1-q\right)S_{\mathrm{q}}\left(R\right)S_{\mathrm{q}}\left(T\right),
\label{eqn_nonext_entropy}
\end{equation}
where the parameter $q$ measures the degree of nonextensivity. The limit $q\rightarrow1$ corresponds to the Maxwell-Boltzmann velocity distribution, $q > 1$ corresponds to subextensive systems with a significant number of low-energy particles, and $q < 1$ corresponds to superextensive systems with a significant number of energetic particles. The $q$-nonextenisve entropy proposed by Tsallis  can be expressed as 
\begin{equation}
 S_{\text{q}} = k_{\text{B}}\frac{1-\sum_{j}(p_{j})^{q}}{(q-1)},
 \label{eqn_Tsallis_entropy}
\end{equation}
where $k_{\text{B}}$ is the Boltzmann constant and $p_{j}$ is the probability of $j{\text{th}}$ microstate. In particular,  for $q\rightarrow1$, one recovers from Eq. (\ref{eqn_Tsallis_entropy}) the B-G entropy, given by,
\begin{equation}
 S_{\text{q}} = -k_{\text{B}}\sum_{j}p_{j}\text{ln}(p_{j}).
 \label{eqn_BG_entropy}
\end{equation}
It is to be mentioned that the spectra of the $q$-nonextensive and $\kappa$-distribution functions exhibit similar behaviors, implying that they are somewhat equivalent. In fact, by applying a formal
transformation, $\kappa=1/(1-q)$ one can find the missing links between these two velocity distribution functions \citep{Leubner2002}.
\par
The velocity distribution of particle species in a thermodynamic equilibrium system represented by Eq. (\ref{eqn_BG_entropy}) obeys the Maxwell-Boltzmann distribution. The $q$-nonextensive velocity distribution function is given by \citep{silva1998maxwellian,bacha2012nonextensive}
\begin{equation}
g_{\mathrm{e}}(\text{v})= A_{\mathrm{q}}\left[1-(q-1)\left(\frac{m_{\mathrm{e}}\text{v}^2}{2k_{\textrm{B}}T_{\textrm{e}}}-\frac{e\phi}{k_{\textrm{B}}T_{\textrm{e}}}\right)\right]^{\frac{1}{q-1}},
\label{eqn_vdfs}
\end{equation}
where $\phi$ is the electrostatic potential, $e$ is the elementary charge, $m_{\mathrm{e}}$ is the electron mass, and $T_e$ is the electron temeperature. Also, the  normalization constant $A_{\mathrm{q}}$ is given by
 \begin{equation} \label{eq-Aq}
A_{\mathrm{q}}=
\begin{cases}
n_{\text{e}0}\frac{3q-1}{2(1-q)}\frac{\Gamma\left( \frac{1}{1-q}\right)}{\Gamma\left(\frac{1}{1-q}-\frac{1}{2}\right)}\left(\frac{m_{\text{e}}(1-q)}{2\pi T_{\text{e}}}\right)^{3/2}, \hspace{0.2cm}1/3<q<1, \\
\\
n_{\mathrm{e0}}\frac{(3q-1)(1+q)}{4(q-1)}\frac{\Gamma \left({\frac{1}{q-1}}+\frac{1}{2}\right)}{\Gamma\left({\frac{1}{q-1}}\right)}\left(\frac{m_{\mathrm{e}}(q-1)}{2\pi k_{\textrm{B}}T_{\textrm{e}}}\right)^{3/2}, \hspace{0.2cm}q>1,
\end{cases}
\end{equation}
where $\Gamma$ denotes the standard gamma function and $n_{\text{e}0}$ is the equilibrium electron number density. For the subextensive electron distribution with $q > 1$, the distribution function, [Eq. (\ref{eqn_vdfs})] exhibits a thermal cut-off at a maximum velocity, given by,
\begin{equation}
{\text{v}}_{\textrm{max}} = \sqrt{{\frac{2k_{\textrm{B}}T_{\textrm{e}}}{m_{\mathrm{e}}}}\left(\frac{1}{q-1}+\frac{e\phi}{k_{\textrm{B}}T_{\textrm{e}}}\right)}.
\label{eqn_maximum_velocity}
\end{equation} 
Integrating Eq. (\ref{eqn_vdfs}) over the three-dimensional velocity space, we obtain the following $q$-nonextensive density distribution of electrons.
\begin{equation}
n_{\mathrm{e}} = \int g_{\mathrm{e}}\left(x,\text{v}\right)d^{3}\text{v} = n_{\mathrm{e}0}\left[1+\left(q-1\right)\frac{e\phi}{k_{\textrm{B}}T_{\textrm{e}}}\right]^{\left(3q-1\right)/\left(2q-2\right)}.
\label{eqn_electron_density}
\end{equation}
Typically, in space and astrophysical plasmas, the particle distribution function has non-Maxwellain high-energy tails. In this context,  Meyer-Vernet introduced the distribution function for superthermal electrons \citep{meyer1982flip}. Later, this work was extended by Rosenberg and Mendis to study the evolution of dust surface potential using a dust charging model\citep{rosenberg1992note}. The choice of electron distribution in dusty plasmas explicitly determines the dust charging process and influences the characteristics of wave propagation \citep{kamran2021dust,abdikian2023role}. It has been shown that in q-nonextensive plasmas,  either compressive or rarefactive DIASWs can exist depending on the values of the parameter $q$ above or below its critical value  \citep{bacha2012dust}. The theory has also been developed in nonplanar geometry to study the propagation characteristics of ion-acoustic and DIA shock waves in the contexts of Saturn's E-ring dusty plasmas with nonextensive cold and hot electrons  \citep{bansal2020nonplanar,bansal2021parametric}. In another investigation, Shahzad \textit{et al.} \citep{shahzad2023nonlinear} showed that the external magnetic field, ion temperature, and dust and ion concentrations can significantly influence the properties of dust-acoustic solitary waves that are relevant in the Earth's lower ionospheric region.  
\par 
In most of the previous investigations in the literature, including those discussed before, the authors either did not consider the influences of pressure anisotropy, ion creation (source term), and ion loss (sink term) in magnetized collisional dusty plasmas or considered them inconsistently. Also, several authors considered the dust-charge number arbitrarily and ignored the collision enhancement current in the dust-charging process. In this work, we incorporate these effects self-consistently in a three-dimensional fluid model to study the oblique propagation of weakly nonlinear dust-ion acoustic solitary waves and shocks in laboratory and space dusty plasmas (Saturn's E-ring). While in laboratory dusty plasmas, the presence of ion creation and ion loss (known as active plasma), as well as the ion-neutral and ion-dust collisions, can have a significant influence on the characteristics of DIA waves, they are insignificant in Saturn's E-ring plasmas due to the presence of a strong magnetic field and low-neutral gas density. In both these environments, the effects of the static magnetic field and dust-charge fluctuations become significant. Unlike previous investigations, we calculate the collision enhancement dust-charge number, in the case of laboratory plasmas, from the dust charging equation by the Newton-Raphson numerical scheme. We show that while in laboratory plasmas, DIA solitary waves can appear as compressive types and get damped due to the effects of collisions and adiabatic dust-charge variation, Saturn's E-ring plasmas can support the existence of both compressive and rarefactive shocks due to the presence of the q-nonextensive electrons.  
\par 
We organize the rest of the paper as follows: Section \ref{sec-model} describes the three-dimensional fluid model with the anisotropic pressure components and the dust charging equation. By using the reductive perturbation technique,  we derive the nonlinear evolution equations for DIA solitary waves and shocks in Sec.\ref{sec-evol-eq} in the cases of laboratory and space plasmas.   In Sec. \ref{sec-numeric}, we numerically calculate the equilibrium dust-charge number by Newton-Raphson method and present the profiles of DIA solitary waves and shocks graphically and analyze their properties with different parameters relevant to laboratory and Saturn's E-ring plasmas. Finally, Sec. \ref{sec-conclu} concludes the results.
\\
\section{MODEL AND BASIC EQUATIONS} \label{sec-model}
We consider the three-dimensional propagation of nonlinear DIA waves in a homogeneous magnetized nonextensive dusty plasma consisting of singly charged positive ions (argon ions for laboratory dusty plasmas and molecular oxygen ions for Saturn's E-ring plasmas), $q$-nonextensive electrons, immobile negatively charged dust grains forming only the background plasma, and neutral particles. At equilibrium, the overall charge neutrality condition gives
\begin{equation}
en_{\mathrm{i0}} = en_{\mathrm{e0}} - q_{d0}n_{\mathrm{d0}}
\end{equation}
where $q_d$ is the charge of dust grains with its equilibrium value $q_{d0}$, $n_{j0}$ is the equilibrium number density of $j$-th species particle, where  $j=e,~i$, and $d$, respectively, stand for electrons, positive ions, and charged dust grains. We consider the static magnetic field along the $z$-direction, i.e., ${\mathbf{B}} = B \hat{z}$.  Due to the presence of a strong magnetic field, an anisotropy in ion plasma pressure occurs, and its components along and perpendicular to the magnetic field can have significant impact on the plasma flows.
\par
The basic equations comprising the ion continuity and ion momentum balance equations, and the Poisson equation are 
\begin{equation}
\frac{\partial n_{\mathrm{i}}}{\partial t} + \nabla\cdot\left(n_{\mathrm{i}}{\textbf{v}_{\mathrm{i}}}\right) = {S_{\mathrm{i}}}' - {S_{\mathrm{l}}}',
\label{ion_continuity}
\end{equation}
\begin{eqnarray}
\begin{aligned}
m_{\mathrm{i}}n_{\mathrm{i}}\left(\frac{\partial {\textbf{v}_{\mathrm{i}}}}{\partial t} + \left({\textbf{v}_{\mathrm{i}}}\cdot\nabla \right){\textbf{v}_{\mathrm{i}}}\right) = &-en_{\mathrm{i}}\nabla\phi + e n_{\mathrm{i}}\left({\textbf{v}_{\mathrm{i}}}\times {\textbf{B}}\right)\\
&- m_{\mathrm{i}}n_{\mathrm{i}}\nu_{\mathrm{i}}{\textbf{v}_{\mathrm{i}}}
- \nabla\cdot{\tilde{\textbf{P}}},
\label{eqn_momentum}
\end{aligned}
\end{eqnarray}
\begin{equation}
{\epsilon_0}\nabla^{2}\phi = en_{\mathrm{e}}-en_{\mathrm{i}}-q_{\mathrm{d}}n_{\mathrm{d0}},
\label{eqn_Poisson}
\end{equation}
where ${S_{\textrm{i}}}'$ and ${S_{\mathrm{l}}}'$ represent the ion creation and ion loss terms  respectively; $m_{\mathrm{i}}$, $n_{\mathrm{i}}$, ${\textbf{v}_{\mathrm{i}}}$, and $\tilde{\textbf{P}}$, respectively, are the mass, number density, velocity, and pressure tensor of positive ion species. Also, $\nu_{\textrm{i}}$ is the sum of ion-neutral and ion-dust collision frequencies. The matrix form of the ion pressure tensor for an anisotropic plasma can be written as
\begin{equation*}
{\tilde{\textbf{P}}} = \left(\begin{array}{ccc}
    p_{\mathrm{\perp}} & 0 & 0  \\
    0 & p_{\mathrm{\perp}} & 0 \\
    0 & 0 &  p_{\mathrm{\parallel}},
\end{array}\right)    \label{eq-P}
\end{equation*}
where $p_{\parallel}$ and $p_{\perp}$, respectively, denote the parallel and perpendicular components of the pressure tensor $\tilde{\textbf{P}}$. The anisotropy pressure form \eqref{eq-P}, associated with the magnetic field, can be rewritten as
 \begin{equation}
\tilde{\textbf{P}} =  p_{\perp}{\tilde{\textbf{I}}} + \left(p_{\parallel}-p_{\perp}\right){\hat{{\textbf{B}}}}{\hat{{\textbf{B}}}},
\label{eqn_pressure_tensor}
\end{equation}
where the unit tensor $\tilde{\mathbf{I}}$ and the dyad form of the magnetic field with $\hat{\textbf{B}}$ denoting the corresponding unit vector are
\begin{equation}
{\tilde{\textbf{I}}} = \left(\begin{array}{ccc}
    1 & 0 & 0  \\
    0 & 1 & 0 \\
    0 & 0 &  1
\end{array}\right),~
 {\hat{\textbf{B}}} {\hat{\textbf{B}}}= \left(\begin{array}{ccc}
    0 & 0 & 0  \\
    0 & 0 & 0 \\
    0 & 0 &  1
\end{array}\right).
\end{equation}
 \par
For plasmas with an isotropic pressure, we have  $p_{\perp} = p_{\parallel}$. The pressure equation for anisotropic and adiabatic systems can be obtained by the double adiabatic theory\citep{chew1956boltzmann}. Since we have considered the static magnetic field, the parallel and perpendicular pressure components, based on  the double adiabatic theory, can be written as\citep{choi2007dust}
\begin{equation}
p_{\parallel} = p_{{\parallel}0} \left(n_{\textrm{i}}/n_{\textrm{i}0}\right)^{3},~p_{\perp} = p_{{\perp}0}  {n_{\mathrm{i}}}/n_{\textrm{i}0},
\end{equation}
where the equilibrium pressure components are $p_{{\parallel}0} = k_{\textrm{B}}n_{\mathrm{i}0} T_{\parallel}$ and $p_{{\perp}0} = k_{\textrm{B}}n_{\mathrm{i}0} T_{\perp}$ with $T_{\parallel}$ and $T_{\perp}$ denoting, respectively, the parallel and perpendicular ion temperatures.
\par 
When spherical sized dust particles of radius $r_{\textrm{d}}$ is immersed in a plasma, they get charged by collecting electrons and positive ions into their surfaces. The dust charging equation due to nonextensive electron and thermal ion flows is given by
\begin{equation}
 \frac{dq_{\textrm{d}}}{dt} = I_{\textrm{e}}+ I_{\textrm{Ti}}, 
 \label{eqn_dustcharging}
\end{equation}
where $I_e$ is the current due to nonthermal nonextensive electron flows and $I_{\textrm{Ti}}$ that of thermal positive ions, which incorporates the ion-neutral collision enhancement current. According to orbital motion limited (OML) theory, the electron current is given by \citep{liu2013bohm} 
\begin{equation}
\begin{split}
I_{\mathrm{e}} = &-\pi r^{2}_{\mathrm{d}}en_{\mathrm{e0}}\sqrt{\frac{8k_{\textrm{B}}T_{\textrm{e}}}{\pi m_{\mathrm{e}}}}B_{\mathrm{q}}\\
&\times\left[1+{\left(q-1\right)}\left(\frac{e\phi}{k_{\textrm{B}}T_{\textrm{e}}}+\frac{eq_{\textrm{d}}}{r_{\textrm{d}}k_{\textrm{B}}T_{\textrm{e}}}\right)\right]^{\frac{2q-1}{q-1}},
\label{eqn_electron_current}
\end{split}
\end{equation}
where 
\begin{equation}
B_{\mathrm{q}} =
\begin{cases}
\frac{\left(1-q\right)^{3/2}}{q\left(2q-1\right)}\frac{\Gamma\left(\frac{1}{1-q}\right)} {\Gamma\left(\frac{1}{1-q}-\frac{3}{2}\right)}, \hspace{0.2cm}\left(3/5<q<1\right)       
\\ \\
\frac{\left(q-1\right)^{3/2}\left(3q-1\right)}{2q\left(2q-1\right)}\frac{\Gamma\left(\frac{1}{q-1}+\frac{3}{2}\right)} {\Gamma\left(\frac{1}{q-1}\right)}, \hspace{0.2cm}\left(q>1\right).
\label{eqn_electron_distribution}
\end{cases}
\end{equation}
From Eqs. \eqref{eq-Aq} and \eqref{eqn_electron_distribution}, it is clear that for the validity of the present dusty plasma model with dust-charge fluctuations,  the parameter $q$ for superextensive electrons should be restricted to $3/5<q<1$.  The ion current ($I_{\mathrm{Ti}}$) flowing into the dust-grain surface is the sum of the ion current due to the thermal motion of ions and ion-neutral collision enhancement current and is written as \citep{khrapak2005particle} 
\begin{equation}
I_{\mathrm{Ti}} = \pi r^{2}_{\mathrm{d}}en_{\mathrm{i}}\sqrt{\frac{8k_{\textrm{B}}T_{\mathrm{i}}}{\pi m_{\mathrm{i}}}}\left[1- \frac{eq_{\textrm{d}}}{r_{\textrm{d}}k_{\textrm{B}}T_{\textrm{i}}}+0.4\left(\frac{eq_{\textrm{d}}}{r_{\textrm{d}}k_{\textrm{B}}T_{\textrm{i}}}\right)^{2}\frac{\lambda_{\textrm{D}}}{\lambda_{\textrm{in}}}\right],
\label{eqn_ion_current}
\end{equation}
where $\lambda_{\mathrm{in}}$ = $1/n_{\mathrm{n}}\sigma_{\mathrm{s}}$ is the mean free path for ion-neutral collisions with  $n_{\mathrm{n}}$ denoting the neutral gas density, $\sigma_{\mathrm{s}}$ the  collision cross section,  and $\lambda_{\textrm{D}}$ the dusty plasma screening length, given by,
\begin{equation}
\lambda_{\mathrm{D}} = \lambda_{\mathrm{De}}\left({\frac{1}{\sigma}}+\delta_{\mathrm{e}}\frac{3q-1}{2}\right)^{-1/2}, \label{eq-LD}
\end{equation}
where $\delta_{\textrm{e}} = n_{\textrm{e}0}/n_{\textrm{i}0}$ is the electron-ion density ratio, $\sigma = T_{\textrm{i}}/T_{\textrm{e}}$ is the ion-electron temperature ratio with $T_i=\sqrt{T_\parallel^2+T_\perp^2}$, and $\lambda_{\mathrm{De}} = \left( \epsilon_0k_{\textrm{B}}T_{\textrm{e}}/n_{\mathrm{i0}}e^{2}\right)^{1/2}$ is the electron Debye length. Note that we have considered the approximate dusty plasma screening length $\lambda_D$ on the assumption that electrons are nonthermal ($q$-nonextensive), ions are at local thermal equilibrium, and the screening potential drops off rapidly with $\lambda_D$ as in the case of dusty plasmas. However, the length scale for the excitation of DIA waves is measured with respect to the electron Debye length and the DIA wavelength is much larger than $\lambda_{\rm{De}}$. 
\par 
From Eq. \eqref{eq-LD}, it is clear that the screening length is modified due to the presence of nonextensive (with parameter $q$) electrons in plasmas and the pressure anisotropy (via $\sigma$). The screening length gets enhanced as the value of $q$ is reduced, i.e., as one enters from thermal ($q\rightarrow1$) to nonthermal plasma states with superextensive electrons $(0.6<q<1)$. Physically, highly energetic electrons will tend to escape from the Debye sphere, and as a result, the inter-particle distance, through which the charged particles get initially separated, will be increased, and the plasma cloud will be less dense. However, an opposite behavior occurs, i.e., the screening length gets reduced when either the electron number density increases or the ion temperature decreases. In either case, the plasma cloud will become denser to neutralize the excess electrons or ions, and the inter-particle distance between the particles will be reduced and the Debye sphere will be more compact. 
\par
 As said before, dust particles immersed in electron-ion plasmas can acquire charge due to electron and ion flows into their surface. Due to the preferential attachments of more mobile electrons than ions, they can be negatively charged. Otherwise, when the ion flow rate into the dust grain surface is higher than electrons, dust particles can be positively charged.  However,  fluctuations in electron (ion) concentration can affect the flux of electrons (ions) sticking to the dust grains and thus can cause the dust charging levels to fluctuate. The fluctuating electron and ion currents are 
\begin{equation}
\begin{split}
{I}_{\mathrm{e}} =& -\pi r^{2}_{\mathrm{d}}en_{\mathrm{e0}}\sqrt{\frac{8k_{\textrm{B}}T_{\textrm{e}}}{\pi m_{\mathrm{e}}}}B_{\mathrm{q}}\\
&\times\left[1+{\left(q-1\right)}\left(\psi+ Z\left(\Delta Q -1\right)\right)\right]^{(2q-1)/(q-1)},
\label{eqn_electron_current}
\end{split}
\end{equation}
\begin{equation}
{I}_{\mathrm{Ti}} = \pi r^{2}_{\mathrm{d}}en_{\mathrm{i}}\sqrt{\frac{8T_{\mathrm{i}}}{\pi m_{\mathrm{i}}}}\left[1- \frac{Z}{\sigma}\left(\Delta Q -1\right) +\frac{0.4\lambda_{\mathrm{D}}}{\lambda_{\mathrm{in}}} \frac{Z^{2}\left(\Delta Q-1\right)^{2}}{\sigma^{2}}\right],
\label{eqn_ion_current}
\end{equation}
where $\psi = e\phi/k_{\textrm{B}}T_{\textrm{e}}$ is the normalized electrostatic potential, $Z = Z_{\textrm{d}0}e^2/r_{\textrm{d}}k_{\textrm{B}}T_{\textrm{e}}$ is the normalized dust-charge parameter, and $\Delta Q = q_{\mathrm{d1}}/eZ_{\mathrm{d}0}$ is the normalized perturbed charge with $Z_{\mathrm{d}0}$ denoting the equilibrium dust-charge number. Typically, ions are lost due to ion-dust collisions.  Also, the total ion current per unit charge (${I}_{\mathrm{Ti}} / e$) determines the flux of ions to the dust grains. Thus, the ion loss rate (${S_{\mathrm{l}}}'$) is determined by the number of ions lost per unit volume per unit time, i.e., ${S_{\mathrm{l}}}' = I_{\mathrm{Ti}}n_{\mathrm{d0}}/e$. The normalized ion loss rate is obtained as
\begin{equation*}
S_{\mathrm{l}} = \frac{{S_{\mathrm{l}}}'}{\omega_{\mathrm{pi}}n_{\mathrm{i0}}} = \frac{I_{\mathrm{Ti}}n_{\mathrm{d0}}}{en_{\mathrm{i0}}\omega_{\mathrm{pi}}},
\end{equation*} 
which yields;
\begin{equation}
S_{\mathrm{l}}=\frac{\nu_{\mathrm{a}}n_{\mathrm{i}}}{n_{\textrm{i}0}}\left[1-\frac{Z\Delta Q}{\beta\sigma}+\frac{0.4Z^2\lambda_{\textrm{D}}}{\sigma^2 \lambda_{\textrm{in}}\beta}\Delta Q\left(\Delta Q-2\right)\right],
\label{eqn_ionloss}    
\end{equation}
where $\beta = 1+Z/\sigma+0.4Z^2\lambda_{\textrm{D}}/\sigma^2\lambda_{\textrm{in}}$, $\nu_{\mathrm{a}} \left(= \nu_{\mathrm{l}}/\omega_{\textrm{pi}}\right)$ is the ion loss rate normalized by the ion-plasma oscillation frequency, $\omega_{\mathrm{pi}}  \left(= n_{\mathrm{i}0}e^{2}/\epsilon_0 m_{\mathrm{i}}\right)^{1/2}$ with $\nu_{\mathrm{l}} \left( = n_{\mathrm{d0}}I_{\mathrm{Ti0}}/en_{\mathrm{i0}}\right)$ satisfying the following:
\begin{equation}
\nu_{\textrm{l}} = \frac{r_{\textrm{d}}}{\sqrt{{2k_{\textrm{B}}T_{\textrm{i}}}/{\pi m_{\textrm{i}}}}}\frac{\sigma}{Z}\omega_{\textrm{pi}}^{2}\beta \left(1-\delta_{\textrm{e}}\right).
\label{eqn_ion_loss_rate}
\end{equation}
We recast Eqs. \eqref{ion_continuity}-\eqref{eqn_Poisson} after making the variables dimensionless and separating the components along the axes as follows:   
\begin{equation}
N_{\mathrm{e}} = \delta_{\mathrm{e}}\left[1+ \left(q-1\right)\psi\right]^{\left(3q-1\right)/2\left(q-1\right)},
\label{eqn_normlaized_electron_density}
\end{equation}
\begin{equation}
\frac{\partial N_{\mathrm{i}}}{\partial t} + \frac{\partial\left(N_{\mathrm{i}}u_{\mathrm{i}x}\right)}{\partial x} + \frac{\partial\left(N_{\mathrm{i}}u_{\mathrm{i}y}\right)}{\partial y}+\frac{\partial\left(N_{\mathrm{i}}u_{\mathrm{i}z}\right)}{\partial z} = S_{\mathrm{i}} -S_{\mathrm{l}},
\label{eqn_conti_normalized}
\end{equation}
\begin{equation}
\begin{aligned}
\frac{\partial{u_{\mathrm {i}x}}}{\partial t} + u_{\mathrm{i}x}\frac{\partial u_{\mathrm{i}x}}{\partial x} + u_{\mathrm{i}y}\frac{\partial u_{\mathrm{i}x}}{\partial y}+u_{\mathrm{i}z}\frac{\partial u_{\mathrm{i}x}}{\partial z} = -\frac{\partial \psi}{\partial x} + \omega_{\mathrm{i}} u_{\mathrm{i}y} &\\ - u_{\mathrm{i}x}k_{\mathrm{i0}} -  \frac{P_{\perp}}{N_{\mathrm{i}}}\frac{\partial N_{\mathrm{i}}}{\partial x},
\label{eqn_xcomp_normalized}
\end{aligned}
\end{equation}
\begin{equation}
\begin{aligned}
\frac{\partial{u_{\mathrm {i}y}}}{\partial t} + u_{\mathrm{i}x}\frac{\partial u_{\mathrm{i}y}}{\partial x} +u_{\mathrm{i}y}\frac{\partial u_{\mathrm{i}y}}{\partial y}+ u_{\mathrm{i}z}\frac{\partial u_{\mathrm{i}y}}{\partial z} =  -\frac{\partial \psi}{\partial y}-\omega_{\mathrm{i}} u_{\mathrm{i}x} &\\-u_{\mathrm{i}y}k_{\mathrm{i0}}- \frac{P_{\perp}}{N_{\mathrm{i}}}\frac{\partial N_{\mathrm{i}}}{\partial y},
\label{eqn_ycomp_normalized}
\end{aligned}
\end{equation}
\begin{equation}
\frac{\partial{u_{\mathrm {i}z}}}{\partial t} + u_{\mathrm{i}x}\frac{\partial u_{\mathrm{i}z}}{\partial x} + u_{\mathrm{i}y}\frac{\partial u_{\mathrm{i}z}}{\partial y}+ u_{\mathrm{i}z}\frac{\partial u_{\mathrm{i}z}}{\partial z} = -\frac{\partial \psi}{\partial z}  - u_{\mathrm{i}z}k_{\mathrm{i0}} - {P_{\parallel }}{N_{\mathrm{i}}}\frac{\partial N_{\mathrm{i}}}{\partial z},
\label{eqn_zcomp_normalized}
\end{equation}
\begin{equation}
\frac{\partial^{2}\psi}{\partial x^{2}} +     \frac{\partial^{2}\psi}{\partial y^{2}}+     \frac{\partial^{2}\psi}{\partial z^{2}} = N_{\mathrm{e}} - N_{\mathrm{i}} - Z_{\mathrm{d0}}\delta_{\mathrm{d}}\Delta Q  + Z_{\mathrm{d0}}\delta_{\mathrm{d}},
\label{eqn_Poisson_Normalized}
\end{equation}
where $S_{\mathrm{i}} ={S_{\mathrm{i}}}'/\omega_{\mathrm{pi}}n_{\mathrm{i}0}$ is the normalized ion creation term, given by, \citep{ghosh2005effects}
\begin{equation*}
S_{\mathrm{i}} = \nu_{\mathrm{a}}\left[1+\frac{\Delta \sigma}{\sigma_0 }\psi +\frac{1}{2\sigma_0}\left(\frac{d^{2}\sigma}{d\psi^{2}}\right)_{0} \psi^{2} + \cdots\right].
\label{eqn_ioncreation_normalized}
\end{equation*}
Here, $\Delta \sigma = \left(d\sigma/d\psi\right)_0$, $\sigma_0$ is the ionization cross-section at $\psi = 0$, $\omega_{\mathrm{i}} = eB/m_{\mathrm{i}}\omega_{\mathrm{pi}}$ is the strength of normalized magnetic field, $\delta_{\textrm{d}} = n_{\textrm{d}0}/n_{\textrm{i}0}$ is the dust-ion number density ratio, and $P_{\perp} = p_{\perp 0}/n_{\mathrm{i0}}k_{\textrm{B}}T_{\textrm{e}}$ and $P_{\parallel} = 3p_{\parallel 0}/n_{\mathrm{i0}}k_{\textrm{B}}T_{\textrm{e}}$ are, respectively, the normalized perpendicular and parallel ion pressure components. Also, $N_{\mathrm{e,i}}$ is the electron (ion) number density normalized by $n_{\mathrm{i0}}$  and $u_{\mathrm{i}x}$,  $u_{\mathrm{i}y}$, $u_{\mathrm{i}z}$ are the velocity components normalized by the ion-acoustic speed $C_{\mathrm{s}}\left(= \sqrt{k_{\textrm{B}}T_{\textrm{e}}/m_{\mathrm{i}}}\right)$. Furthermore, the time coordinate  $t$ is normalized by $\omega_{\mathrm{pi}}^{-1}$ and the space coordinates $x$, $y$, and $z$ by the electron Debye length $\lambda_{De}$. The term $k_{\mathrm{i0}}$ $\left(=k_{\mathrm{in0}} + k_{\mathrm{id0}}\right)$ is the sum of ion-neutral and ion-dust collision frequencies normalized by $\omega_{\mathrm{pi}}$.
\par
The normalized form of Eq. (\ref{eqn_dustcharging}) can be obtained as
\begin{equation}
\frac{1}{\sqrt{\mu_{\mathrm{d}}\left(1-\delta_{\mathrm{e}}\right)}} \frac{\omega_{\mathrm{pd}}}{\nu_{\mathrm{ch}}}\frac{\partial \Delta Q}{\partial t} = \frac{1}{\nu_{\mathrm{ch}}eZ_{\mathrm{d0}}} \left(I_{\mathrm{Ti}} + I_{\mathrm{e}}\right),
\label{eqn_dust_charging_normalized}
\end{equation}
where $\mu_{\textrm{d}} = Z_{\textrm{d}0}m_{\textrm{i}}/m_{\textrm{d}}$ with $\omega_{\mathrm{pd}}$ denoting the dust-plasma  oscillation frequency. The dust-charging frequency $\nu_{\mathrm{ch}}$ is given by\citep{tsytovich1997dust} 
\begin{equation}
\nu_{\mathrm{ch}} = -\frac{1}{eZ_{\mathrm{d}0}}\left[\frac{\partial \left(I_{\mathrm{Ti}} + I_{\mathrm{e}}\right)}{\partial \Delta Q}\right]_{\psi = 0,\hspace{0.1cm} \Delta Q = 0}, 
\end{equation}
which yields  
\begin{equation}
\nu_{\mathrm{ch}} = \frac{r_{\mathrm{d}}\omega^{2}_{\mathrm{pi}}}{\sqrt{2\pi k_{\textrm{B}}T_{\textrm{i}}/m_{\textrm{i}}}}\left(1+\chi+{Z}'+\frac{0.4\lambda_{\mathrm{D}}Z\left(2+Z'\right)}{\lambda_{\mathrm{in}}\sigma}\right),
\end{equation}
where $\chi = \sigma q_{\textrm{z}}$ and ${Z}' = Zq_{\textrm{z}}$ with $q_{\textrm{z}} = \left(2q-1\right)/\left(1-Z\left(q-1\right)\right)$. In particular, for dusty plasmas with Maxwell-Boltzmann distributed electrons and in the absence of collision enhancement ion current, one recovers the dust-charging frequency as in the previous work \citep{ghosh2000nonlinear}. 
\section{NONLINEAR EVOLUTION EQUATIONS} \label{sec-evol-eq}
To derive the evolution equation for DIA waves in a collisional dusty plasma with dust-charge fluctuation, we employ the standard reductive perturbation technique \citep{washimi1966propagation} in which the space and time coordinates are stretched as
\begin{eqnarray}
\begin{aligned}
\xi &= \epsilon^{1/2}\left(l_{x}x +l_{y}y  + l_{z}z - \text{v}_{\textrm{s}} t\right), \\
\tau &= \epsilon^{3/2}t,
\label{eqn_stretched_coodinates}
\end{aligned}
\end{eqnarray}
where $\text{v}_{\textrm{s}}$ is the frame velocity or the wave phase velocity normalized by the ion-acoustic speed $C_s$ and $\epsilon$ ($0<\epsilon<1$) is a scaling parameter measuring the weakness of the amplitudes of perturbations. Also, $l_x,~l_y$, and $l_z$ are the direction cosines of the wave vector such that $l_x^2+l_y^2+l_z^2=1$. By applying the transformations, Eqs. (\ref{eqn_conti_normalized}) - (\ref{eqn_dust_charging_normalized}) reduce to
\begin{equation}
\begin{aligned}
-\epsilon^{1/2}\text{v}_{\textrm{s}}\frac{\partial N_{\mathrm{i}}}{\partial \xi} + \epsilon^{3/2}\frac{\partial N_{\mathrm{i}}}{\partial\tau} + \epsilon^{1/2}l_{{x}}\frac{\partial}{\partial \xi}\left(N_{\mathrm{i}}u_{\mathrm{i}x}\right) + \epsilon^{1/2}l_{{y}}\frac{\partial}{\partial \xi}\left(N_{\mathrm{i}}u_{\mathrm{i}y}\right) \\ 
+ \epsilon^{1/2}l_{{z}}\frac{\partial}{\partial \xi}\left(N_{\mathrm{i}}u_{\mathrm{i}z}\right) 
=  S_{\mathrm{i}} - S_{\mathrm{l}},
\label{eqn_conti_stretched}
\end{aligned}
\end{equation}
\begin{equation}
\begin{aligned}
-\epsilon^{1/2}\text{v}_{\textrm{s}} \frac{\partial u_{\mathrm{i}x}}{\partial \xi} + \epsilon^{3/2}\frac{\partial u_{\mathrm{i}x}}{\partial \tau} + \epsilon^{1/2}l_{x}u_{\mathrm{i}x}\frac{\partial u_{\mathrm{i}x}}{\partial \xi}+ \epsilon^{1/2}l_{y}u_{\mathrm{i}y}\frac{\partial u_{\mathrm{i}x}}{\partial \xi} \\ +  \epsilon^{1/2}l_{z}u_{\mathrm{i}z}\frac{\partial u_{\mathrm{i}x}}{\partial \xi} = -\epsilon^{1/2}l_{x}\frac{\partial \psi}{\partial \xi} +  \omega_{\mathrm{i}} u_{\mathrm{i}y} -  u_{\mathrm{i}x}k_{\mathrm{i0}}- \epsilon^{1/2}\frac{l_{x}P_{\perp}}{N_{\mathrm{i}}}\frac{\partial N_{\mathrm{i}}}{\partial \xi},
\label{eqn_xcomp_stretched}
\end{aligned}
\end{equation}
\begin{equation}
\begin{aligned}
-\epsilon^{1/2}\text{v}_{\textrm{s}} \frac{\partial u_{\mathrm{i}y}}{\partial \xi} + \epsilon^{3/2}\frac{\partial u_{\mathrm{i}y}}{\partial \tau} + \epsilon^{1/2}l_{x}u_{\mathrm{i}x}\frac{\partial u_{\mathrm{i}y}}{\partial \xi}+ \epsilon^{1/2}l_{y}u_{\mathrm{i}y}\frac{\partial u_{\mathrm{i}y}}{\partial \xi} \\+ \epsilon^{1/2}l_{z}u_{\mathrm{i}z}\frac{\partial u_{\mathrm{i}y}}{\partial \xi} =  -\epsilon^{1/2}l_{y}\frac{\partial \psi}{\partial \xi}- \omega_{\mathrm{i}} u_{\mathrm{i}x} - u_{\mathrm{i}y}k_{\mathrm{i0}}- \epsilon^{1/2}\frac{l_{y}P_{\perp}}{N_{\mathrm{i}}}\frac{\partial N_{\mathrm{i}}}{\partial \xi},
\label{eqn_ycomp_stretched}
\end{aligned}
\end{equation}
\begin{equation}
\begin{aligned}
-\epsilon^{1/2}\text{v}_{\textrm{s}} \frac{\partial u_{\mathrm{i}z}}{\partial \xi} + \epsilon^{3/2}\frac{\partial u_{\mathrm{i}z}}{\partial \tau} + \epsilon^{1/2}l_{x}u_{\mathrm{i}x}\frac{\partial u_{\mathrm{i}z}}{\partial \xi}+ \epsilon^{1/2}l_{y}u_{\mathrm{i}y}\frac{\partial u_{\mathrm{i}z}}{\partial \xi}\\ +  \epsilon^{1/2}l_{z}u_{\mathrm{i}z}\frac{\partial u_{\mathrm{i}z}}{\partial \xi} = -\epsilon^{1/2}l_{z}\frac{\partial \psi}{\partial \xi} -   u_{\mathrm{i}z}k_{\mathrm{i0}} -  \epsilon^{1/2}{l_{z}P_\parallel N_{\mathrm{i}}}\frac{\partial N_{\mathrm{i}}}{\partial \xi},  
\label{eqn_ycomp_stretched}
\end{aligned}
\end{equation}
\begin{equation}
\epsilon \frac{\partial^{2}\psi}{\partial\xi^{2}} = N_{\mathrm{e}} - N_{\mathrm{i}} - Z_{\mathrm{d0}}\delta_{\mathrm{d}}\Delta Q + Z_{\mathrm{d0}}\delta_{\mathrm{d}},
\label{eqn_Poisson_stretched}
\end{equation}
\begin{equation}
\begin{aligned}
\frac{1}{\sqrt{\mu_{\mathrm{d}}\left(1-\delta_{\mathrm{e}}\right)} } \frac{\omega_{\mathrm{pd}}}{\nu_{\mathrm{ch}}}\left(-\epsilon^{1/2}\text{v}_{\textrm{s}}\frac{\partial \Delta Q}{\partial \xi} + \epsilon^{3/2}\frac{\partial \Delta Q}{\partial \tau}\right) \\= \frac{1}{Z_{\mathrm{d0}}e\nu_{\mathrm{ch}}}\left(I_{\mathrm{Ti}} + I_{\mathrm{e}}\right).
\label{eqn_dust_charging_stretched}  
\end{aligned}
\end{equation}
The dependent variables are expanded as \citep{adnan2014effect}
\begin{eqnarray}
\begin{aligned}
 N_{\mathrm{i}} &= 1 + \epsilon N_{\mathrm{i}}^{\left(1\right)} +\epsilon^{2}N_{\mathrm{i}}^{\left(2\right)} + \cdots, \\
 u_{\mathrm{i}x,\mathrm{i}y}& = \epsilon^{3/2}u_{\mathrm{i}x,\mathrm{i}y}^{\left(1\right)} + \epsilon^{2}u_{\mathrm{i}x,\mathrm{i}y}^{\left(2\right)} + \cdots, \\
 u_{\mathrm{i}z} & = \epsilon u_{\mathrm{i}z}^{\left(1\right)} + \epsilon^{2}u_{\mathrm{i}z}^{\left(2\right)} + \cdots, \\
 \psi& = \epsilon \psi^{\left(1\right)} + \epsilon^{2}\psi^{\left(2\right)} + .....\\
 \Delta Q& = \epsilon \Delta Q^{\left(1\right)} + \epsilon^{2}\Delta Q^{\left(2\right)} + \cdots
 \label{eqn_dependent_variables}
 \end{aligned}
\end{eqnarray}
In the presence of a magnetic field, the ion-gyro motion is treated as a higher-order effect and so the perturbations for the transverse components of the ion fluid velocity $u_{\textrm{i}x}$ and $u_{\textrm{i}y}$ in Eq. \eqref{eqn_dependent_variables} are considered at higher order of $\epsilon$ than the parallel component  $u_{\textrm{i}z}$. To make the nonlinear perturbations consistent with Eqs. (\ref{eqn_stretched_coodinates}) and (\ref{eqn_dependent_variables}) based on observed values \citep{wang2001ionization}, we further consider the following ordering for  $k_{\textrm{i}0}$ and $\nu_{\textrm{a}}$.

\begin{eqnarray}
\begin{aligned}
 k_{\mathrm{i}0} = \epsilon^{3/2}k_{\mathrm{i}}, \\
 \nu_{\mathrm{a}} = \epsilon^{3/2}g_{\mathrm{a}},
 \label{eqn_collision_stretched}
\end{aligned}
\end{eqnarray}
where $k_{\mathrm{i}} (= k_{\mathrm{in}}+k_{\mathrm{id}})$ and $g_a$ are of the order of unity.
\par 
In what follows, we substitute the expansions (\ref{eqn_dependent_variables}) and (\ref{eqn_collision_stretched}) into Eqs. (\ref{eqn_conti_stretched})-(\ref{eqn_Poisson_stretched}) and equate different powers of $\epsilon$. In the lowest order of $\epsilon$, we obtain
\begin{equation}
N_{\mathrm{i}}^{\left(1\right)} = \frac{l_z}{\text{v}_{\textrm{s}}}u_{\mathrm{i}z}^{\left(1\right)},
\label{eqn_ion_perturbed}
\end{equation}
\begin{equation}
u_{\mathrm{i}x}^{\left(1\right)} = -\frac{l_{y}}{\omega_{\mathrm{i}}}\left(\frac{\partial \psi^{\left(1\right)}}{\partial\xi} + P_\perp\frac{\partial N_{\mathrm{i}}^{\left(1\right)}}{\partial \xi}\right),
\label{eqn_xcomp_velocity_perturbed}
\end{equation}
\begin{equation}
u_{\mathrm{i}y}^{\left(1\right)} = \frac{l_{x}}{\omega_{\mathrm{i}}}\left(\frac{\partial \psi^{\left(1\right)}}{\partial\xi} + P_\perp\frac{\partial N_{\mathrm{i}}^{\left(1\right)}}{\partial \xi}\right),
\label{eqn_ycomp_velocity_perturbed}
\end{equation}
\begin{equation}
u_{\mathrm{i}z}^{\left(1\right)} = \frac{l_z}{\text{v}_{\textrm{s}}}\left(\psi^{\left(1\right)} + P_\parallel N_{\mathrm{i}}^{\left(1\right)}\right),
\label{eqn_zcomp_velocity_perturbed}
\end{equation}
\begin{equation}
0 = \frac{3q-1}{2}\delta_{\mathrm{e}}\psi^{\left(1\right)} - N_{\mathrm{i}}^{\left(1\right)}-Z_{\mathrm{d0}}\delta_{\mathrm{d}}\Delta Q^{\left(1\right)}.
\label{eqn_Poisson_perturbed}
\end{equation}
Equations (\ref{eqn_ion_perturbed}) and (\ref{eqn_zcomp_velocity_perturbed}) give the following relation between the first-order perturbations of the ion density and the electrostatic potential.
\begin{equation}
N_{\mathrm{i}}^{\left(1\right)} = \frac{{l_{\textrm{z}}}^2}{\text{v}_{\textrm{s}}^2-{l_{\textrm{z}}}^2 P_{\parallel}}\psi^{(1)}
\label{eqn_ion_perturbed_1}
\end{equation}
In the next order of $\epsilon$, i.e., $\epsilon^{2}$,  we obtain from Eqs. (\ref{eqn_conti_stretched})-(\ref{eqn_Poisson_stretched}) the following equations.
\begin{equation}
\begin{split}
&-\text{v}_{\textrm{s}}\frac{\partial N_{\mathrm{i}}^{\left(2\right)}}{\partial\xi} + \frac{\partial N_{\mathrm{i}}^{\left(1\right)}}{\partial\tau} + l_x\frac{\partial u_{\mathrm{i}x}^{\left(2\right)}}{\partial\xi}\\
&+ l_y\frac{\partial u_{\mathrm{i}y}^{\left(2\right)}}{\partial\xi} +  l_z\frac{\partial u_{\mathrm{i}z}^{\left(2\right)}}{\partial\xi} + l_z\frac{{\partial \left(N_{\mathrm{i}}^{\left(1\right)}u_{\mathrm{i}z}^{\left(1\right)}\right)}}{\partial \xi}  \\
&= g_{\mathrm{a}}\left[\frac{\Delta \sigma}{\sigma_0}\psi^{\left(1\right)}-N_{\mathrm{i}}^{\left(1\right)}+{\frac{Z}{\beta\sigma}\Delta Q^{\left(1\right)}+\frac{0.8Z^{2}\lambda_{\mathrm{D}}}{\beta\sigma^{2}\lambda_{\mathrm{in}}}}\Delta Q^{\left(1\right)}\right],
\label{eqn_momentum_1}
\end{split}
\end{equation}
\begin{equation}
u_{\mathrm{i}x}^{\left(2\right)} =\frac{\text{v}_{\textrm{s}}}{\omega_{\mathrm{i}}}\frac{\partial u_{\mathrm{i}y}^{\left(1\right)}}{\partial \xi},
\label{eqn_momentum_2}
\end{equation}
\begin{equation}
u_{\mathrm{i}y}^{\left(2\right)} =-\frac{\text{v}_{\textrm{s}}}{\omega_{\mathrm{i}}}\frac{\partial u_{\mathrm{i}x}^{\left(1\right)}}{\partial \xi},
\label{eqn_momentum_2a}
\end{equation}
\begin{equation}
\begin{aligned}
-\text{v}_{\textrm{s}}\frac{\partial u_{\mathrm{i}z}^{\left(2\right)}}{\partial\xi} + \frac{\partial u_{\mathrm{i}z}^{\left(1\right)}}{\partial \tau} + l_z u_{\mathrm{i}z}^{\left(1\right)}\frac{\partial u_{\mathrm{i}z}^{\left(1\right)}}{\partial\xi} = -l_z\frac{\partial \psi^{\left(2\right)}}{\partial \xi} -k_{\mathrm{i}}u_{\mathrm{i}z}^{\left(1\right)}\\ - {l_{z}P_\parallel}N_{\mathrm{i}}^{\left(1\right)}\frac{\partial N_{\mathrm{i}}^{\left(1\right)}}{\partial \xi} - {l_{z}P_\parallel}\frac{\partial N_{\mathrm{i}}^{\left(2\right)}}{\partial \xi},
\label{eqn_momentum_3}
\end{aligned}
\end{equation}
\begin{equation}
\begin{aligned}
\frac{\partial^{2}\psi^{\left(1\right)}}{\partial \xi^{2}} = \delta_{\mathrm{e}}\frac{3q-1}{2}\psi^{\left(2\right)} + \delta_{\mathrm{e}}\frac{\left(3q-1\right)\left(q+1\right)}{8}\psi^{\left(1\right)2}& \\  - N_{\mathrm{i}}^{\left(2\right)} 
- Z_{\mathrm{d0}}\delta_{\mathrm{d}}\Delta Q^{\left(2\right)}.
\label{eqn_Poisson_2nd_order}
\end{aligned}
\end{equation} 
\par
In the following two subsections \ref{sec-sub-evol-eq-lab} and \ref{sec-sub-evol-eq-space}, we derive the evolution equations for DIA waves in two different plasma environments, namely laboratory and space (Saturn's E-ring) plasmas in the limits of adiabatic and non-adiabatic dust-charge fluctuations. We note that while the collisional effects and the adiabatic dust-charge variation become significant in laboratory plasmas, the collision effects in space plasmas are rather insignificant but non-adiabatic dust-charge variation may be relevant there.  
\subsection{{Laboratory collisional magnetized dusty plasma}} \label{sec-sub-evol-eq-lab}
\vspace{-0.5cm}
In this subsection, we consider two extreme limits for the dust-charge variation effects. When $\nu_{\mathrm{ch}}\ll\omega_{\mathrm{pd}}$, i.e., the dust-charging time is much longer than the hydrodynamic time scale, Eq.   \eqref{eqn_dust_charging_normalized} gives $\partial \Delta Q/\partial t\approx0$, yielding $\Delta Q$ a constant. In this approximation, the plasma is effectively a two-component electron-ion plasma and thus is of less interest to the present study. In the opposite limit, i.e.,  $\nu_{\mathrm{ch}} \gg \omega_{\mathrm{pd}}$, the dust-charge variation may be neglected compared to the time scale of plasma oscillations, and Eq.   \eqref{eqn_dust_charging_normalized} reduces to     
\begin{equation}
 I_{\mathrm{Ti}} + I_{\mathrm{e}} = 0.
    \label{eqn_dustcharge_equilibrium}
\end{equation}
 This approximation is well-known as the adiabatic dust-charge variation. Thus, equating the lowest order of $\epsilon$, from Eq. (\ref{eqn_dust_charging_stretched}), we get 
\begin{equation}
\Delta Q^{\left(1\right)} = \beta_{\mathrm{d}}\left(N_{\mathrm{i}}^{\left(1\right)}-q_{\mathrm{z}}\psi^{\left(1\right)}\right),
    \label{eqn_dust_charge_fluctuation_1}
\end{equation}
where $\beta_{\textrm{d}}$ is given by
\begin{eqnarray*}
\beta_{\mathrm{d}} = \chi_{1}\left({Z+\sigma+{0.4Z^{2}\lambda_{\mathrm{D}}}/{\sigma \lambda_{\mathrm{in}}}}\right),
\end{eqnarray*}
with 
\begin{eqnarray*}
\chi_{1} = \left[{Z\left(1+\chi+Z'+\frac{0.4Z\lambda_{\mathrm{D}}\left(2+Z'\right)}{\sigma\lambda_{\mathrm{in}}}\right)}\right]^{-1}.
\end{eqnarray*}
From Eqs. (\ref{eqn_Poisson_perturbed}), (\ref{eqn_ion_perturbed_1}), and (\ref{eqn_dust_charge_fluctuation_1}), we obtain the following expression for the phase velocity of long-wavelength DIA perturbations. 
\begin{equation}
\text{v}_{\textrm{s}} = l_z \left[P_{\parallel}+\frac{1+Z_{\textrm{d}0}\beta_{\textrm{d}}\delta_{\textrm{d}}}{\delta_{\textrm{e}}(3q-1)/2+Z_{\textrm{d}0}\beta_{\textrm{d}}\delta_{\textrm{d}}q_{\textrm{z}}}\right]^{1/2}.
\label{eqn_dispersion}
\end{equation}
From Eq. \eqref{eqn_dispersion}, it is evident that the expression for the phase velocity is modified by the plasma pressure (See $P_\parallel$ and the contribution of $P_\perp$ through $\sigma$ in $\beta_d$), the nonextensive distribution of electrons $(q)$, the contribution from charged dust grains $(Z_{d0})$, and the obliqueness of wave propagation $(l_z=\cos\theta)$. Here, $\theta$ is the angle between the magnetic field and the wave vector. It is also noted that in the absence of charged dust grains and ion thermal pressure, one can recover the phase velocity $(v_s)$ as the ion-acoustic velocity $(C_s)$ of ion-plasma oscillations propagating along the magnetic field in a Maxwellian electron-ion plasma ($q\rightarrow1$). Furthermore, the adiabatic dust-charge variation introduces an additional term proportional to $Z_{d0}$ that couples to the contributions from the ion pressure and the nonthermality of electrons.  We will discuss more about the characteristics of $v_s$ in Sec. \ref{sec-numeric}.
\par 
By considering the coefficients  of $\epsilon^{2}$, from Eq. (\ref{eqn_dust_charging_stretched}) we obtain
\begin{equation}
\begin{aligned}
\Delta Q^{\left(2\right)} =  \beta_{\mathrm{d}}\left(N_{\mathrm{i}}^{\left(2\right)}- q_{\mathrm{z}}\psi^{\left(2\right)}\right)  -  \chi_2 N_{\mathrm{i}}^{\left(1\right)}\Delta Q^{\left(1\right)}  - \chi_3 \psi^{\left(1\right)^{2}} \\ -\chi_4 \psi^{\left(1\right)}\Delta Q^{\left(1\right)}  + \chi_5 \Delta Q^{\left(1\right)^{2}},
\label{eqn_dust_charge_fluctuation_2}
\end{aligned}
\end{equation}
where the coefficients $\chi_{2}$, $\chi_{3}$, $\chi_{4}$, and $\chi_{5}$ are given by
\begin{align*}
\chi_{2} = \frac{Z\beta_{\mathrm{d}}\left(1+{0.8Z\lambda_{\mathrm{D}}}/{\sigma\lambda_{\mathrm{in}}}\right)}{\sigma+Z+{0.4Z^{2}\lambda_{\mathrm{D}}}/{\sigma\lambda_{\mathrm{in}}}},\hspace{0.2cm}
\chi_{3} = \frac{0.5qq_{\mathrm{z}}\beta_{\textrm{d}}}{\left(1-Z(q-1)\right)},    
\end{align*}
\begin{align*}
\chi_{4} = \frac{qq_{\textrm{z}}Z\beta_{\mathrm{d}}}{\left(1-Z\left(q-1\right)\right)},
\end{align*}
\begin{align*}
\chi_{5} = Z^{2}\chi_{1}  
\left[\frac{0.4\lambda_{\mathrm{D}}}{\sigma\lambda_{\mathrm{in}}}-\left(Z'+\chi+\frac{0.4\lambda_{\mathrm{D}}ZZ'}{\sigma\lambda_{\mathrm{in}}}\right)\frac{0.5}{1-\left(q-1\right)Z}\right].
\end{align*}
Finally, eliminating  ${\partial u_{\mathrm{i}x}^{\left(2\right)}}/{\partial\xi}$, ${\partial u_{\mathrm{i}y}^{\left(2\right)}}/{\partial\xi}$, ${\partial u_{\mathrm{i}z}^{\left(2\right)}}/{\partial\xi}$, ${\partial\psi^{\left(2\right)}}/{\partial\xi}$, ${\partial N_{\mathrm{i}}^{\left(2\right)}}/{\partial\xi}$, and ${\partial \Delta Q^{\left(2\right)}}/{\partial\xi}$, and using Eq. (\ref{eqn_Poisson_2nd_order}), we obtain the following modified damped KdV equation for the evolution of DIA solitary waves in a magnetized collisional dusty plasma with adiabatic dust-charge variation. 
\begin{equation}
\frac{\partial\psi^{\left(1\right)}}{\partial\tau} + A_{1} \psi^{\left(1\right)}\frac{\partial\psi^{\left(1\right)}}{\partial\xi} + B_{1}\frac{\partial^{3}\psi^{\left(1\right)}}{\partial \xi^{3}} + C_{1}\psi^{\left(1\right)} = 0,
\label{eqn_kdv}
\end{equation}
where the nonlinear, dispersion, and damping coefficients, respectively, are $A_{1} = b_{1}/a_{1},~
B_{1} = c_{1}/a_{1}$, and $C_{1} = d_{1}/a_{1}$. Typically, while the dispersion effect causes wave broadening, the nonlinear term gives rise to wave steepening. The damping term appears due to the effects of ion creation and ion loss as well as ion-dust and ion-neutral collisions. The expressions for $a_1,~b_1,~c_1$, and $d_1$ are
\begin{eqnarray}
a_{1} = 2l_z^{2}\textrm{v}_{\text{s}}\frac{\left(1+Z_{\mathrm{d0}}\beta_{\mathrm{d}}\delta_{\mathrm{d}}\right)}{{\left(\textrm{v}_{\text{s}}^{2}-{l_{z}^{2}}P_\parallel\right)^{2}}},
\end{eqnarray}
\begin{eqnarray}
b_{1} = -2\chi_6 + l_z^{4}\left(1+Z_{\mathrm{d0}}\beta_{\mathrm{d}}\delta_{\mathrm{d}}\right)\frac{\left(3\textrm{v}_{\text{s}}^{2}+P_\parallel l_z^{2}\right)}{\left(\textrm{v}_{\text{s}}^{2}-l_z^{2}P_{\parallel}\right)^{3}},
\end{eqnarray}
\begin{equation}
\begin{split}
c_{1} = 1+ \left(1+Z_{\mathrm{d0}}\beta_{\mathrm{d}}\delta_{\mathrm{d}}\right)\left(1-l_{z}^{2}\right)&\left(\frac{\textrm{v}_{\text{s}}}{\omega_{\textrm{i}}\left(\textrm{v}_{\text{s}}^{2}-l_z^{2}P_\parallel\right)}\right)^{2}\\
&\times\left(\textrm{v}_{\text{s}}^{2}+l_z^{2}(P_\perp-P_\parallel)\right),
\end{split}
\end{equation}
\begin{equation}
\begin{split}
d_{1} = &\frac{\textrm{v}_{\text{s}}\left(1+Z_{\mathrm{d0}}\beta_{\mathrm{d}}\delta_{\mathrm{d}}\right)}{\left(\textrm{v}_{\text{s}}^{2}-{l_{z}^{2}}P_\parallel\right)}\\
&\times\left[g_{\mathrm{a}}\left(-\frac{\Delta\sigma}{\sigma_0}+\frac{l_z^{2}}{\textrm{v}_{\text{s}}^{2}-l_z^{2}P_\parallel}- \chi_7 \right)+\frac{l_z^{2}}{\text{v}_{\textrm{s}}^{2}-l_z^{2}P_{\parallel}}k_{\textrm{i}}\right],
\end{split}
\end{equation}
\begin{equation}
\begin{split}
\chi_{6} &= \frac{\delta_{\textrm{e}}\left(3q-1\right)\left(q+1\right)}{8}\\
& + Z_{\textrm{d}0}\delta_{\textrm{d}}\chi_{3} + Z_{\textrm{d}0}\delta_{\textrm{d}}\beta_{\textrm{d}}\left(\frac{l_z^{2}}{\textrm{v}_{\text{s}}^{2}-l_z^{2}P_\parallel}-q_{\textrm{z}}\right)\\  
&\times\left[\frac{\chi_2 l_z^{2}}{\textrm{v}_{\text{s}}^{2}-l_z^{2}P_\parallel}+\chi_4-\beta_{\textrm{d}}\chi_{5}\left(\frac{l_z^{2}}{\textrm{v}_{\text{s}}^{2}-l_z^{2}P_\parallel}-q_{\textrm{z}}\right)\right], 
\end{split}
\end{equation}
\begin{equation}
\chi_{7} = \frac{Z\beta_{\textrm{d}}}{\beta\sigma}\left(\frac{l_z^{2}}{\textrm{v}_{\text{s}}^{2}-l_z^{2}P_\parallel}-q_{\textrm{z}}\right)\left(1+\frac{0.8Z\lambda_{\textrm{D}}}{\sigma\lambda_{\textrm{in}}}\right).
\end{equation}
\par 
A traveling wave (solitary) solution of Eq. (\ref{eqn_kdv}) can be obtained by transforming the independent variables $\xi$ and $\tau$ to $\eta = \xi - M\tau$  with $M$ ($M=M_0$ at $\tau = \tau_{0}$,  the initial time) denoting the Mach number normalized by the ion-acoustic speed $C_{\mathrm{s}}$, and using the 
boundary conditions, namely, $\psi^{\left(1\right)}$ $\to 0$, $d\psi^{\left(1\right)} / d\eta$ $\to$ 0, and $d^{2}\psi^{\left(1\right)} / d\eta^{2}$ $\to$ 0 as $\eta \to \pm \infty$, as 
  \citep{chatterjee2018analytical}
\begin{equation}
\psi^{\left(1\right)} \left(\xi, \tau\right) = \psi_{\mathrm{m}}^{\left(1\right)} \left(\tau \right) {\text{sech}}^{2}\left(\frac{\xi-M\left(\tau\right)\tau}{\Delta_{\mathrm{w}}\left(\tau\right)}\right),
\label{eqn_analytical_presence}
\end{equation}
where the time-dependent amplitude, the width, and the Mach number of the solitary solution are
\begin{equation}
\psi_{\mathrm{m}}^{\left(1\right)} \left(\tau\right) = 3M\left(\tau\right)/A_{1},
\label{eqn_amplitude_timedependent}
\end{equation}
\begin{equation}
\Delta_{\mathrm{w}}\left(\tau\right) = \sqrt{4B_{1}/M\left(\tau\right)},
\label{eqn_width_timedependent}
\end{equation}
\begin{equation}
M\left(\tau\right) = M_0\left(\tau_0\right){\text{exp}}\left(-\frac{4}{3}C_{1}\left(\tau-\tau_0\right)\right).
\label{eqn_timedependent_mach}
\end{equation}
The soliton energy and the perturbed electric field are  given by
\begin{equation}
I = \int_{-\infty}^{\infty}\left[\psi^{\left(1\right)} \left(\xi, \tau \right)\right]^{2} d\xi=\frac{4}{3}\left[\psi_{\mathrm{m}}^{\left(1\right)}\right]^2\Delta_{\mathrm{w}},
\label{eqn_energy_Integral}
\end{equation}
\begin{equation}
\frac{d\psi^{\left(1\right) }\left(\xi, \tau\right)}{d\xi} = \frac{2}{\Delta_{\mathrm{w}}\left(\tau\right)}\psi^{\left(1\right)}\left(\xi, \tau\right) \text{tanh}\left(\frac{\xi-M\left(\tau\right)\tau}{\Delta_{\mathrm{w}}\left(\tau\right)}\right).
\label{eqn_perturbed_electric_field}
\end{equation}
We find that the Mach number $(M)$ decays with time $\tau>\tau_0$, leading to the decay of the soliton amplitude but an increase in the width. Thus, due to the effects of the adiabatic dust-charge variation and collisions, the DIA solitons will tend to lose energy as time progresses. Since $M$ is directly proportional to the soliton amplitude, but inversely to the width, solitons moving with higher (lower) velocities may be bigger (smaller) and narrower (wider). Also, since the soliton energy  $I$ is directly dependent on the squared amplitude and the width, DIA solitons with higher amplitudes/widths would propagate with higher energies in collisional nonthermal dusty magnetoplasmas at a given time. We also find that for the parameters relevant to laboratory plasmas, the nonlinear coefficient $A_1$ is always positive, implying that the damped DIA solitary waves will evolve as compressive types with positive potential. A detailed study on the characteristics of damped DIA solitons will be carried out in Sec. \ref{sec-numeric}.
\subsection{Magnetized Saturn's E-ring dusty plasma} \label{sec-sub-evol-eq-space}
In contrast to laboratory collisional dusty plasmas, the deviation of the extreme limits of adiabatic dust-charge variation may be more relevant in space plasma environments. In this situation, one can assume either $\nu_{\mathrm{ch}}/\omega_{\mathrm{pd}}<1$   or  $\omega_{\mathrm{pd}}/\nu_{\mathrm{ch}}<1$. In the former case, the dust-charge variation can be assumed to be a higher-order effect, i.e., it may not contribute to the linear dispersion relation, but to the wave damping in the nonlinear evolution of DIA waves. In the latter case, the dust-charge does not instantaneously reach its equilibrium value and it can plays a dissipative role on the propagation of DIA waves. This approximation is known as the non-adiabatic dust-charge variation. Depending on the order of smallness of the above frequency ratios, one can obtain different evolution equations for DIA waves. Typically, the ordering  $\nu_{\mathrm{ch}}/\omega_{\mathrm{pd}}\sim\epsilon^{3/2}$ is used to obtain the damped KdV equation, while the KdV Burgers equation can be obtained with $\omega_{\mathrm{pd}}/\nu_{\mathrm{ch}}\sim\epsilon^{1/2}$ [See, e.g., Ref. \onlinecite{gupta2001effect}]. To be consistent with the present perturbation expansion scheme,  
we consider Case I: ${{\nu_{\mathrm{ch}}}\sqrt{\mu_{\mathrm{d}}\left(1-\delta_{\mathrm{e}}\right)}}/{\omega_{\mathrm{pd}}} \sim{\nu_{\text{d}}}\epsilon^{3/2}$ to obtain the modified damped KdV equation and Case II: ${\omega_{\mathrm{pd}}}/{{\nu_{\mathrm{ch}}}\sqrt{\mu_{\mathrm{d}}\left(1-\delta_{\mathrm{e}}\right)}}  \sim  {\nu_{\text{d}}}\epsilon^{1/2}$ for the KdV Burgers equation. Here, the parameter ${\nu_{\text{d}}}\sim o(1)$  measures the dust-charge fluctuation rate. 
\subsubsection{Case I: Damped KdV equation} 
To obtain the modified damped KdV equation, we consider the ordering as in Case I. Thus, from Eq. (\ref{eqn_dust_charging_stretched}), we obtain  
\begin{equation}
\left(-\epsilon^{1/2}\text{v}_{\mathrm{s}}\frac{\partial \Delta Q}{\partial \xi} + \epsilon^{3/2}\frac{\partial \Delta Q}{\partial \tau}\right) = \frac{\epsilon^{3/2}\nu_{\text{d}}}{Z_{\mathrm{d0}}e\nu_{\mathrm{ch}}}\left(I_{\mathrm{Ti}} + I_{\mathrm{e}}\right)
\label{eqn_mkdv_dust_charge}
\end{equation}
Next, equating the coefficient of $\epsilon$, from Eq. (\ref{eqn_mkdv_dust_charge}), we get 
\begin{equation}
\Delta Q^{(1)} = 0
\label{eqn_nonadiabatic_first}
\end{equation}
Thus, the reduced frame velocity of DIA solitary waves in Saturn's E-ring dusty plasmas can be obtained from Eq. (\ref{eqn_Poisson_perturbed}) as 
\begin{equation}
\text{v}_{\textrm{s}} = l_z \sqrt{P_{\parallel} + \frac{2}{\delta_{\textrm{e}}\left(3q-1\right)}}. 
\label{eqn_frame_mkdv}
\end{equation}
By comparing Eq. \eqref{eqn_frame_mkdv} with Eq. \eqref{eqn_dispersion}, we see that the charged dust particles do not have any influence on the frame velocity $\rm{v}_s$. Setting $Z_{d0}=0$ in Eq. \eqref{eqn_dispersion}, gives the same result as in \eqref{eqn_frame_mkdv}.   
\par  
Next, by equating the coefficients of $\epsilon^{2}$ from equation (\ref{eqn_mkdv_dust_charge}), we obtain
\begin{equation}
\frac{\partial \Delta Q^{(2)}}{\partial \xi} = -\frac{\nu_{\textrm{d}}\left(Z+\sigma\right)}{\text{v}_{\textrm{s}}Z\left(1+\chi+Z'\right)}\left(\frac{l_z^{2}}{\text{v}_{\textrm{s}}^{2}-l_z^{2}P_{\parallel}}-q_{\textrm{z}}\right)\psi^{(1)}.
\label{eqn_nonadiabatic_second}
\end{equation}
Proceeding in the same way as for Eq. \eqref{eqn_kdv}, we obtain the following damped KdV equation for the evolution of DIA solitary waves.
\begin{equation}
\frac{\partial\psi^{\left(1\right)}}{\partial\tau} + A_{2} \psi^{\left(1\right)}\frac{\partial\psi^{\left(1\right)}}{\partial\xi} + B_{2}\frac{\partial^{3}\psi^{\left(1\right)}}{\partial \xi^{3}} +C_{2}\psi^{\left(1\right)} = 0,
\label{eqn_mkdv_nonadiabatic}
\end{equation}
where $A_{2} = a_{2}/b_{2}$, $B_{2} = c_{2}/b_{2}$, and $C_{2} = d_{1}/b_{2}$ with
\begin{equation}
a_{2} = -\frac{\delta_{\textrm{e}}(3q-1)(q+1)}{4} + \frac{l_z^{4}(3\text{v}_{\textrm{s}}^{2}+l_z^{2}P_{\parallel})}{\left(\text{v}_{\textrm{s}}^{2}-l_z^{2}P_{\parallel}\right)^{3}},
\end{equation}
\begin{equation}
b_{2} = \frac{2l_z^{2}\text{v}_{\textrm{s}}}{\left(\text{v}_{\textrm{s}}^{2}-l_z^{2}P_{\parallel}\right)^{2}}, 
\end{equation}
\begin{equation}
c_{2} = 1+(1-l_z^{2})\left(\frac{\text{v}_{\textrm{s}}}{\omega_{\textrm{i}}(\text{v}_{\textrm{s}}^{2}-l_z^{2}P_{\parallel})}\right)^{2}\left(\text{v}_{\textrm{s}}^{2}+l_z^{2}(P_{\perp}-P_{\parallel})\right),
\end{equation}
\begin{equation}
d_{2} = -\frac{Z_{\textrm{d}0}\delta_{\textrm{d}}\nu_{\textrm{d}}(Z+\sigma)}{\text{v}_{\textrm{s}}Z(1+Z'+\chi)}\left(\frac{l_z^{2}}{\text{v}_{\textrm{s}}^{2}-l_z^{2}P_{\parallel}}-q_{\textrm{z}}\right).    
\end{equation}
We note that although Eq. \eqref{eqn_mkdv_nonadiabatic} has the same form as Eq. \eqref{eqn_kdv},  the nonlinear and the dispersion coefficients get modified due to the deviation from the extreme limiting condition of the adiabatic dust-charge variation. Such modifications will also modify the characteristics of damped DIA solitons compared to the case of adiabatic dust-charge variation. By comparing the expressions for $a_1$ and $b_2$, we also note that the damping coefficient $C_2$ is modified in the absence of the term proportional to $Z_{d0}$ in $b_2$ (Compare with $a_1$). It follows that the damping rate of DIA waves may be higher compared to the results due to adiabatic dust-charge variation. We find that in contrast to Eq. \eqref{eqn_kdv}, the nonlinear coefficient $A_2$ may assume both positive and negative values depending on the parameter $q$. Typically, $A_2<0$ corresponding to the superextensive region $(0.6<q<1)$ and  $A_2>0$ as $q\rightarrow1$. Thus, deviation from the extreme limit of the adiabatic dust-charge variation may result into the existence of both compressive and rarefactive damped DIA solitons in nonthermal nonextensive plasmas. 
\par 
In what follows, we obtain a similar analytic (traveling wave)  solution of Eq. ({\ref{eqn_mkdv_nonadiabatic}}) as
\begin{equation}
\psi^{\left(1\right)} \left(\xi, \tau\right) = \psi_{\mathrm{sm}}^{\left(1\right)} \left(\tau \right) {\text{sech}}^{2}\left(\frac{\xi-M_{\textrm{s}}\left(\tau\right)\tau}{\Delta_{\mathrm{sw}}\left(\tau\right)}\right),
\label{eqn_analytical_presencea}
\end{equation}
where the time-dependent modified amplitude and width of the solitary wave solution are 
\begin{equation}
\psi_{\mathrm{sm}}^{\left(1\right)} \left(\tau\right) = 3M_{\textrm{s}}\left(\tau\right)/A_{2},
\label{eqn_amplitude_nonadiabatic}
\end{equation}
\begin{equation}
\Delta_{\mathrm{sw}}\left(\tau\right) = \sqrt{4B_{2}/M_{\textrm{s}}\left(\tau\right)}.
\label{eqn_width_nonadiabatic}
\end{equation}
The temporal evolution of Mach number is found to be
\begin{equation}
M_{\textrm{s}}\left(\tau\right) = M_0\left(\tau_0\right){\text{exp}}\left(-\frac{4}{3}C_{2}\left(\tau-\tau_0\right)\right).
\label{eqn_nonadiabatic_mach}
\end{equation}
As in Sec. \ref{sec-sub-evol-eq-lab}, the Mach number $(M)$ decays with time $\tau>\tau_0$, but the decay rate can be higher than that predicted for Eq. \eqref{eqn_kdv} due to the larger damping coefficient $C_2~(>C_1)$. In this case, one can also have a similar decay of the soliton amplitude but an increase in the width, and decay of the soliton energy as time progresses.   A detailed study on the characteristics of the damped DIA solitons will be given in Sec. \ref{sec-numeric}.
\subsubsection{Case II: KdV Burgers equation} 
We consider the non-adiabatic dust-charge variation (Case II) to obtain an evolution equation (KdV Burgers) for DIA waves. Thus, from Eq. (\ref{eqn_dust_charging_stretched}), we have  
\begin{equation}
\nu_{\text{d}}\epsilon^{1/2} \left(-\epsilon^{1/2}\text{v}_{\mathrm{s}}\frac{\partial \Delta Q}{\partial \xi} + \epsilon^{3/2}\frac{\partial \Delta Q}{\partial \tau}\right) = \frac{1}{Z_{\mathrm{d0}}e\nu_{\mathrm{ch}}}\left(I_{\mathrm{Ti}} + I_{\mathrm{e}}\right).
\label{eqn_nonadiabatic_dust_charge}
\end{equation} 
Next, equating the coefficients of $\epsilon$ and $\epsilon^{2}$, from Eq. (\ref{eqn_nonadiabatic_dust_charge})  we get 
\begin{equation}
\Delta Q^{\left(1\right)} = \beta_{\mathrm{d}}\left(N_{\mathrm{i}}^{\left(1\right)}-q_{\textrm{z}}\psi^{\left(1\right)}\right),  
\end{equation}
\begin{equation}
\begin{aligned}
\Delta Q^{\left(2\right)} = \text{v}_{\mathrm{s}} \nu_{\text{d}} \frac{\partial \Delta Q^{\left(1\right)}}{\partial \xi}+\beta_{\mathrm{d}}\left(N_{\mathrm{i}}^{\left(2\right)}- q_{\mathrm{z}}\psi^{\left(2\right)}\right)  -  \chi_2 N_{\mathrm{i}}^{\left(1\right)}\Delta Q^{\left(1\right)}  \\ -  \chi_3 \psi^{\left(1\right)^{2}} - \chi_4 \psi^{\left(1\right)}\Delta Q^{\left(1\right)}  + \chi_5 \Delta Q^{\left(1\right)^{2}}.
\label{eqn_dust_fluctuation_nonadiabatic}
\end{aligned}
\end{equation}
After eliminating ${\partial u_{\mathrm{i}x}^{\left(2\right)}}/{\partial\xi}$, ${\partial u_{\mathrm{i}y}^{\left(2\right)}}/{\partial\xi}$, ${\partial u_{\mathrm{i}z}^{\left(2\right)}}/{\partial\xi}$, ${\partial\psi^{\left(2\right)}}/{\partial\xi}$, ${\partial N_{\mathrm{i}}^{\left(2\right)}}/{\partial\xi}$, and ${\partial \Delta Q^{\left(2\right)}}/{\partial\xi}$, and using Eq. (\ref{eqn_Poisson_2nd_order}), we obtain the following modified damped KdV Burgers equation for DIAWs.
\begin{equation}
\frac{\partial\psi^{\left(1\right)}}{\partial\tau} + A_{1}\psi^{\left(1\right)} \frac{\partial\psi^{\left(1\right)}}{\partial\xi} + B_{1}\frac{\partial^{3}\psi^{\left(1\right)}}{\partial \xi^{3}} + C_{1}\psi^{\left(1\right)} = D_{1} \frac{\partial^{2}\psi^{\left(1\right)}}{\partial \xi^{2}},
\label{eqn_mkdv}
\end{equation}
where the coefficients $A_{1}$, $B_{1}$ and $C_{1}$ are given by Eq. (\ref{eqn_kdv}), and the dissipation coefficient, $D_{1}$ (Coefficient of the Burgers term) is given by
\begin{equation}
D_{1} = -\frac{Z_{\mathrm{d0}}\delta_{\mathrm{d}}\beta_{\mathrm{d}} \nu_{\text{d}}}{2l_z^{2}\left(1+Z_{\textrm{d}0}\beta_{\textrm{d}}\delta_{\textrm{d}}\right)}\left(\frac{l_z^{2}}{\text{v}_{\mathrm{s}}^{2}-l_z^{2}P_\parallel}-q_{\textrm{z}}\right)\left(\text{v}_{\mathrm{s}}^{2}-l_z^{2}P_\parallel\right)^{2}.
\label{eqn_dissipation}
\end{equation}
From Eq. \eqref{eqn_dissipation}, it is evident that because of the explicit dependence of $D_1$ on $Z_{d0}$ and $\nu_d$, the Burgers term appears due to the non-adiabatic dust-charge variation and gives rise to the evolution of DIA shocks. 
Typically, in space dusty plasmas (especially Saturn's E-ring plasmas), the effects of ion creation and ion loss, and the ion-neutral and ion-dust collisions are insignificant compared to the dissipation effect ($\propto D_1$) due to the non-adiabatic dust-charge fluctuations \citep{yaroshenko2007dust}. Thus, neglecting the term proportional to $C_{1}$, Eq. (\ref{eqn_mkdv}) reduces to
\begin{equation}
\frac{\partial\psi^{\left(1\right)}}{\partial\tau} + A_{1}\psi^{\left(1\right)} \frac{\partial\psi^{\left(1\right)}}{\partial\xi} + B_{1}\frac{\partial^{3}\psi^{\left(1\right)}}{\partial \xi^{3}} = D_{1} \frac{\partial^{2}\psi^{\left(1\right)}}{\partial \xi^{2}}.
\label{eqn_burger}
\end{equation}
\par
An analytic  shock wave solution of the modified Burgers equation (\ref{eqn_burger}) can be obtained by transforming the independent variables $\xi$ and $\tau$ to $\eta = \gamma \left(\xi - V\tau\right)$, where $\gamma^{-1} \left( = \Delta_{\mathrm{s}}\right)$ is the shock spatial width and $V$ is the shock speed. Under the transformation Eq. (\ref{eqn_burger})  reduces to
\begin{eqnarray}
-V\frac{d\psi^{\left(1\right)}}{d\eta} + A_{1}\psi^{\left(1\right)}\frac{d\psi^{\left(1\right)}}{d\eta} + B_{1}\gamma^{2}\frac{d^{3}\psi^{\left(1\right)}}{d\eta^{3}} = D_{1}\gamma\frac{d^{2}\psi^{\left(1\right)}}{d\eta^{2}}.
\label{eqn_transform}
\end{eqnarray}
Next, using the boundary conditions $\psi^{\left(1\right)}$ $\to 0$, $d\psi^{\left(1\right)} / d\eta$ $\to$ 0, and $d^{2}\psi^{\left(1\right)} / d\eta^{2}$ $\to$ 0 as $\eta \to \pm \infty$, and integrating Eq. (\ref{eqn_transform}) once, we get 
\begin{equation}
-V\psi^{\left(1\right)} + \frac{A_{1}}{2}\psi^{\left(1\right)^{2}} + B_{1}\gamma^{2}\frac{d^{2}\psi^{\left(1\right)}}{d\eta^{2}} = D_{1}\gamma \frac{d\psi^{\left(1\right)}}{d\eta}.
\label{eqn_modified}
\end{equation}
By employing the hyperbolic tangent (tanh) method, a traveling wave solution of Eq. (\ref{eqn_modified}) can be obtained in the following form \citep{demiray2004travelling}
\begin{equation}
\psi^{\left(1\right)} \left(\xi, \tau\right) = m_{0} + m_{1}W - m_{2}W^{2},
\label{eqn_hyperbolic}
\end{equation}
where $W = \tanh\eta$, and $m_{0}$, $m_{1}$, and $m_{2}$ are the constants to be determined from the solution of equation (\ref{eqn_modified}). 
\par 
Next, we Differentiate Eq. (\ref{eqn_hyperbolic}) with respect to $\eta$ and substitute the result into Eq. (\ref{eqn_modified}). After setting the coefficients of the same powers of $W$ to zero, we obtain the following DIA shock wave solution. 
\begin{eqnarray}
\psi^{\left(1\right)} \left(\xi, \tau\right) = \psi^{\left(1\right)}_{\mathrm{s}} \left[1-
\frac{1}{4}\left(1+{\mathrm{tanh}}\left(\frac{\xi-V\tau}{{\Delta_{\mathrm{s}}}}\right)\right)^{2}\right],
\label{eqn_hypertangent}
\end{eqnarray}
where the shock amplitude, shock width, and shock speed are given by
\begin{equation}
\psi^{\left(1\right)}_{\mathrm{s}} = \frac{12}{25}\frac{D_{1}^{2}}{A_{1}B_{1}}, \hspace{0.1cm} \Delta_{\mathrm{s}} = 10\frac{B_{1}}{D_{1}}\hspace{0.1cm}\textrm{and}\hspace{0.1cm}V = \frac{6}{25}\frac{D_{1}^{2}}{B_{1}}.
\label{eqn_burger_solution}
\end{equation} 
\par 
In Sec. \ref{sec-numeric}, we numerically investigate the characteristics of the phase speed or the frame speed $v_s$, the profiles of the damped DIA solitary waves and DIA shocks for parameters relevant for laboratory and space plasma environments. 
\section{NUMERICAL RESULTS AND DISCUSSIONS} \label{sec-numeric}
\subsection{Laboratory collisional magnetized dusty plasma} \label{sec-sub-numeric-lab}
Typical plasma parameters relevant for laboratory collisional magnetized dusty plasmas are \citep{wang2001ionization, ferdousi2015oblique, choi2007dust,pramanik2002experimental, vladimirov1999ion}: the ion plasma density  $n_{\mathrm{i}0}$ = 10$^{16}$ m$^{-3}$, the static magnetic field  $B= 0.43-4.33$ T, the dust number density  $n_{\mathrm{d}0} = 10^{10}$ m$^{-3}$, the mass density of dust particle $\rho_{\mathrm{d}}= 2000$ kg/m$^{3}$, the electron temperature  $k_BT_{\textrm{e}} = 2.0$ eV, the mass of an argon ion $m_{\mathrm{i}} = 40.0$ amu, the positive ion temperature  $k_BT_{\textrm{i}} = 0.40$ eV, the neutral gas pressure $= 1.0-56.0$ Pa, the normalized ion-dust collision rate  $k_{\text{id}} = 0.01-0.6$, the collision cross-section of an argon gas $\sigma_{\mathrm{s}} = 42.1 \times 10^{-20}$ m$^{2}$, the obliqueness of wave propagation  $\theta$ = 0 - 30$^{\circ}$, the degree of electron nonextensivity $q = 0.65 -0.999$ [$q\rightarrow1$ corresponds to Boltzmann distribution (BD)], and $\Delta \sigma / \sigma_0 = 10$. The normalized time $\tau$ is retained fixed ($\tau = 1$) for the plots of DIA profiles unless mentioned. To obtain the ion-neutral collision enhancement equilibrium dust-charge number $Z_{\mathrm{d}0}$ for a given set of physical parameters, we have solved the dust-charging equation $I_{\mathrm{Ti0}} + I_{\mathrm{e0}} = 0$ by using the Newton-Raphson method. 
\par
\subsubsection{Dust-charge number $(Z_{d0})$}
\begin{figure}[!h]
\centering
\includegraphics[width=8.5cm]{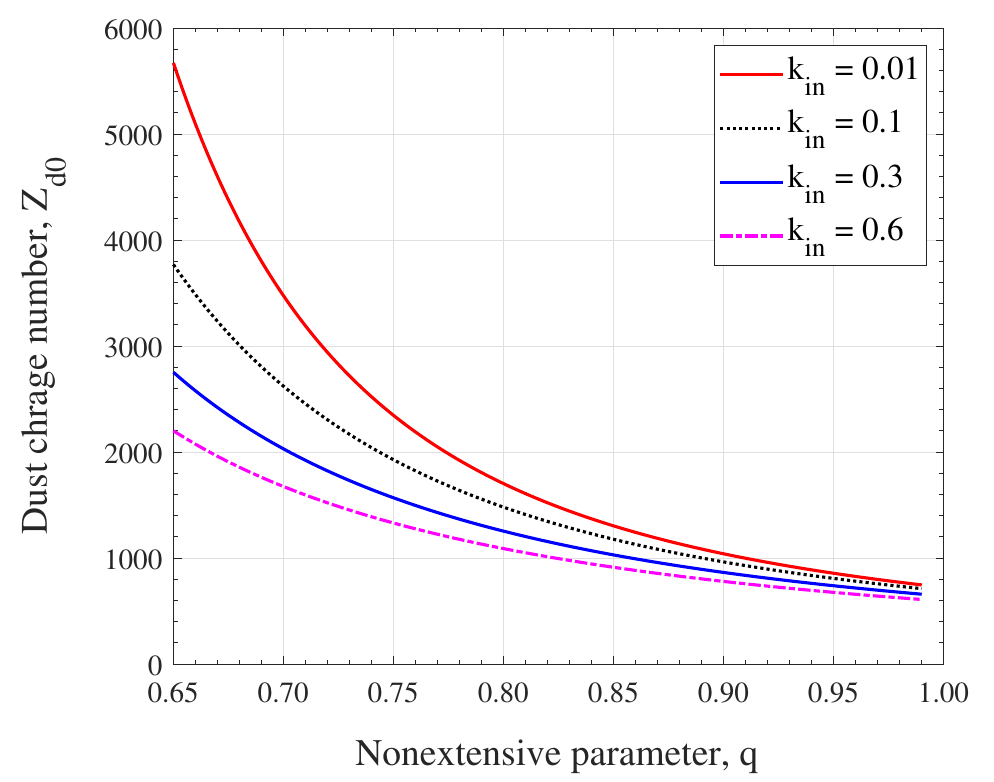}
\caption{The variation of the equilibrium dust-charge number [Eq. (\ref{eqn_dustcharge_equilibrium})] is shown against the nonextensive parameter $q$ for different values of the ion-neutral collision rate as in the legend. The fixed values are $P_\parallel = 0.1$ and $P_\perp = 0.2$.} 
\label{fig_1}
\end{figure}

The ion-neutral collisions in plasmas influence the flux of positive ions flowing into the charged dust grain surfaces and thus influence the evolution of dust charge. The effects of the ion-neutral collision and the nonextensive parameter $q$ on the equilibrium dust-charge number are shown in Fig. \ref{fig_1}. On increasing the nonextensive parameter $q$ of electrons, i.e., as one approaches from a region of nonthermal superextensive electrons to a region with thermal electrons [with $q\rightarrow1$, corresponding to the Boltzmann distribution (BD) of electrons], the dust-charge number is found to be reduced. In the superextensive region with retaining the parameter $q~(<1)$ fixed, as the ion-neutral collision frequency is increased, the dust-charge number is also significantly reduced.   Physically, on increasing the ion-neutral collision, the positive ions suffer several collisions with neutral atoms, which slow down the flow of fast ions to the dust grain surfaces, resulting in the reduction of the dust-charge number (See the line for $k_{in}=0.6$). The decreasing rate of dust-charge number is relatively slow in the region of increasing $q$ towards $1$ (i.e., approaching to the Boltzmann distribution). 
\par
\begin{figure}[!h]
\centering
\includegraphics[width=8.5cm]{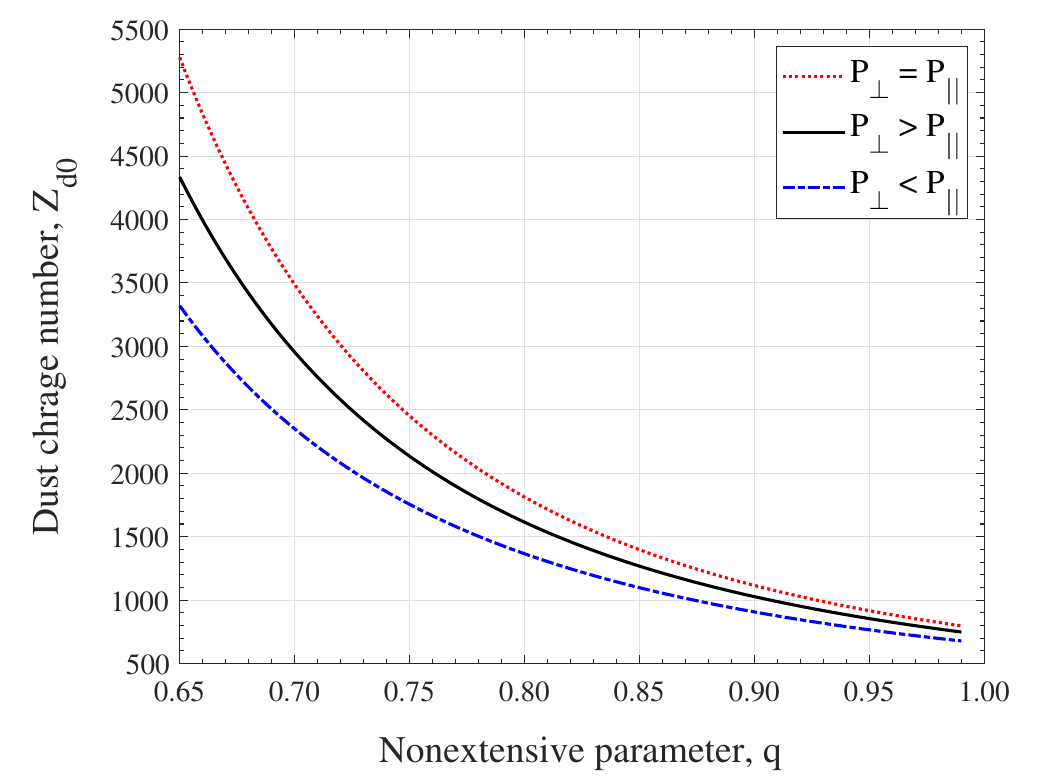}
\caption{The variation of the equilibrium dust charge number [Eq. (\ref{eqn_dustcharge_equilibrium})] is shown against the nonextensive parameter $q$ for three different cases of ion pressure: $P_\perp = P_\parallel = 0.5$, $P_\perp\hspace{0.1cm}(0.3) > P_\parallel\hspace{0.1cm}(0.1)$, and $P_\perp \hspace{0.1cm}(0.1) < P_\parallel\hspace{0.1cm}(0.3)$ with a constant ion-neutral collision rate $k_{\mathrm{in}} = 0.1$.} 
\label{fig_2}
\end{figure}

 
To exhibit the influences of the plasma pressure anisotropy on the dust-charge number, we note that for strongly magnetized plasmas, the plasma particles can have two possible motions: (i) motion along the magnetic field (or parallel motion) and (ii) motion transverse to the magnetic field (or perpendicular motion). At a strong magnetic field, the thermal pressure of positive ions becomes anisotropic and can have two components:  parallel ion pressure $(P_{\parallel})$ and perpendicular ion pressure $(P_{\perp})$. Moreover, the magnetic field restricts the free streaming of positive ions by forcing them to gyrate in the Larmor's orbit. Consequently, the ion flux towards the dust grain surface gets affected and so is the dust-charge state. The effects of anisotropic ion thermal pressure on the evolution of dust-charge are shown in Fig. \ref{fig_2}.  It is found that the dust-charge number achieves a maximum value in the isotropic case $(P_{\parallel}=P_{\perp})$ compared to the case of anisotropic pressure $(P_{\parallel}\neq P_{\perp})$, and a minimum value when the parallel pressure dominates over the perpendicular one. For example, when the parameter $q$ approaches from the region of nonthermal superextensive electrons to thermal Boltzmann distributed electrons $(q\rightarrow1)$, the equilibrium dus-charge number is found to be $797$ (approx) for an isotropic ion pressure, while it is noted to be approximately $748$ and $678$ for anisotropic cases.
\vspace{-0.2cm}
\subsubsection{Phase velocity $(\rm{v}_s)$}
\begin{figure}[!h]
\centering
\includegraphics[width=8.5cm]{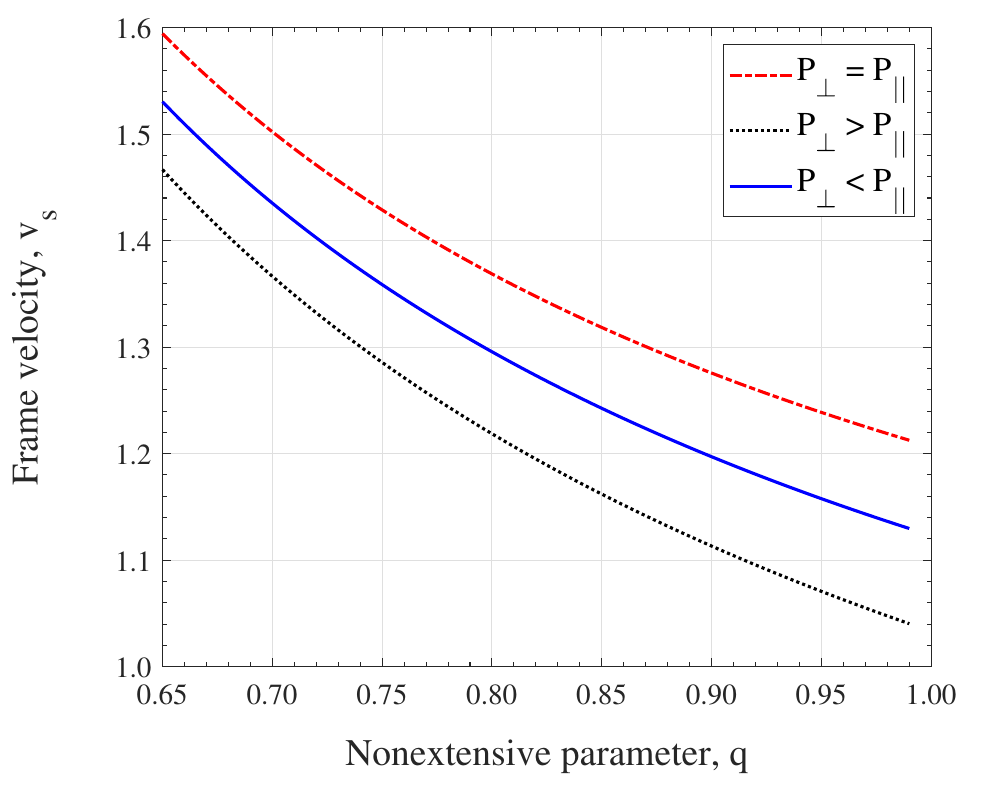}
\caption{The variation of the phase velocity [Eq. (\ref{eqn_dispersion})] is shown against the nonextensive parameter $q$ for three different cases of the ion pressure: $P_{\perp}$ = $P_{\parallel} = 0.5$, $P_\perp\hspace{0.1cm}(0.3) > P_\parallel\hspace{0.1cm}(0.1)$, and $P_\perp \hspace{0.1cm}(0.1) < P_\parallel\hspace{0.1cm}(0.3)$. The fixed parameter values are   $k_{\mathrm{in}} = 0.1$,  
$\omega_{\mathrm{i}}$ = 0.1,   $q$ = 0.7,  $\theta$ = 10$^{\circ}$,   $k_{\mathrm{id}} = 0.01$,   $M_{0} = 0.05$, and   $g_{\mathrm{a}} = \nu_{\mathrm{l}}/\omega_{\mathrm{pi}}$.} 
\label{fig_3}
\end{figure}

 The dust-charge number and its variation with different parameters (See Figs. \ref{fig_1} and \ref{fig_2}) explicitly determines and modifies the frame velocity or the phase velocity of DIA solitary waves. The effects of the anisotropic ion pressure on the frame velocity as a function of the nonextensive parameter $q$ is depicted in Fig. \ref{fig_3}. It is found that the frame velocity lies in the supersonic domain in the whole ranges of the nonextensive parameter $q$, and it decreases as $q$ approaches from superextensive region  $(0.6<q<1)$ to the Maxwellian one $(q\rightarrow1)$ in the cases of both isotropic and anisotropic ion pressures. From Eq. (\ref{eqn_dispersion}), the frame velocity has a linear relationship with the parallel pressure and so it increases with an increase of the parallel pressure. The frame velocity attains a minimum value for the anisotropic case,  and it reaches a maximum value when the isotropic thermal pressure is considered. For example, the maximum value of $v_s$ (for isotropic case) decreases from about $1.59$ to $1.21$ as the parameter $q$ increases from $0.65$ to $0.999$ (close to BD).
\subsubsection{Soliton energy and characteristics of damped solitons}
\begin{figure}[!h]
\centering
\includegraphics[width=8.5cm]{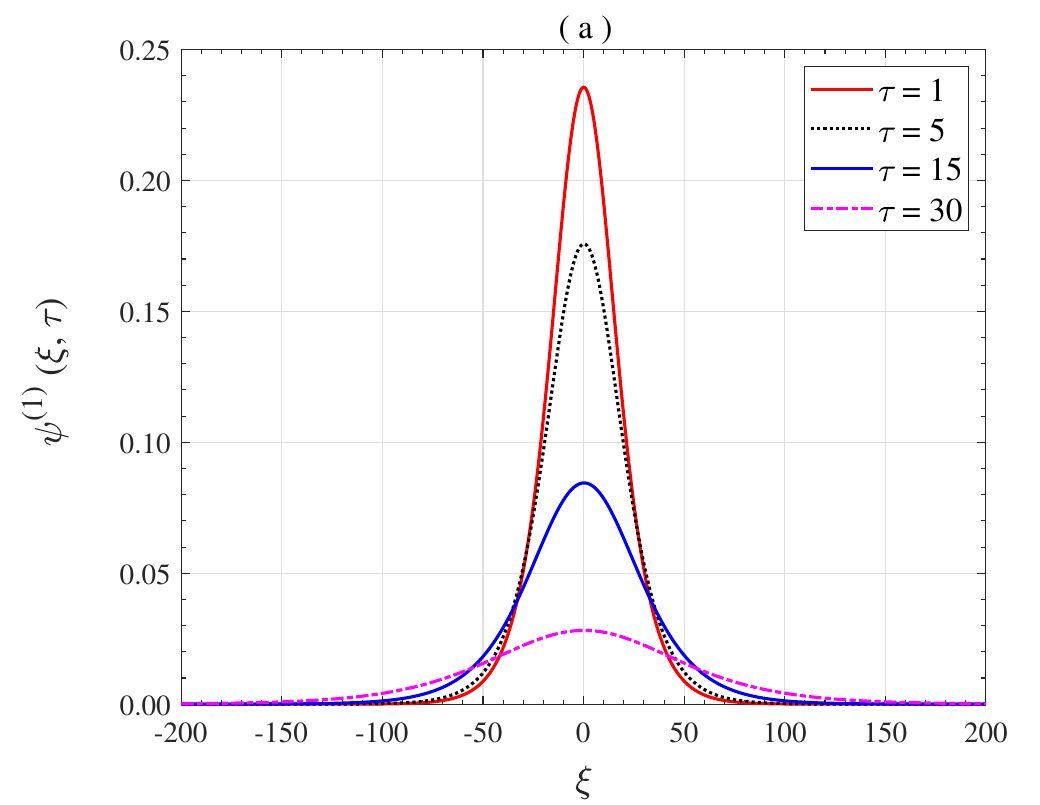}
\quad
\includegraphics[width=8.5cm]{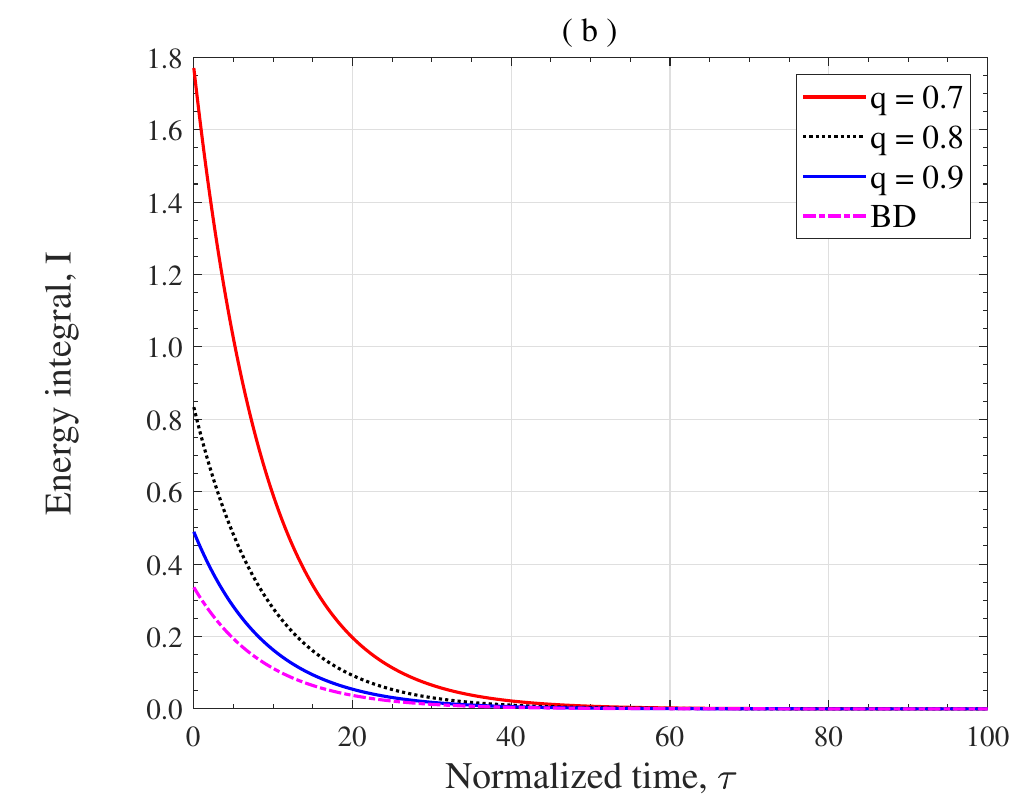}
\caption{Profiles of the DIA soliton [subplot (a); Eq. \eqref{eqn_analytical_presence}] and the soliton energy [subplot (b); Eq. \eqref{eqn_energy_Integral}] are shown at different times $(\tau)$ and for different values of $q$ as in the legends.  The fixed parameter values are $P_\parallel = 0.1$ and   $P_\perp = 0.2$. The other parameter values are the same as for Fig. \ref{fig_3}.} 
\label{fig_4}
\end{figure}

The profiles of the DIA soliton at different times $\tau$ and the effects of the nonextensive parameter $q$ on the soliton energy are shown in Fig. \ref{fig_4}. It is found that due to the effects of the adiabatic dust-charge variation, the ion creation and ion loss, as well as the ion-neutral and ion-dust collisions, the DIA solitons get damped with a reduction of the wave amplitude as time progresses [See subplot (a)]. Consequently, the soliton energy also decreases with time [See subplot (b)]. Furthermore, the soliton energy reaches maximum and minimum values in the regions of the superextensive and Maxwellian distribution of electrons.   
\par 
From Eq. (\ref{eqn_amplitude_timedependent}), we find that when the nonlinear coefficient $A_{1}$ tends to vanish ($A_{1} \sim0$), the amplitudes of solitary waves tend to become extremely large. In this situation, the KdV equation fails to describe the evolution of small-amplitude DIA perturbations. Thus, $A_{1} = 0$ gives the critical value of the plasma parameters, above which $A_{1} > 0$, yielding DIA solitons with positive potential (compressive type) and below which $A_{1} < 0$, i.e., one can have DIA solitons with negative potential (rarefactive type). However,  for laboratory collisional magnetized dusty plasma parameters, we always have $A_{1} > 0$ for which the damped DIA solitons are always of the compressive types.
\par
\begin{figure}[!h]
\centering
\includegraphics[width=8.5cm]{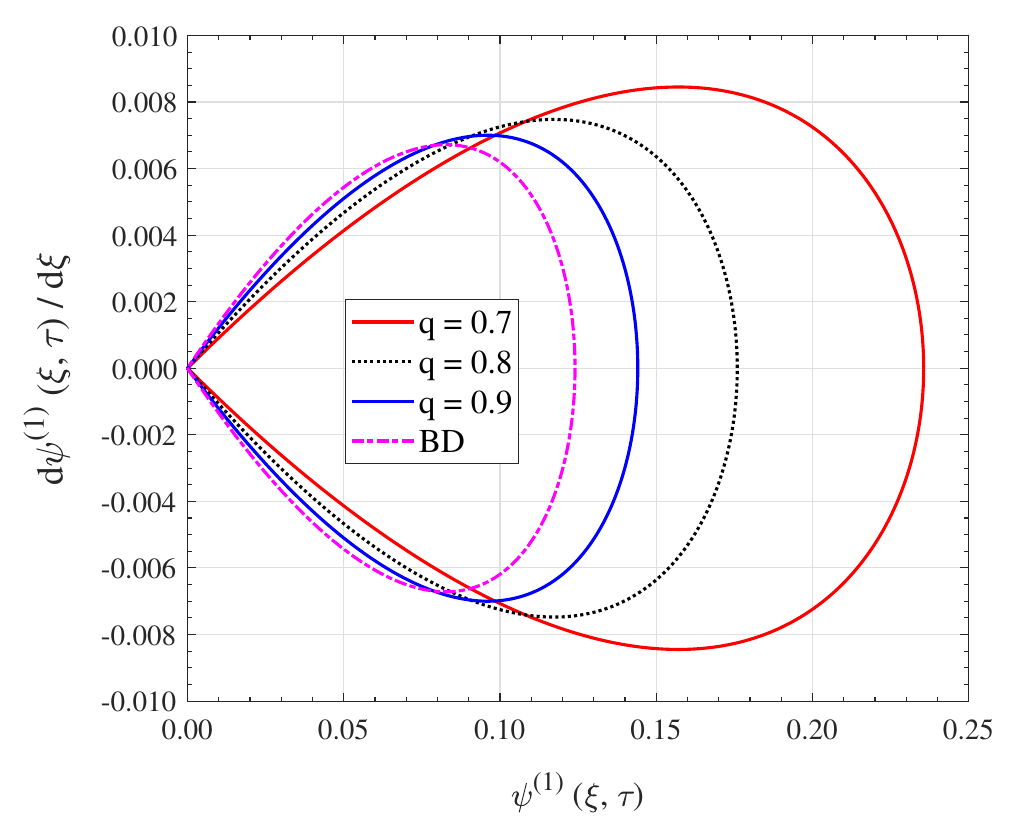}
\caption{Phase portraits are shown for different values of the nonextensive parameter $q$ with   $P_\parallel = 0.1$ and  $P_\perp = 0.2$. The other parameter values are the same as for Fig. \ref{fig_3}.} 
\label{fig_5}
\end{figure}

 
\textit{Phase portraits}: The phase portraits (The electrostatic field vs the potential) of the DIA solitary waves are shown in Fig. \ref{fig_5}. We find that the repetition of the phase curves occurs along the same trajectory when $d\psi^{\left(1\right)} \left(\xi, \tau\right) / d\xi$ $\neq0$. It is also found that the trajectory of the system in physical space is a homoclinic orbit, which may be useful for studying the dynamic behaviors of DIA solitons. Initially, the phase curve starts from the saddle equilibrium point and encircles the positive $\psi^{\left(1\right)} \left(\xi,\tau\right)$ -axis in the anti-clockwise direction. The trajectory then stops at the saddle equilibrium point in entering from the upper side. 
\par
\begin{figure}[!h]
\centering
\includegraphics[width=8.5cm]{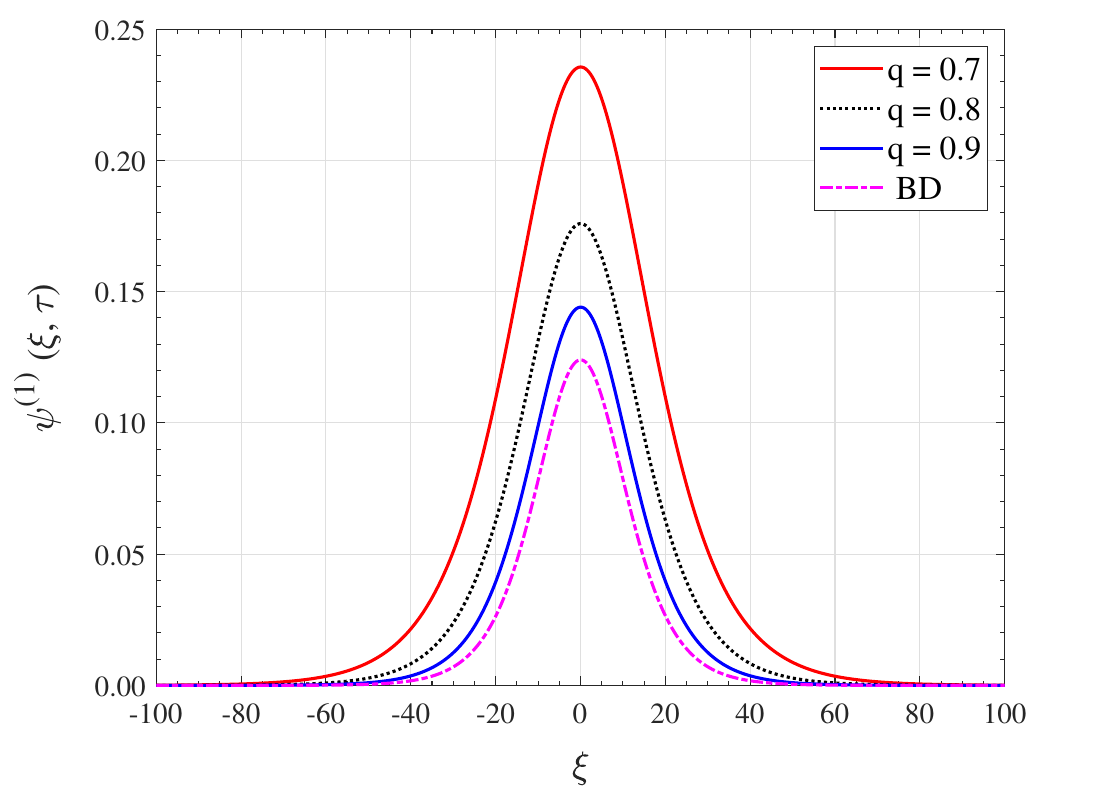}
\caption{Profiles of the damped DIA soliton [Eq. (\ref{eqn_analytical_presence})] are shown for different values of the nonextensive parameter $q$ with $P_\parallel = 0.1$ and   $P_\perp = 0.2$. The other parameter values are the same as for Fig. \ref{fig_3}.} 
\label{fig_6}
\end{figure}

 
\textit{Effects of nonextensive parameter $q$}: The effects of the nonextensive parameter $q$ on the profiles of compressive damped DIA solitons are shown in Fig. \ref{fig_6}.
From Eqs. (\ref{eqn_amplitude_timedependent}) and (\ref{eqn_width_timedependent}), it is evident that both the amplitude and widths of DIA solitons are inversely dependent on $q$, implying that they are reduced as the values of $q~(0<q<1)$ increase from the values corresponding to nonextensive electron distribution to the values corresponding to the Boltzmann distribution of electrons. For example,  the amplitude and width of solitary waves tend to decrease from about $0.24$ to $0.12$ and from $21.46$ to $14.34$ respectively, when the nonextensive parameter $q$ increases from $0.7$ to a value close to unity. It follows that the DIA solitons can evolve with higher energies in nonthermal plasmas than thermal ones.  
\par
\begin{figure}[!h]
\centering
\includegraphics[width=8.5cm]{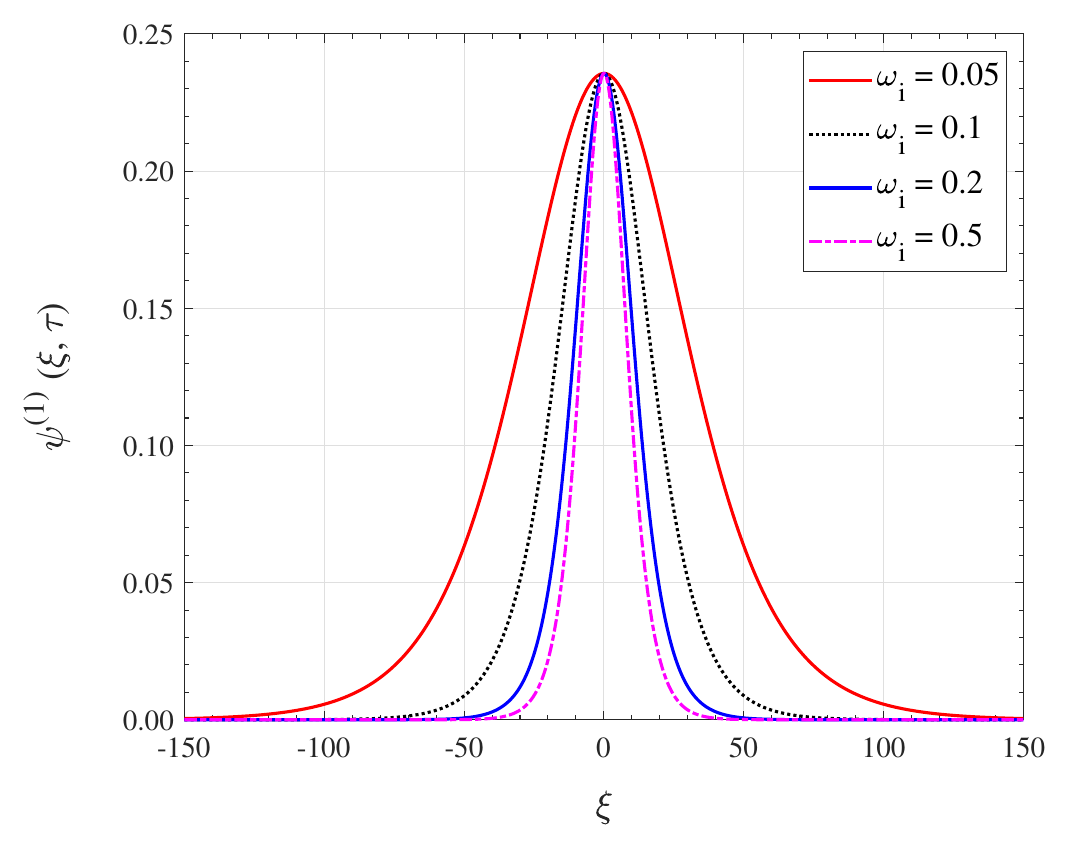}
\caption{Profiles of the damped DIA soliton [Eq. (\ref{eqn_analytical_presence})] are shown  for different values of the ion gyrofrequency $\omega_i$ as in the legend with  $P_\parallel = 0.1$, and  $P_\perp = 0.2$. The other parameter values are the same as for Fig. \ref{fig_3}.} 
\label{fig_7}
\end{figure}

\textit{Effects of $\omega_i$ and $\theta$}:  From Eqs. (\ref{eqn_amplitude_timedependent}) and (\ref{eqn_width_timedependent}), it is seen that while the solitary wave amplitude $\psi_{\mathrm{m}}^{\left(1\right)} \left(\tau\right)$ is proportional to the Mach number $M\left(\tau\right)$, the width   $\Delta_{\mathrm{w}}\left(\tau\right)$ varies inversely with $M$. Also, since the magnetic field contributes only to the wave dispersion, the amplitude remains independent of $\omega_i$, but the width varies inversely with it.   The effects of the magnetic field on the profiles of the damped DIA solitons are shown in Fig. \ref{fig_7}. It is seen that on increasing the normalized ion gyrofrequency $\omega_i$ from $\omega_i=0.05$ to $\omega_i=0.5$, the amplitude remains constant while the width decreases from $39.21$ to $10.77$.  Thus, it follows that the stronger the magnetic field strength narrower the DIA solitons, which move faster with lower energies than those in the region of weak magnetic fields at a fixed time. 
\par
\begin{figure}[!h]
\centering
\includegraphics[width=8.5cm]{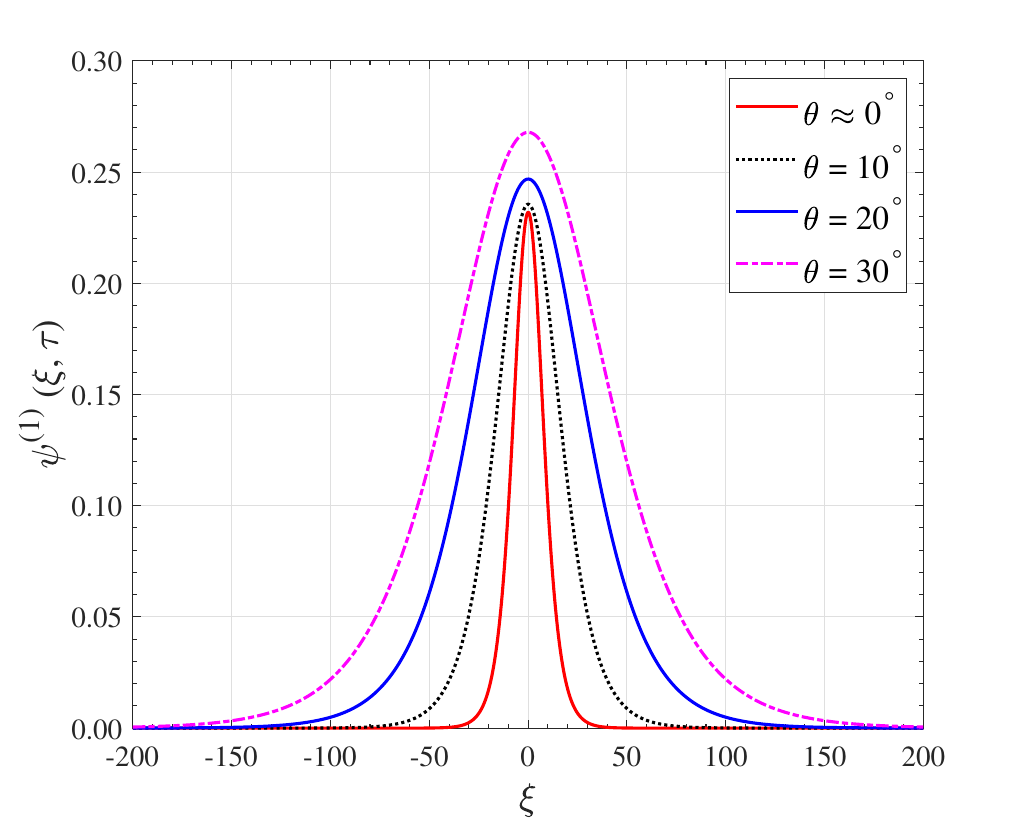}
\caption{Profiles of the damped DIA soliton [Eq. (\ref{eqn_analytical_presence})] are shown for different values of the obliqueness of wave propagation $\theta$ as in the legend with $P_\parallel = 0.1$  and  $P_\perp = 0.2$. The other parameter values are the same as for Fig. \ref{fig_3}.} 
\label{fig_8}
\end{figure}

 
The obliqueness ($\theta$) of wave propagation relative to the magnetic field direction also plays a crucial role in the characteristics of DIASWs. The results are shown in Fig. \ref{fig_8}. Since the amplitude and width of DIA solitons have an inverse relation with the direction cosine  $l_z = \cos\theta$, both of them increase with increasing the propagation angle $\theta$. For example, the amplitude and width  increase from about $0.23$ to $0.27$ and $10.16$ to $52.02$ respectively  as $\theta$ increases from $0$ to $30^{\circ}$. Thus, DIA waves propagating along (perpendicular to) the magnetic field will evolve as DIA solitons with the lowest (highest) energy decreasing (increasing) both the amplitude and width significantly.    
\par
\begin{figure}[!h]
\centering
\includegraphics[width=8.5cm]{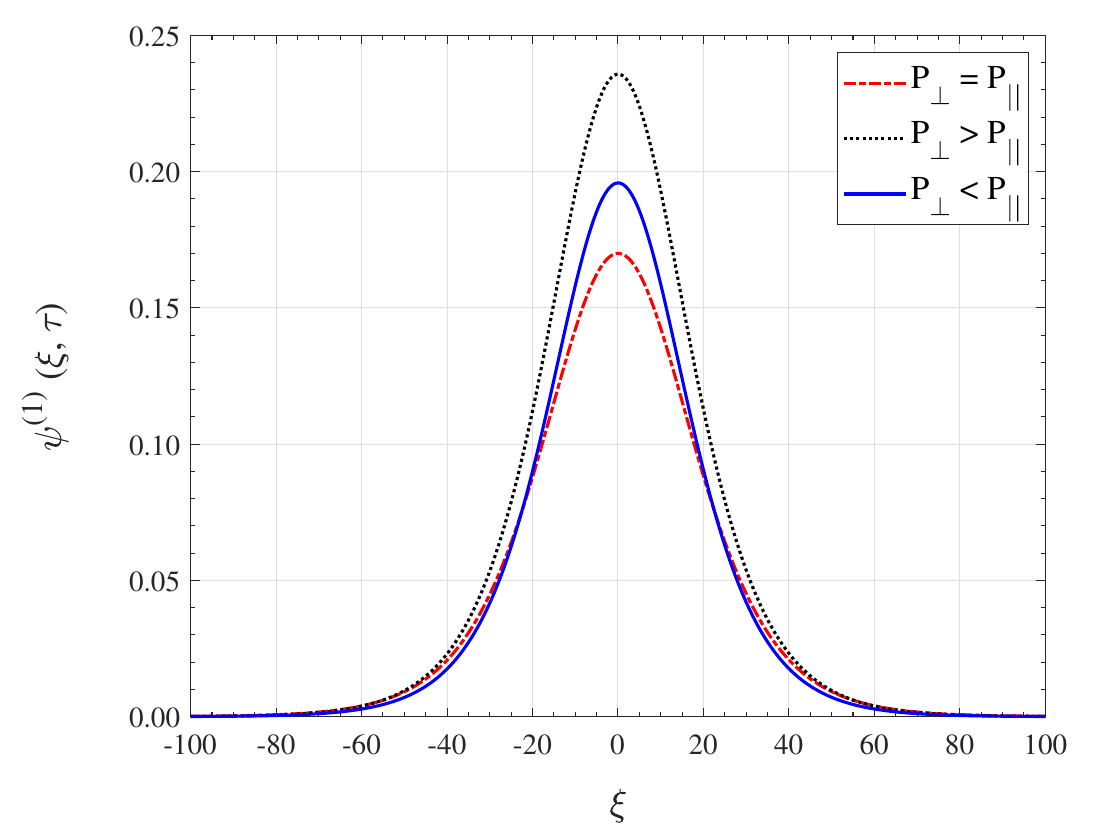}
\caption{Profiles of the DIA soliton [Eq. (\ref{eqn_analytical_presence})] are shown for different cases of the anisotropic pressure: $P_\perp = P_\parallel = 0.5$, $P_\perp\hspace{0.1cm}(0.3) > P_\parallel\hspace{0.1cm}(0.1)$, and $P_\perp \hspace{0.1cm}(0.1) < P_\parallel\hspace{0.1cm}(0.3)$. The other parameter values are the same as for Fig. \ref{fig_3}.} 
\label{fig_9}
\end{figure}

\begin{figure}[!h]
\centering
\includegraphics[width=8.5cm]{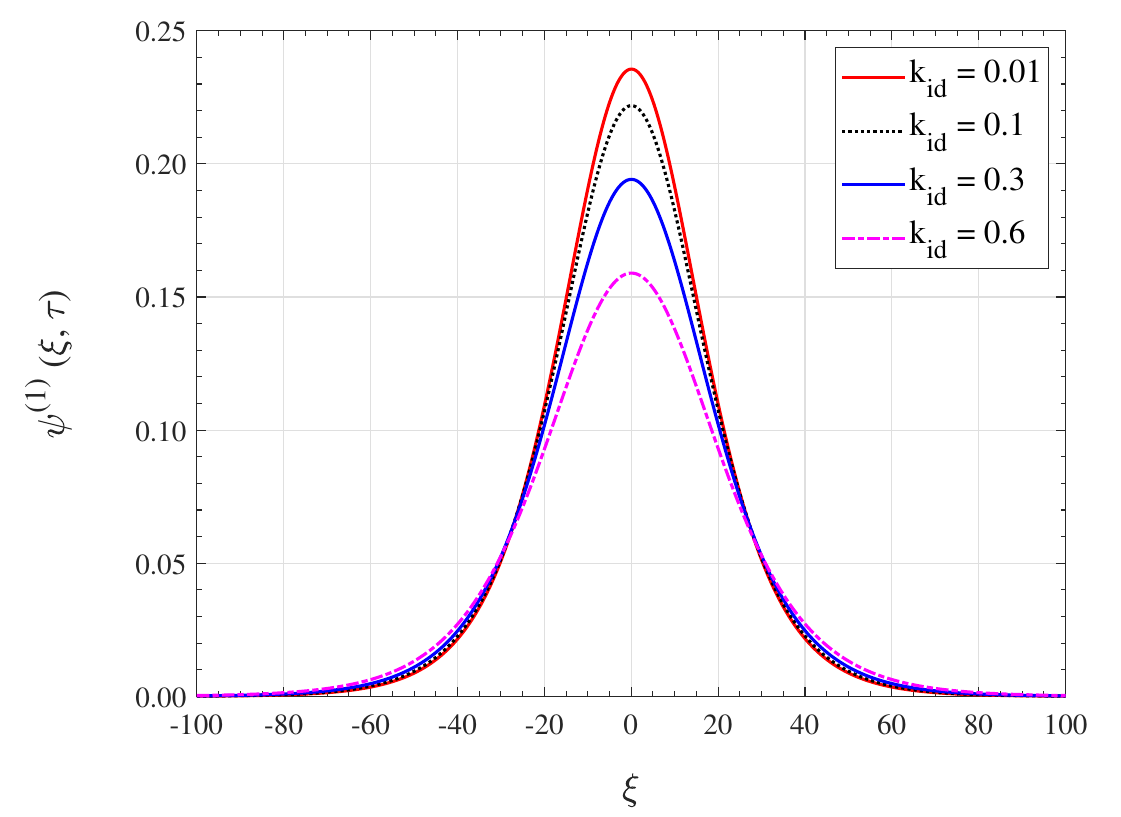}
\caption{Profiles of the DIA soliton [Eq. (\ref{eqn_analytical_presence})] are shown for different values of the ion-dust collision frequency $k_{\mathrm{id}}$ as in the legend with $P_\parallel = 0.1$ and $P_\perp = 0.2$. The other parameter values are the same as for Fig. \ref{fig_3}.} 
\label{fig_10}
\end{figure}

 
\textit{Effects of anisotropic pressure and ion-dust collision}:
We study the characteristics of the damped DIA soliton for different cases of anisotropic pressure and for different values of the ion-dust collision frequency. The results are shown in Figs. \ref{fig_9} and \ref{fig_10}. From Fig. \ref{fig_9}, we find that the amplitude of DIA solitary waves reaches maximum when the perpendicular pressure component remains higher than the parallel pressure (anisotropic) and it assumes the lowest value when the parallel and perpendicular pressure components are equal in the isotropic case. The presence of ion-dust collision also affects the width and amplitude of DIASWs (See Fig. \ref{fig_10}).  We find that an increase in the ion-dust collision frequency decreases the Mach number, which causes to decrease in both the wave amplitude and the width. For example, on increasing the ion-dust collision frequency from $ 0.01$ to $0.6$, the amplitude decreases from $0.24$ to $0.16$, while the width increases from $21.46$ to $26.13$. Thus, it may be concluded that the ion pressure anisotropy plays a significant role in increasing the soliton amplitude and hence the soliton energy. In contrast, DIA solitons evolve with lower energies in regimes with higher collision frequencies.   
\par
\begin{figure}[!h]
\centering
\includegraphics[width=8.5cm]{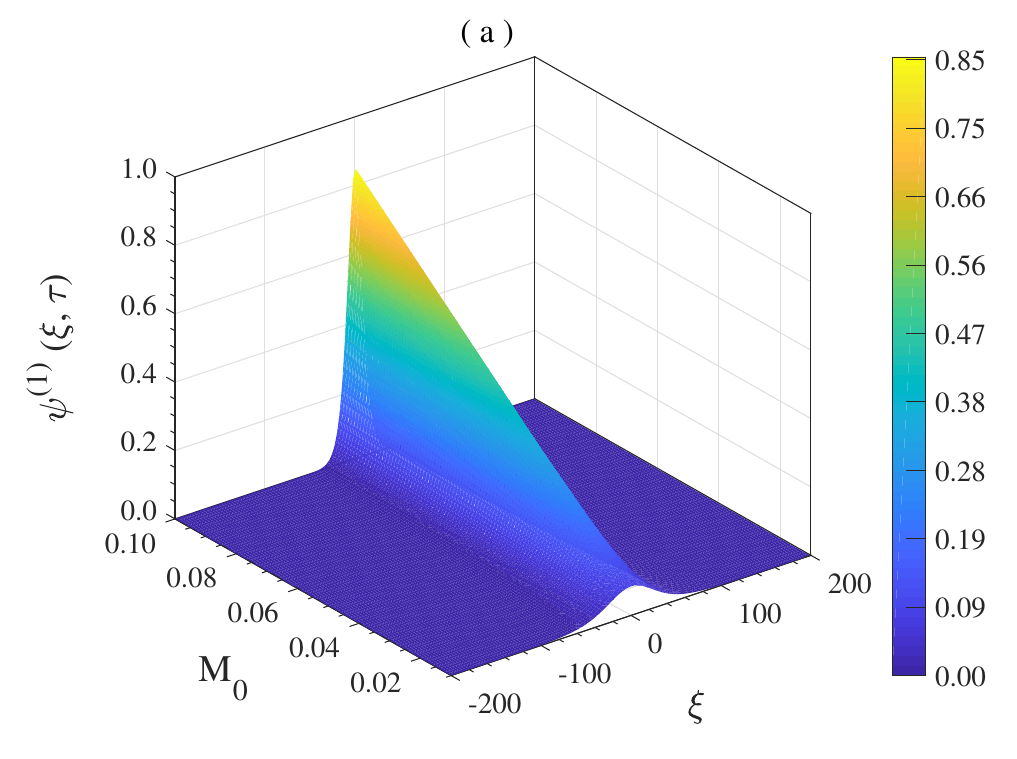}
\quad
\includegraphics[width=8.5cm]{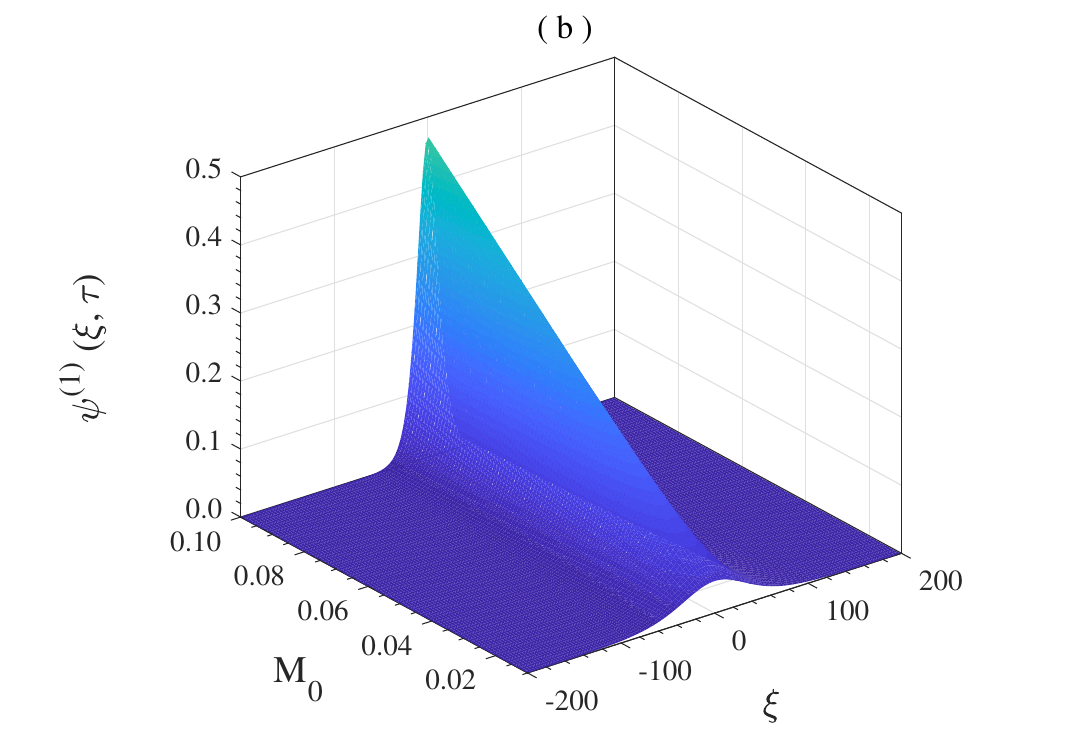}
\caption{Dust-ion acoustic solitons [Eq. (\ref{eqn_analytical_presence})] in two different plasma states: (a) active plasma (upper panel) and (b) quiescent plasma (lower panel) with  $P_\parallel = 0.1$  and  $P_\perp = 0.2$. The other parameter values are the same as for Fig. \ref{fig_3}.} 
\label{fig_11}
\end{figure}

\textit{Cases of active and quiescent dusty plasmas}:
In dusty plasmas, the presence of source and sink terms can have significant roles in exploring the features of various nonlinear phenomena. A comparative study of the wave dynamics in active and quiescent dusty plasmas is done and the results are displayed in Fig. \ref{fig_11}. It is seen that on increasing the Mach number $M_{0}$, the wave amplitude increases, but the width decreases in both active and quiescent states. As the plasma system changes from the active to a quiescent state, the amplitude of solitary waves decreases, whereas the width increases.  At $M_{0} = 0.1$, the amplitude becomes maximum, i.e., $\psi^{\left(1\right)}_{\mathrm{m}} \left(\tau\right) = 0.85$ and the width becomes minimum, i.e., $\Delta_{\mathrm{w}} \left(\tau\right) = 11.29$ for active plasmas [Fig. \ref{fig_11} (a)]. On the other hand, for quiescent plasmas, the maximum amplitude is $\psi^{\left(1\right)}_{\mathrm{m}} \left(\tau\right) = 0.47$ and the minimum width is $\Delta_{\mathrm{w}} \left(\tau\right) = 15.18$ at the same value of $M_0$ [Fig. \ref{fig_11}(b)].
\\
\par
\begin{figure}[!h]
\centering
\includegraphics[width=8.5cm]{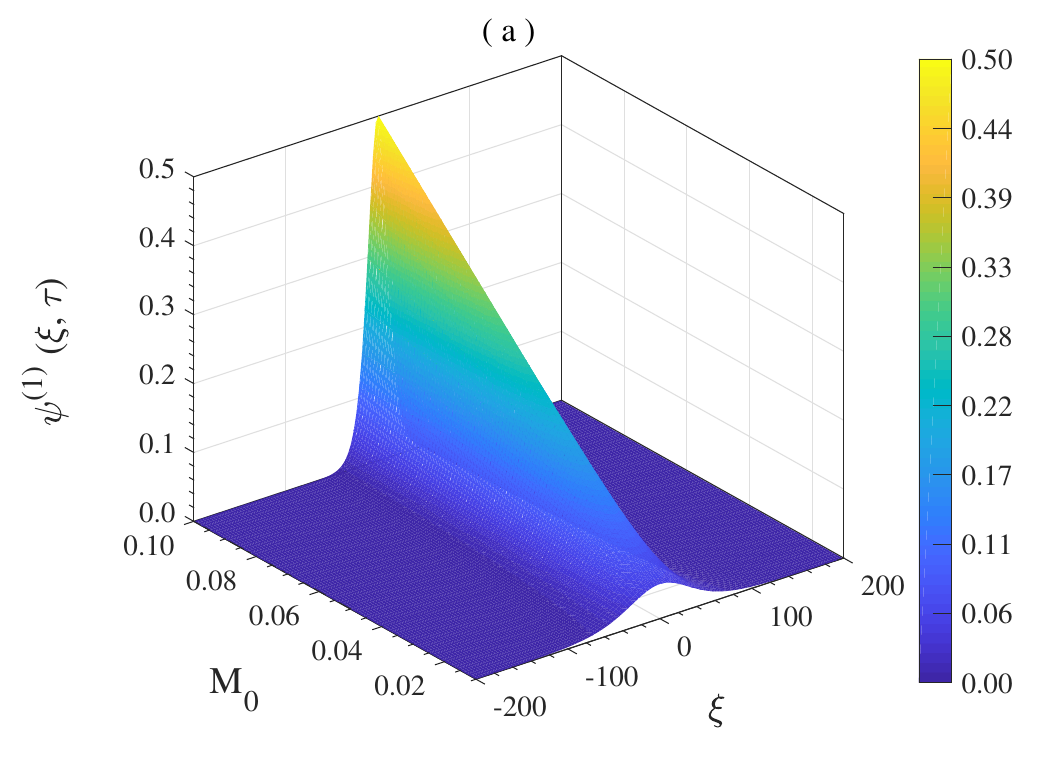}
\quad
\includegraphics[width=8.5cm]{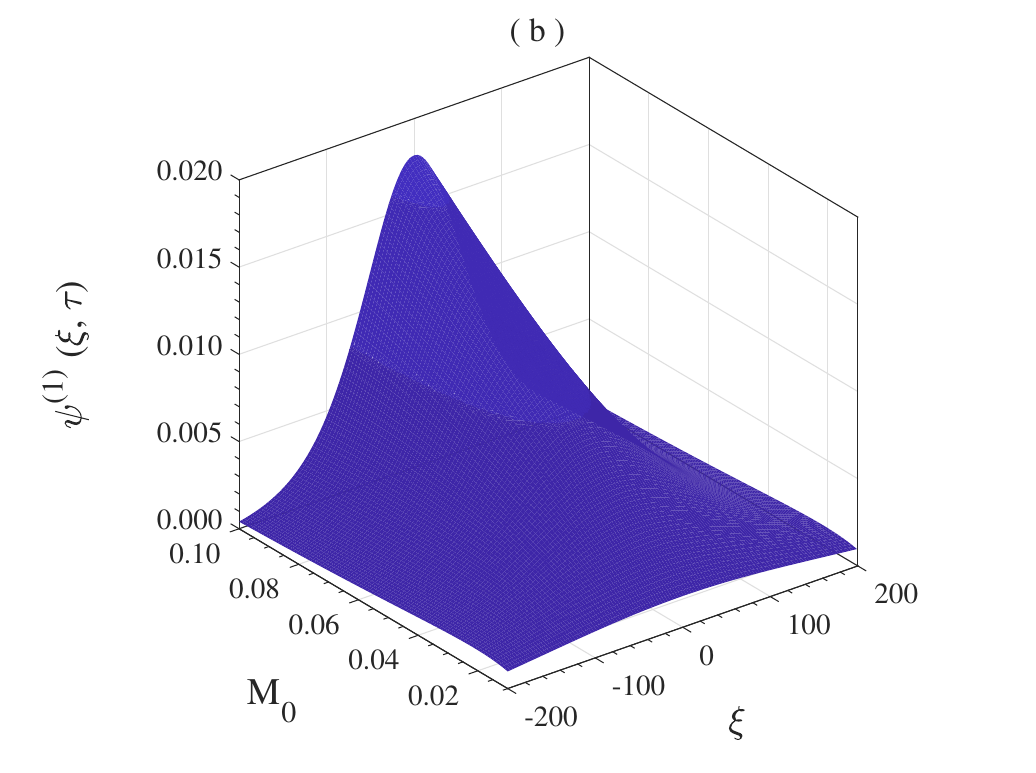}
\caption{Profiles of damped DIA solitons [Eq. (\ref{eqn_analytical_presence})] are shown for two different cases: weakly collisional [subplot (a)] and strongly collisional [subplot (b)] regimes with $P_\parallel = 0.1$  and   $P_\perp = 0.2$. The other parameter values are the same as for Fig. \ref{fig_3}.} 
\label{fig_12}
\end{figure}

 
\textit{Weakly and strongly collisional regimes}: The relative influence of the ion-neutral collision frequency in comparison with the ion-cyclotron frequency can be measured by defining a parameter $\Omega = {2\pi m_{\textrm{i}}\nu_{\textrm{in}}}/{eB}$, where $\nu_{\textrm{in}}$ is the ion-neutral collision frequency. The limits $\Omega \gg 1$  and $\Omega \ll 1$, respectively, correspond to strongly collisional (or weakly magnetized)  and weakly collisional (or strongly magnetized) plasmas. The profiles of the DIA soliton with respect to the space-time coordinate $\xi$ and the Mach number $M_0$ are shown in Fig. \ref{fig_12} in two different cases: weakly collisional [subplot (a)] and strongly collisional [subplot (b)] regimes. It is found that in weakly (strongly) collisional plasmas, the amplitude (width) of the solitary wave becomes higher than that in strongly (weakly) collisional plasmas. It follows that the soliton moving faster gets amplified and evolves with higher energies when the magnetic force dominates over the collisional force and becomes slower and widened in the case when the collisional force dominates over the magnetic force.
\par
\begin{figure}[!h]
\centering
\includegraphics[width=8.5cm]{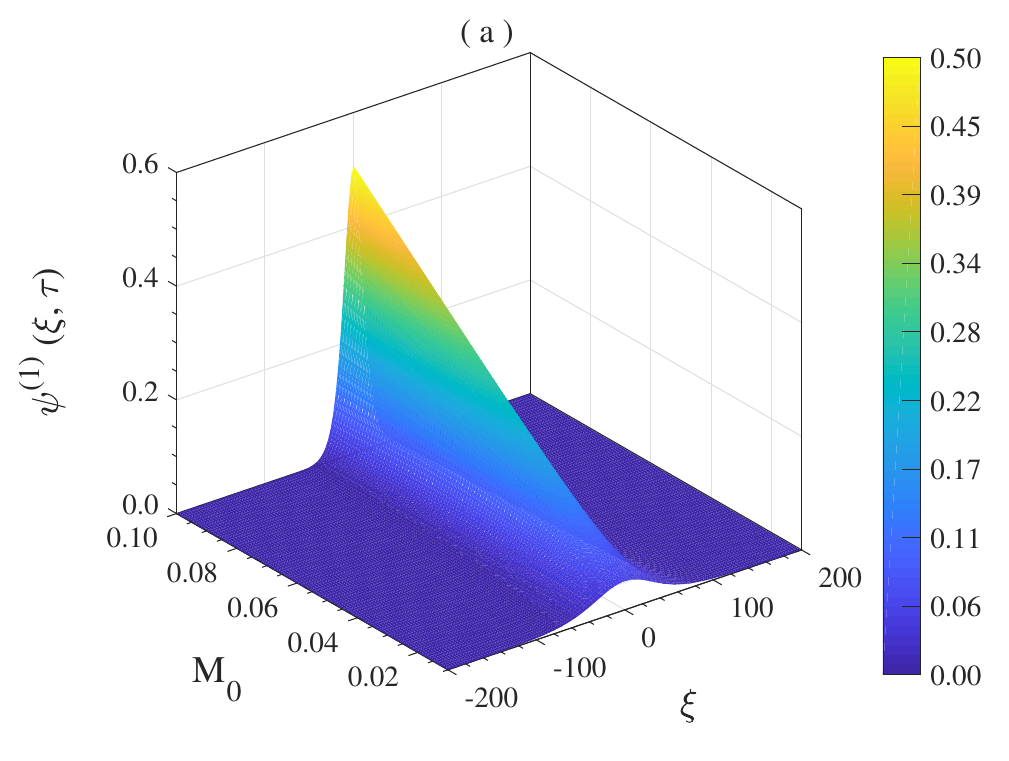}
\quad
\includegraphics[width=8.5cm]{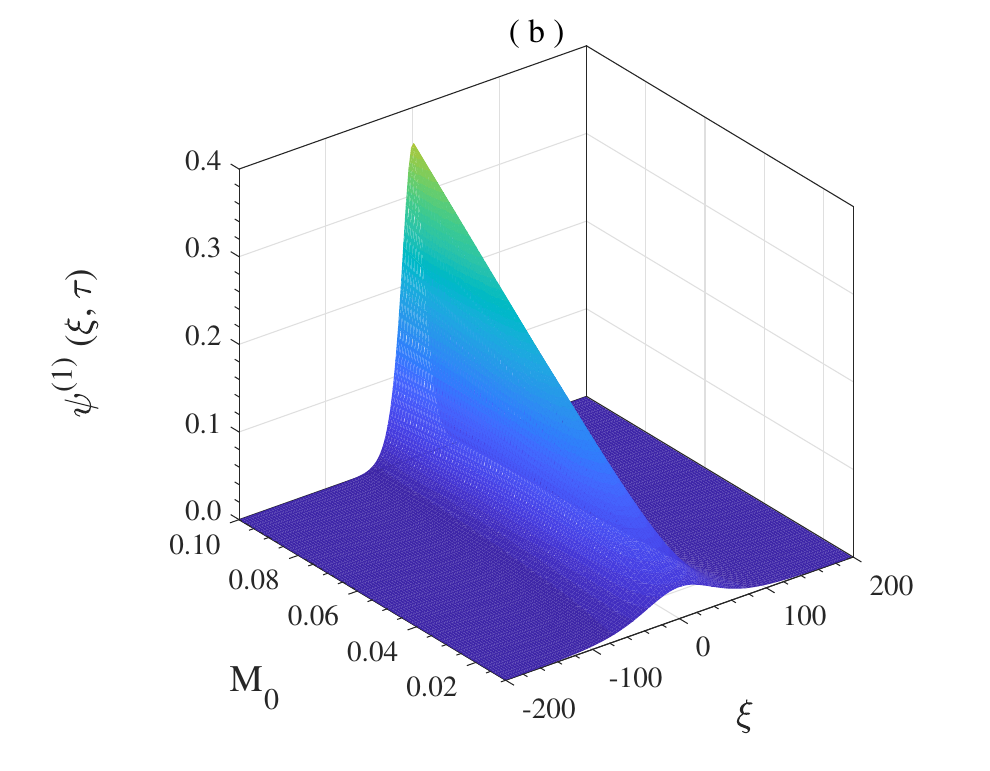}
\caption{Profiles of the damped DIA solitons [Eq. (\ref{eqn_analytical_presence})] are shown in the absence [subplot (a)] and presence [subplot (b)] of collision enhancement ion current with   $P_\parallel = 0.1$  and  $P_\perp = 0.2$. The other parameter values are the same as for Fig. \ref{fig_3}.} 
\label{fig_13}
\end{figure}

 
On the other hand, the ion-neutral collisions considerably affect the dust-charging mechanism\citep{khrapak2005particle}. So, their influences on the properties of DIA solitons can be of interest. Figure \ref{fig_13} shows the profiles of DIA solitons in the absence [subplot (a)] and presence [subplot (b)] of collision enhancement ion current in the dust-charging process.  By incorporating the ion-neutral collision enhancement in the self-consistent dust charge process, we find that the wave amplitude decreases, but the soliton widens in increasing its width. The maximum values of the amplitude and width are $\psi_{\mathrm{m}}^{\left(1\right)} \left(\tau\right) = 0.51$ and $\Delta_{\mathrm{w}}\left(\tau\right) = 46.47$ in absence of the collision enhancement ion current [Fig. \ref{fig_13}(a)]. However, in the presence of the collision enhancement ion current [Fig. \ref{fig_13}(b)], the corresponding maximum values of the amplitude and width are $\psi_{\mathrm{m}}^{\left(1\right)} \left(\tau\right) = 0.36$ and $\Delta_{\mathrm{w}}\left(\tau\right) = 54.79$. It follows that the properties of DIA solitons get significantly modified by the effects of ion-neutral collision enhancement current in dusty plasmas.   
\subsection{Magnetized Saturn's E-ring dusty plasmas} \label{sec-sub-numeric-space}
We numerically examine the characteristics of DIA damped solitons and shocks for the plasma parameters that are relevant in the environments of Saturn's E-ring \citep{yaroshenko2007dust}. We consider the plasma density $n_{\mathrm{i}0}$ = 10$^{7}$ m$^{-3}$, magnetic field $B$ = 0.098- 2.21 G, number density of dust particles $n_{\mathrm{d}0}$ = 3.0$\times$10$^{3}$ m$^{-3}$, mass density of dust particle $\rho_{\mathrm{d}}$ = 2650 kg/m$^{3}$, electron temperature $k_{\textrm{B}}T_{\textrm{e}}$ = 8.0 eV,  mass of molecular oxygen ion ($\text{O}_{2}^{+}$) $m_{\mathrm{i}}$ = 32 amu, positive ion temperature $k_{\textrm{B}}T_{\textrm{i}}$ = 1.62 eV,  nonextensive parameter for electron distribution $q$ = 0.7 -0.999 (Boltzmann distribution, BD), and the obliqueness of wave propagation $(0<\theta<\pi/2)$.  Since the space dusty plasma (Saturn's E-ring) can be considered collisionless, the equilibrium ion current ($I_{\mathrm{i}0}$) due to the thermal motion of positive ions is given by 
\begin{equation}
I_{\mathrm{i}0} = \pi r^{2}_{\mathrm{d}}en_{\mathrm{i}0}\sqrt{\frac{8T_{\mathrm{i}}}{\pi m_{\mathrm{i}}}}\left(1- \frac{eq_{\textrm{d}0}}{r_{\textrm{d}}k_{\textrm{B}}T_{\textrm{i}}}\right).
\label{eqn_ion_current_2nd}
\end{equation}
Next, to calculate the equilibrium dust-charge $q_{\mathrm{d}0}$ for the parameter values stated above, we solve the equilibrium dust-charge equation, $I_{\mathrm{i}0} + I_{\mathrm{e0}} = 0$ numerically by the Newton-Raphson method. Similar to the analysis for laboratory dusty plasmas, the normalized time ($\tau$) is held constant, i.e. $\tau = 1$ for simplicity. Since for Saturn's E-ring dusty plasmas, the dust-charging frequency is lower than the natural-dust collision frequency, we have the case of non-adiabatic dust-charge variation and the evolution of DIA damped solitary waves and shocks. From the expressions of the amplitudes of solitons and shocks, i.e.,   $\psi_{\textrm{sm}}^{(1)} \left(\tau\right) = 3M_{0}/A_{2}$ and  $\psi_{\textrm{s}}^{(1)}$  = $12D_{1}^{2}/25A_{1}B_{1}$, it can be assessed that the profiles of both solitons and shocks can appear with positive or negative potentials corresponding to $A_{1,2} > 0$ or $A_{1,2}< 0$  in space dusty plasmas.  The case of $A_{1,2} = 0$ at which the KdV or KdV Burgers equation fails to describe the evolution of DIA solitons or shocks is not considered for the present investigation.  We find that the nonextensive parameter $q$ and (or) the dust-charge number may be responsible for the existence of both rarefactive and compressive DIA solitons and shocks. Typically, for the plasma parameter values as above, the DIA solitons can be of compressive or rarefactive types according to when $q>0.98$ or $q\lesssim0.98$. On the other hand, the DIA shocks can appear as compressive or rarefactive types for $q>0.75$ or $q\lesssim0.75$. Thus, there are two different regimes of $q$ for nonextensive electrons for the existence of DIA solitons and shocks.
\subsection*{Damped DIA solitons}
\begin{figure}[!h]
\centering
\includegraphics[width=8.5cm]{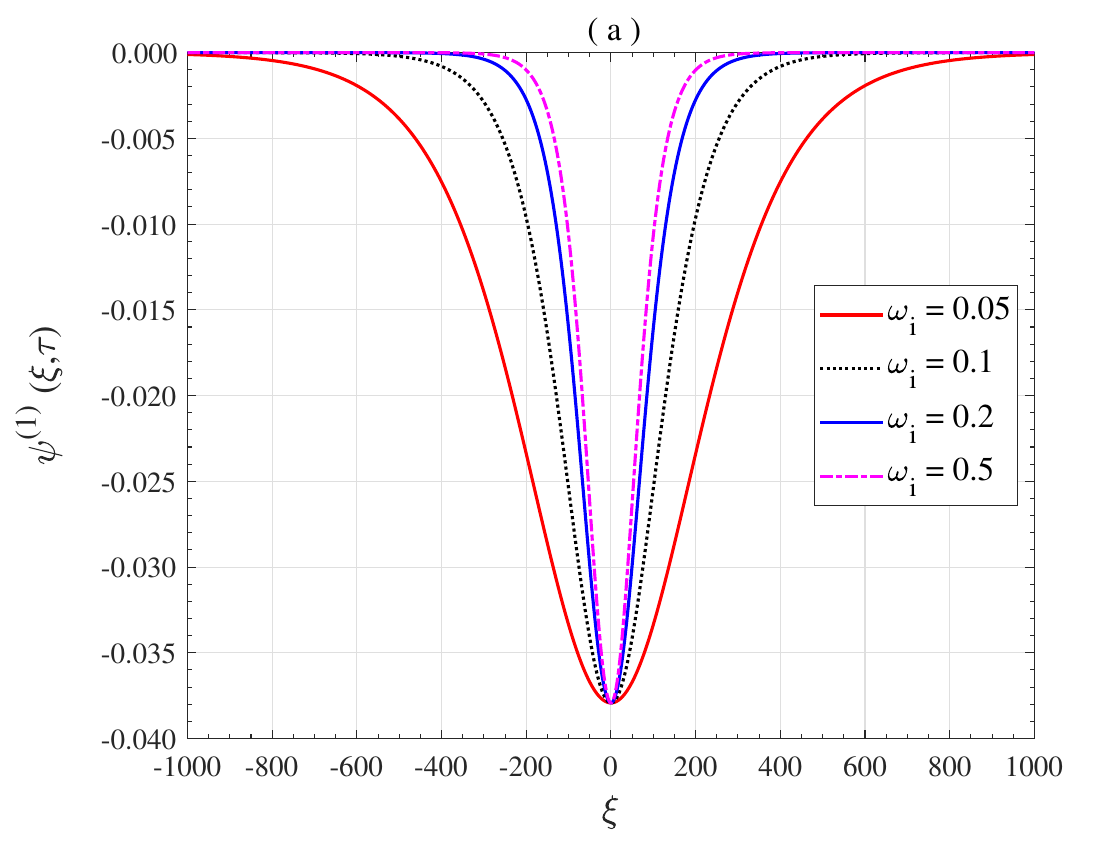}
\includegraphics[width=8.5cm]{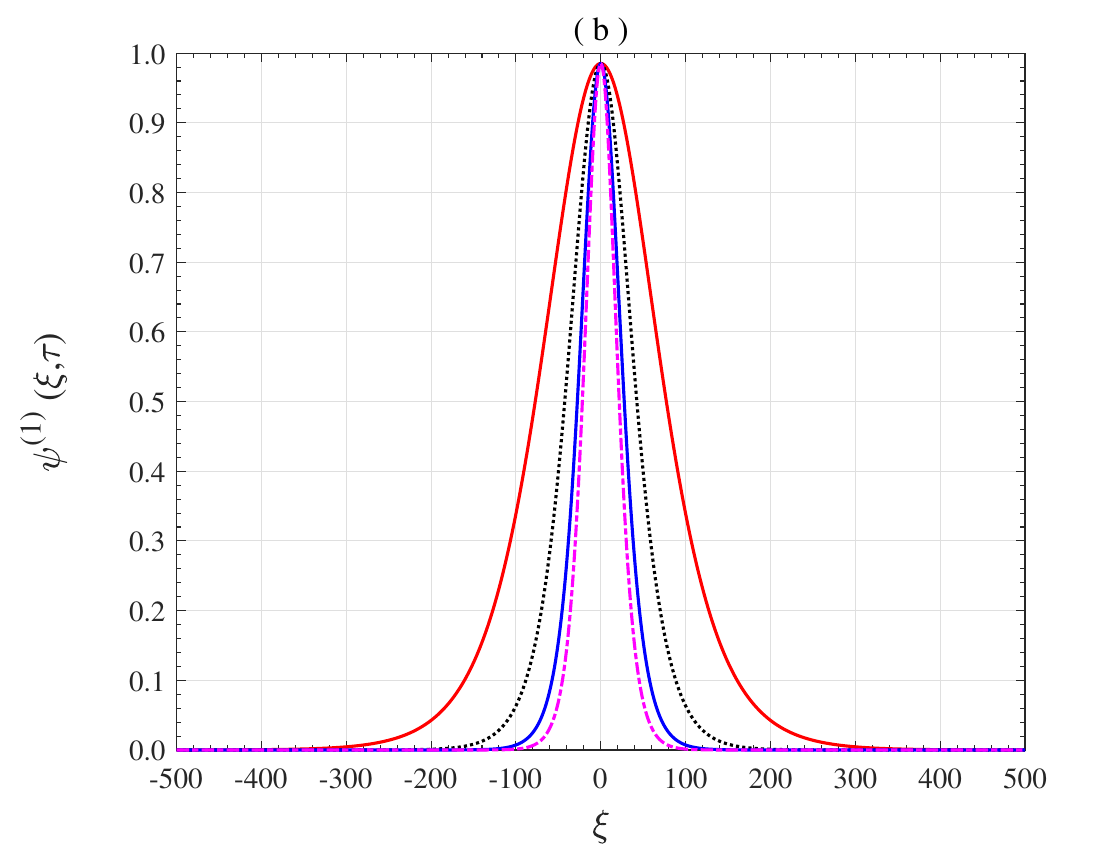}
\caption{Profiles of the rarefactive [Subplot (a) for $q = 0.8$] and compressive [Subplot (b) for $q = 0.98$] DIA solitons [Eq. (\ref{eqn_analytical_presencea})] are shown for different values of the ion gyrofrequency $\omega_i$ as in the legend. The fixed parameter values are   $P_\parallel = 0.1$,  $P_\perp  = 0.2$,  $r_{\mathrm{d}}$ = 0.15 $\mu$m,   $M_0$ = 0.02,   $\nu_{\text{d}} = 0.5$, and   $\theta$ = 10$^{\circ}$.}
\label{fig_14}
\end{figure}

\textit{Effects of magnetic field}: 
The effects of the magnetic field strength on the profiles of rarefactive [Subplot (a) for $q = 0.8$] and compressive [Subplot (b) for $q = 0.98$] solitons are shown in Fig. \ref{fig_14} for different values of the ion gyrofrequency $(\omega_i)$ as in the legend. Since the magnetic field contributes only to the wave dispersion (inversely proportional), the amplitude remains unchanged, while the width tends to decrease with increasing values of $\omega_i$.   Thus, we have similar features as for laboratory plasmas.
\par
\begin{figure}[!h]
\centering
\includegraphics[width=8.5cm]{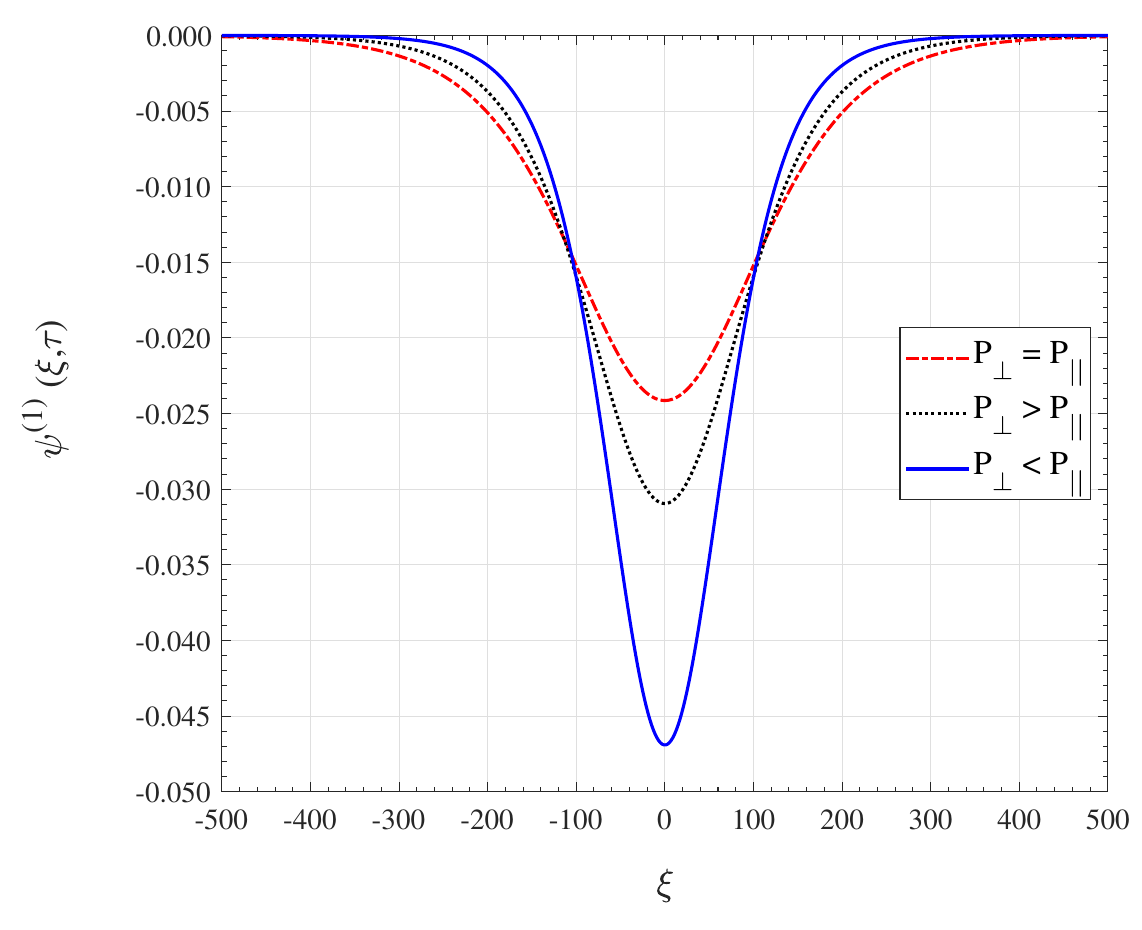}
\caption{Effects of the anisotropic pressure on the profiles of rarefactive DIA  soliton [Eq. (\ref{eqn_analytical_presencea})] are  shown when $P_\perp = P_\parallel = 0.5$, $P_\perp\hspace{0.1cm}(0.3) > P_\parallel\hspace{0.1cm}(0.1)$, and $P_\perp \hspace{0.1cm}(0.1) < P_\parallel\hspace{0.1cm}(0.3)$. The fixed parameter values are $q$ = 0.8, $\omega_{\mathrm{i}}$ = 0.2, and others as for Fig. \ref{fig_14}.}
\label{fig_15}
\end{figure}

\begin{figure}[!h]
\centering
\includegraphics[width=8.5cm]{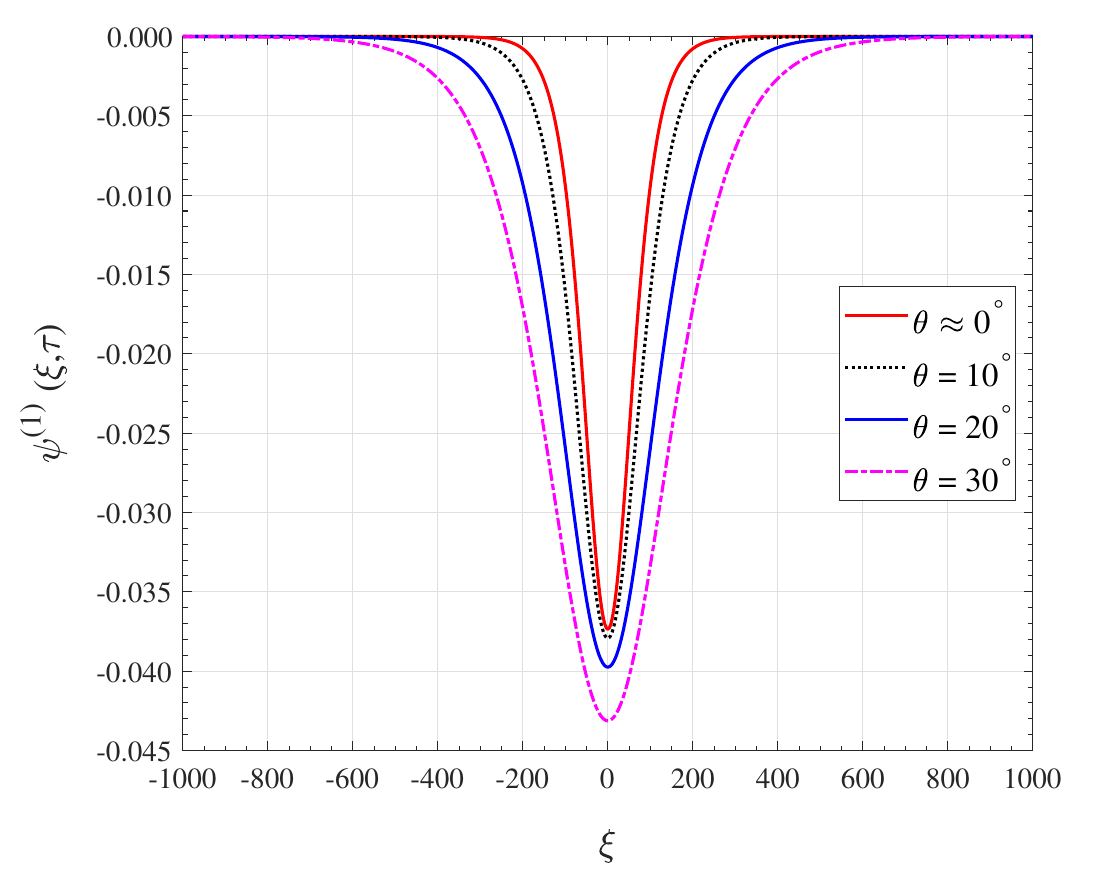}
\caption{Effects of the obliqueness of wave propagation on the profiles of rarefactive DIA solitons [Eq. (\ref{eqn_analytical_presencea})] are shown for different values of $\theta$. The fixed parameter values are $q = 0.8$, $\omega_{\mathrm{i}}$ = 0.2, and the others are the same as for Fig. \ref{fig_14}.}
\label{fig_16}
\end{figure}

 
\textit{Effects of anisotropic pressure and obliqueness of wave propagation}: 
The profiles of the damped DIA solitons in cases of isotropic and anisotropic pressures are shown in Fig. \ref{fig_15}. It is seen that in contrast to laboratory plasmas, the soliton amplitude decreases with an increase in the perpendicular component of ion pressure, and the width is found to be maximum in the isotropic case in which the perpendicular and parallel pressure components are equal. Like laboratory dusty plasmas, a similar effect of $\theta$ on the rarefactive DIA solitons can be seen for its different values as exhibited in Fig. \ref{fig_16}. It is noted that as the angle increases, the magnitudes of both the amplitude and width increase.
\par
\begin{figure}[!h]
\centering
\includegraphics[width=8.5cm]{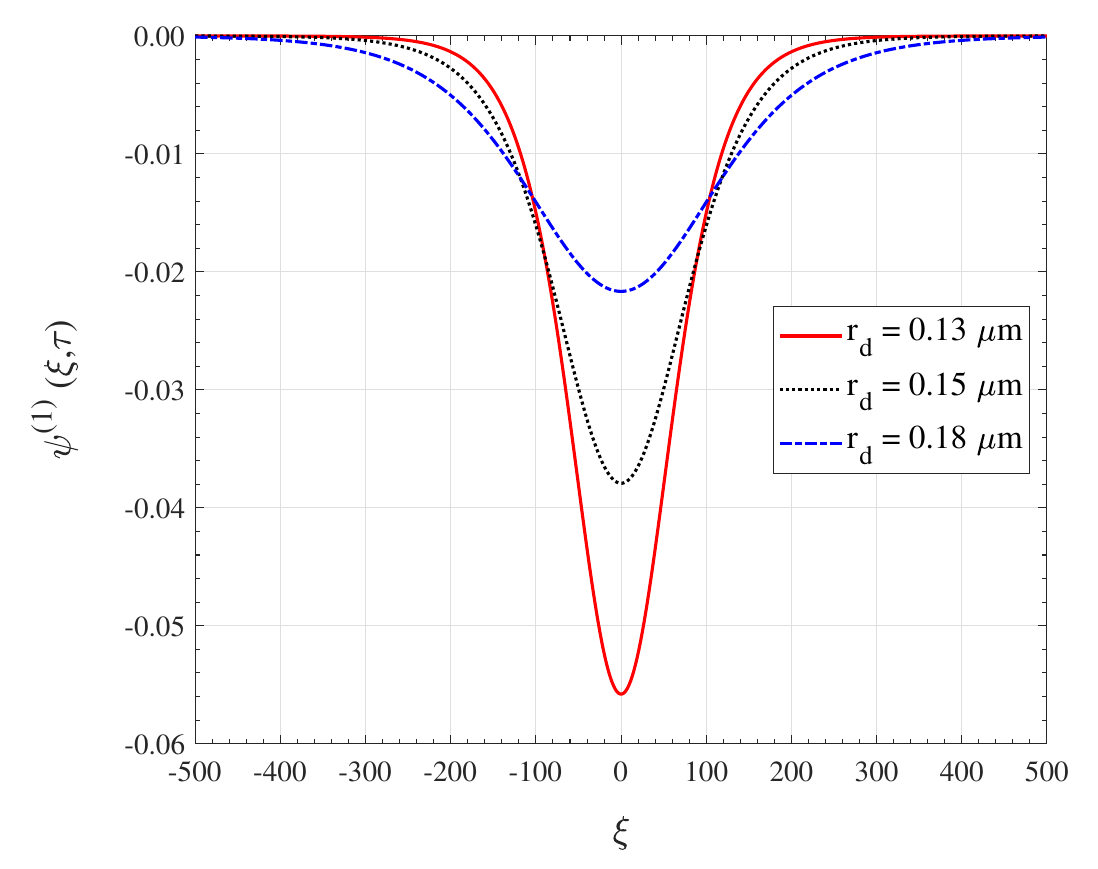}
\caption{Effects of different dust-grain radius on the profiles of rarefactive DIA solitons [Eq. (\ref{eqn_analytical_presencea})] are shown for different values of $r_{\textrm{d}}$ as in the legend. The fixed parameter values are $q = 0.8$, $\omega_{\mathrm{i}}$ = 0.2, and the others are the same as for Fig. \ref{fig_14}.}
\label{fig_17}
\end{figure}

\textit{Effects of dust grain radius $r_d$}:
Different sizes (with different radii) of spherical charged dust grains can considerably affect the evolution of equilibrium dust charge. As the radius increases, the flow of electrons into the surface of dust grains increases. As a result, the dust-charge number increases. For example,  the dust-charge number increases from about $2809$ to $3028$ as the radius of dust grains increases from $0.13$ to $0.18$ $\mu$m. The effects of $r_d$ on the rarefactive DIA solitons are shown in Fig. \ref{fig_17}. It is seen that on increasing the size of the dust grains, a decrease (increase) of the magnitude of the amplitude (width) is noted. 
\subsection*{DIA shocks}
From the expression of shock solution [Eq. (\ref{eqn_burger_solution})], it can be seen that shock amplitude, shock width, and shock speed all depend on the nonextensive parameter $q$, the magnetic field, and the dust-charge fluctuations rate $\nu_{\text{d}}$. We show only the monotonic shocks due to stronger dissipation effects than the dispersion. While the magnetic field has an inverse relation with the shock amplitude and a direct relation with the shock width, the dust-charge fluctuation rate has a direct relation with the shock amplitude and an inverse relation with the shock width. Besides the shock amplitude and width, the shock speed also has an inverse (direct) relation with the magnetic field (dust charge fluctuations rate) [Eq. (\ref{eqn_burger_solution})]. 
\par
\begin{figure}[!h]
\centering
\includegraphics[width=8.5cm]{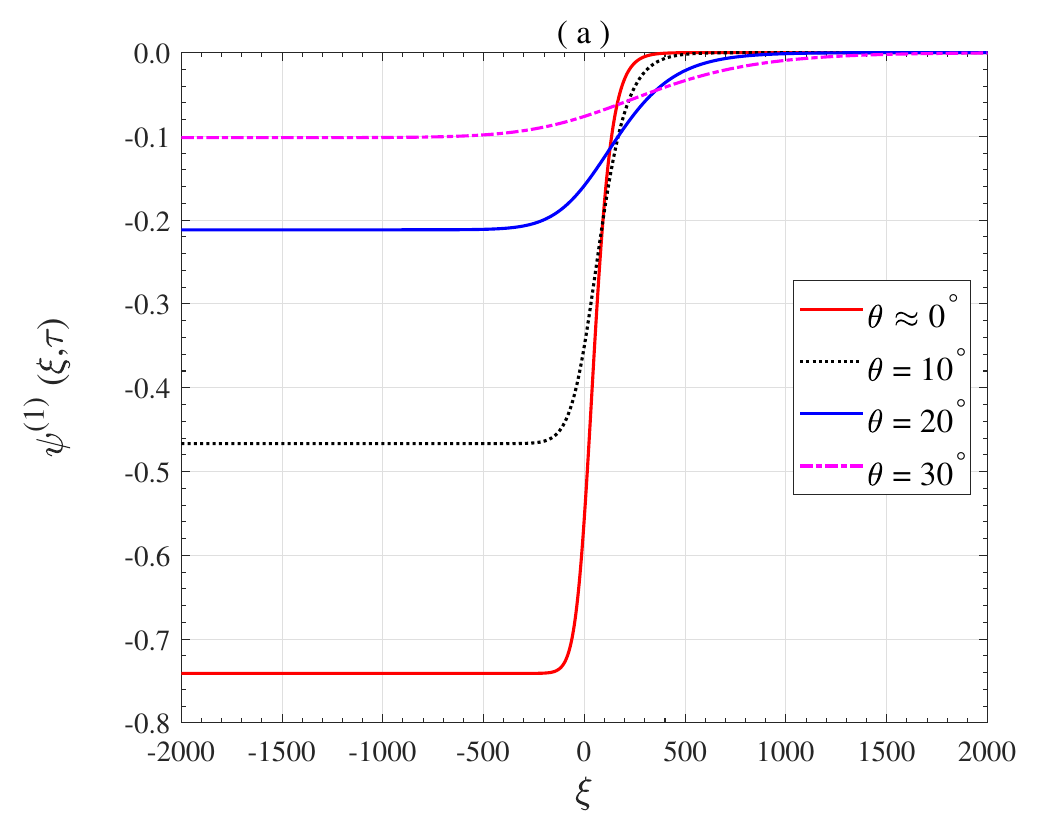}
\quad
\includegraphics[width=8.5cm]{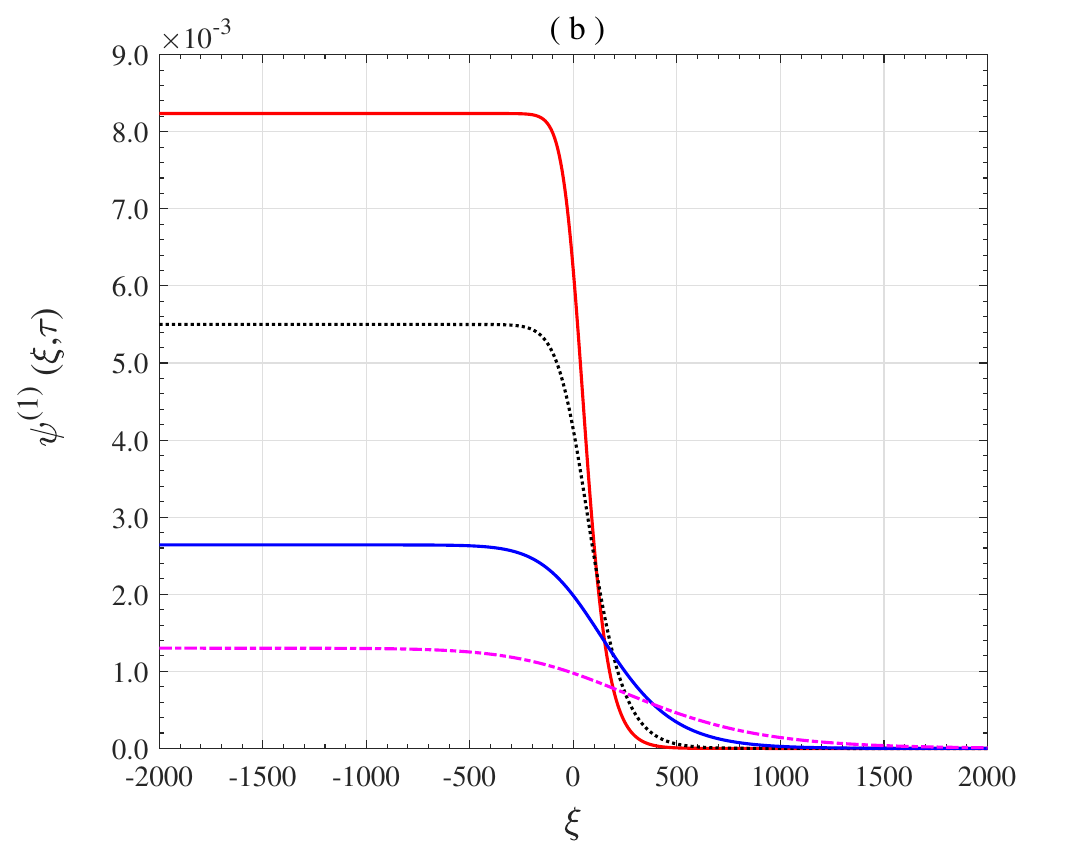}
\caption{Shock profiles [Eq. (\ref{eqn_hypertangent})] are shown for two different electron distributions: (a) $q = 0.7$  and (b) $q\rightarrow 1$ (BD) with $\omega_{\mathrm{i}} = 0.3$,  $r_{\text{d}} = 0.15$ $\mu$m,   $\nu_{\text{d}} = 0.5$,  $P_\parallel = 0.1$, and $P_\perp = 0.2$.}
\label{fig_18}
\end{figure}

\textit{Effects of the obliqueness parameter $\theta$}:
The effects of $\theta$ on the profiles of rarefactive [subplot (a) corresponding to $q = 0.7$] and compressive [subplot (b) corresponding to $q \rightarrow1$]  DIA shocks are shown in Fig. \ref{fig_18}. Subplot (a) shows that on increasing the obliqueness of wave propagation from $\theta=0 $ to $\theta=30^{\circ}$, the magnitude of shock amplitude decreases from $0.74$ to $0.1$, while the width increases from $105$ to $661$. On the other hand, as one approach from the region of nonthermal nonextensive electrons to thermal Maxwellian electrons (i.e., $q\rightarrow1$), a transition from rarefactive to compressive shocks occurs.  The amplitudes of these compressive shocks get reduced with increasing values of the angle $\theta$. However, the shock width increases, e.g.,  from $129$ to $707$ as $\theta$ increases from $\theta\approx0 $ to $\theta=30^{\circ}$[subplot (b)]. 
\par
\begin{figure}[!h]
\centering
\includegraphics[width=8.5cm]{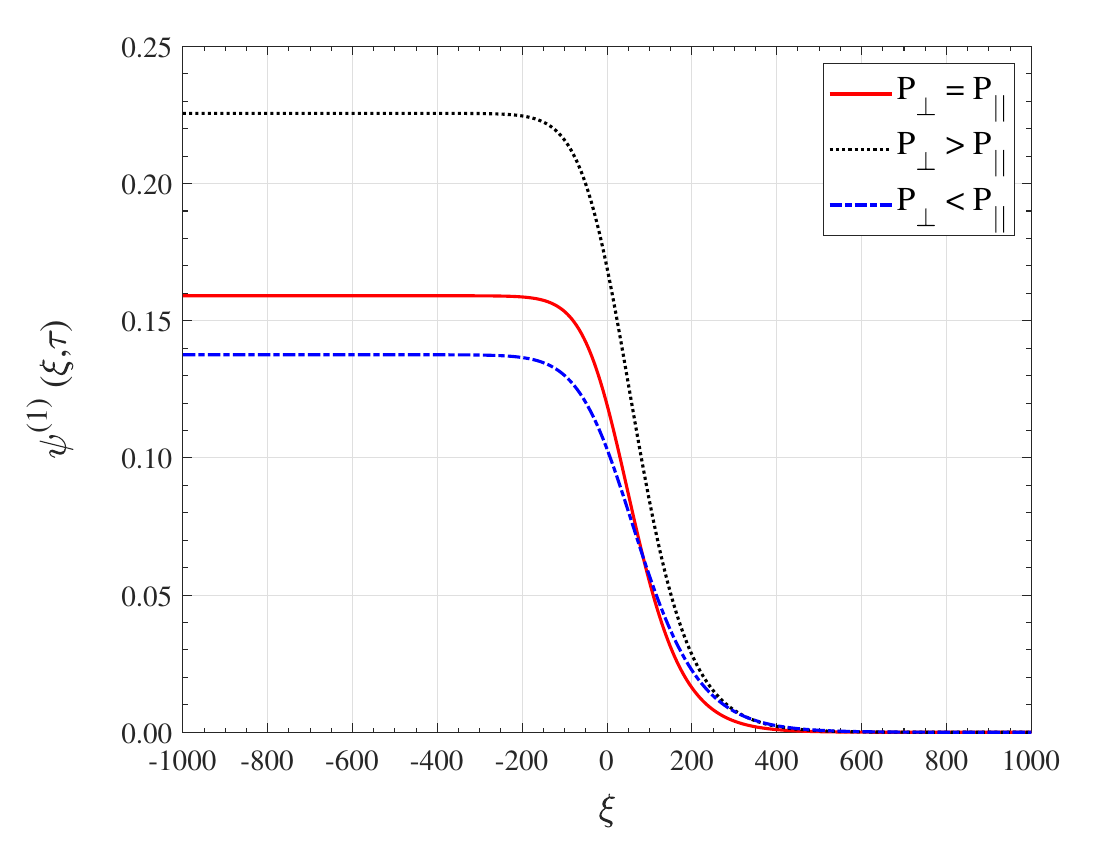}
\caption{Monotonic shock profiles [Eq. (\ref{eqn_hypertangent})] are shown for the cases of isotropic and anisotropic pressures: $P_\perp = P_\parallel = 0.5$, $P_\perp\hspace{0.1cm}(0.3) > P_\parallel\hspace{0.1cm}(0.1)$, and $P_\perp \hspace{0.1cm}(0.1) < P_\parallel\hspace{0.1cm}(0.3)$. The fixed parameter values are $q = 0.8$ and others are the same as for Fig. \ref{fig_18}.}
\label{fig_19}
\end{figure}

 
\textit{Effects of anisotropic pressure and different dust size $(r_d)$}: 
The anisotropic ion thermal pressure can significantly affect the profiles of DIA shocks. The influences of the isotropic and anisotropic pressures on the DIA shock solution are shown in Fig. \ref{fig_19}. We find that the shock amplitude becomes higher when the perpendicular pressure dominates over the parallel pressure (i.e., $P_\perp>P_\parallel$) compared to the case with $P_\perp<P_\parallel$. In the latter, the shock width is relatively higher than the former. However, in the case of isotropic pressure ($P_\perp=P_\parallel$), while the shock amplitude lies in between the values corresponding to anisotropic pressures, the width remains almost the same as the case $P_\perp<P_\parallel$.   The maximum shock amplitude and width can be calculated as $0.22$ (for $P_{\perp}>P_{\parallel}$) and $169$ ( for $P_{\perp}<P_{\parallel}$)  respectively.
 \par 
 \par
\begin{figure}[!h]
\centering
\includegraphics[width=8.5cm]{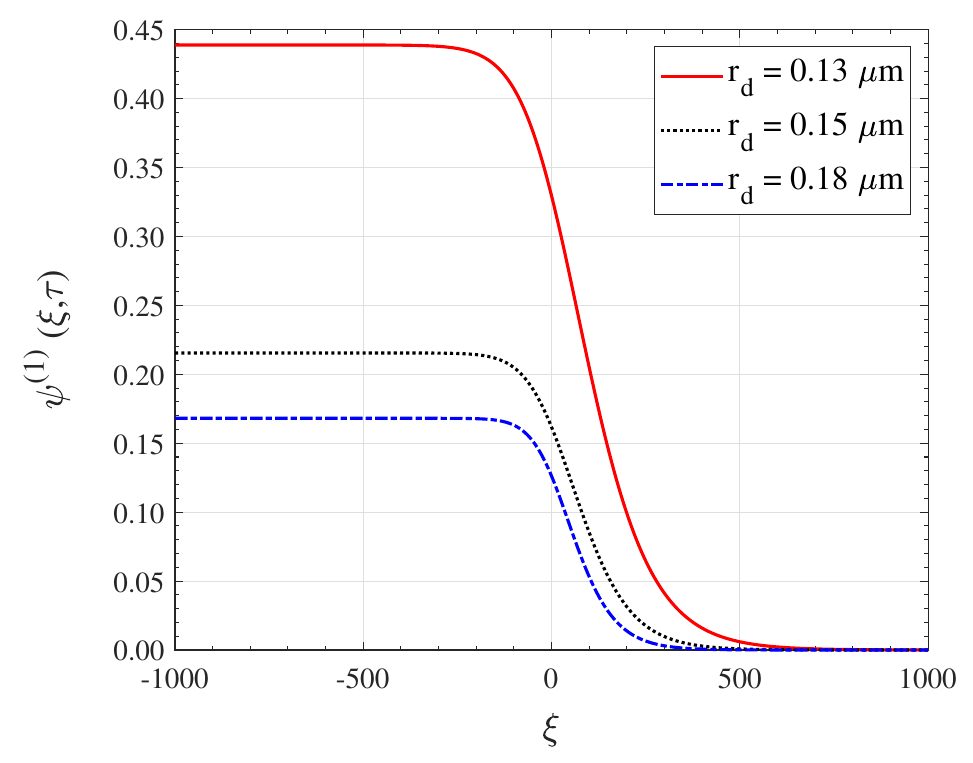}
\caption{Monotonic shock profiles [Eq. (\ref{eqn_hypertangent})] are shown for different values of the dust-grain radius $r_{\mathrm{d}}$ with $q = 0.8$. The other parameter values are the same as for Fig. \ref{fig_18}.}
\label{fig_20}
\end{figure}

Different sizes of dust grains with different radii can influence the dust-charge state and thus have significant influence on the profiles of DIA shocks. The results with different values of the dust grain radius $r_d$ are shown in Fig. \ref{fig_20}. We note that as the values of $r_d$ increase, the magnitudes of both the nonlinear and dissipation coefficients ($A_{1}$ and $D_{1}$) also increase reducing both the shock amplitude and width. For example, as $r_d$ increases from $0.13$ to $0.18$ $\mu$m, the amplitude and width, respectively, decrease from about $0.44$ to $0.17$ and from $201$ to $128$.
\par
\begin{figure}[!h]
\centering
\includegraphics[width=8.5cm]{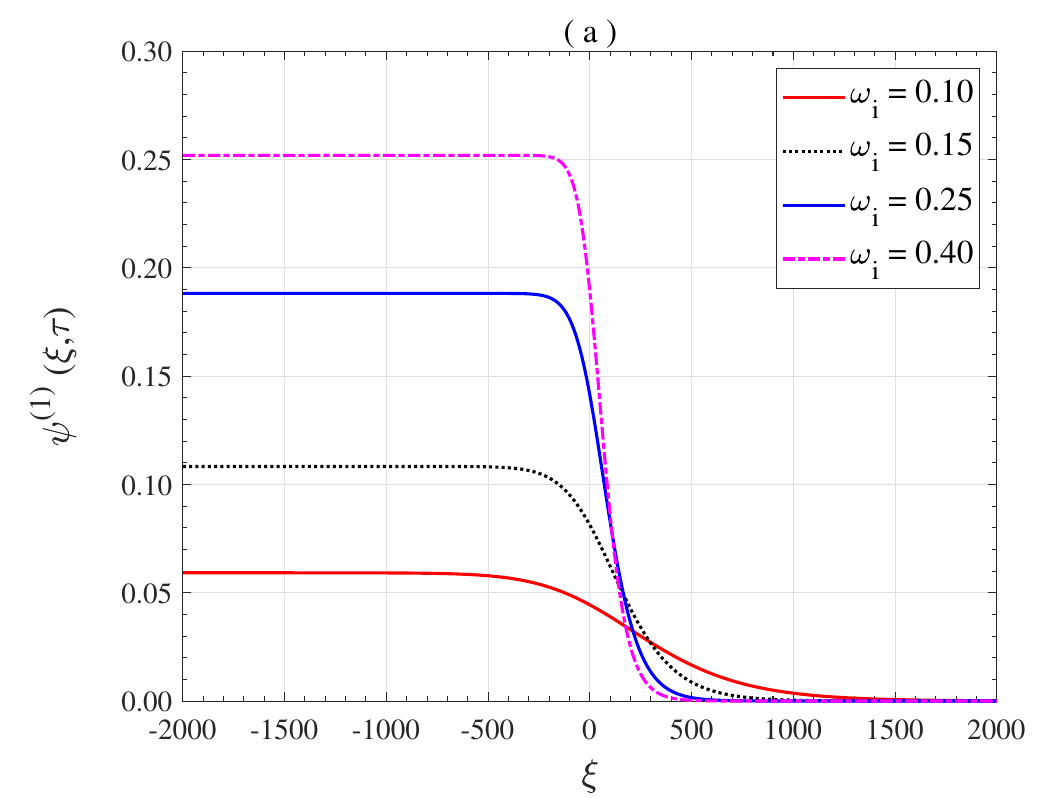}
\quad
\includegraphics[width=8.5cm]{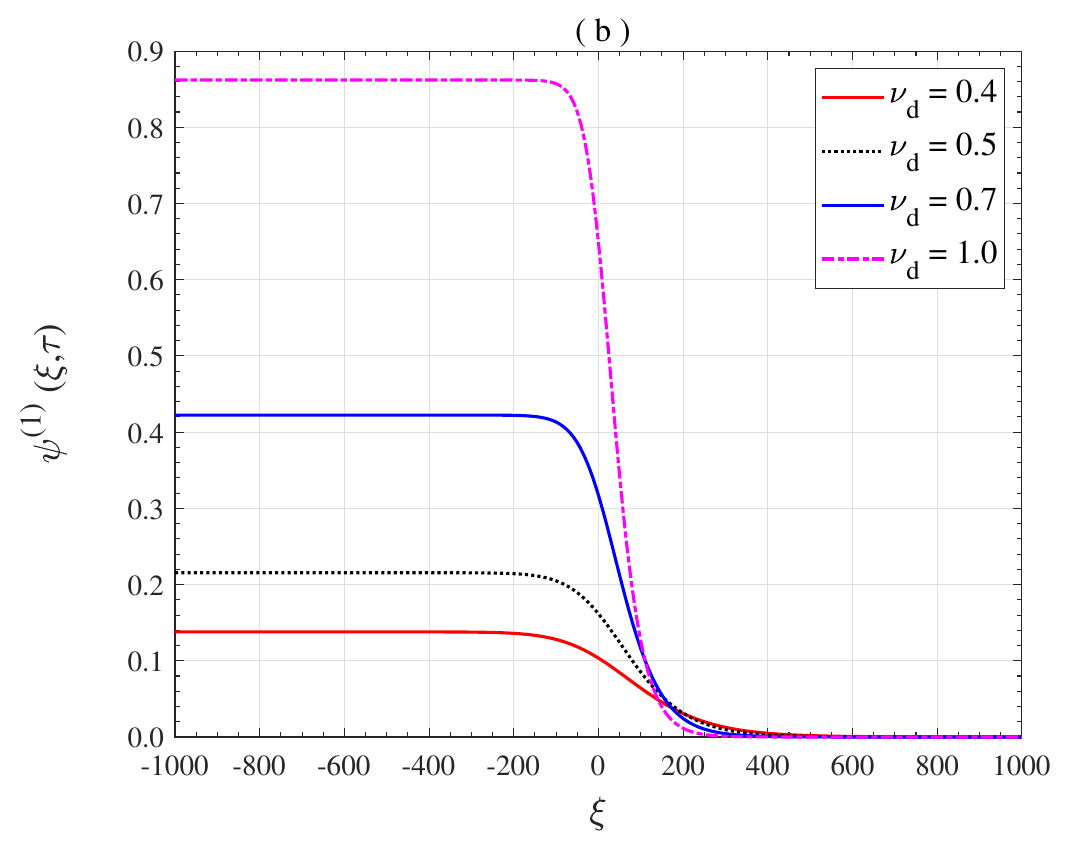}
\caption{Monotonic shock profiles [Eq. (\ref{eqn_hypertangent})] are shown for different values of the ion gyrofrequency, $\omega_i$ [subplot (a)] and the dust-charge fluctuation rate, $\nu_{\text{d}}$ [subplot (b)] with $q = 0.8$. The other parameter values are the same as for Fig. \ref{fig_18}.}
\label{fig_21}
\end{figure}

 
\textit{Effects of magnetic field and dust-charge fluctuation rate}:
The effects of the static magnetic field [via $\omega_{\text{i}}$, subplot (a)] and the dust-charge fluctuation rate [$\nu_{\text{d}}$, subplot (b)] on the profiles of DIA monotonic shocks are shown in Fig. \ref{fig_21}. In both cases, the shock amplitude (width) gets increased (decreased) for increasing values of $\omega_i$ and $\nu_d$. For example, as $\omega_i$ increases from $0.1$ to $0.4$,   the shock amplitude increases from $0.06$ to $0.25$ and the width decreases from $581$ to $136$ [subplot (a)]. On the other hand, when $\nu_d$ increases from $0. 4$ to $1.0$, the shock amplitude increases from about $0.14$ to $0.86$ and the width decreases from $199.5$ to $79.8$ [subplot (b)]. 
\section{Summary and Conclusion} \label{sec-conclu}
We have studied the propagation characteristics of weakly nonlinear obliquely propagating dust-ion-acoustic (DIA) solitary waves (SWs) and shocks in nonthermal collisional magnetized dusty plasmas with the effects of dust-charge fluctuations, the q-nonextensive distribution of electrons, and the anisotropic ion pressure that are relevant in laboratory and space (Saturn's E-ring) plasmas. We have considered the effects of ion creation (source) and ion loss (sink), the ion-neutral and ion-dust collisions, and the anisotropic ion pressure to the fluid model and calculated the collisional enhancement dust-charge number from the dust-charging equation self-consistently. Thus, the previous theory in the literature is generalized and advanced by these effects. Using the standard reductive perturbation technique, we have derived the KdV and KdV Burgers equations, which describe the evolution of small but finite amplitude DIA solitary waves and shocks. The traveling wave solutions of these evolution equations are obtained and analyzed separately for parameters relevant to collisional laboratory and collisionless space plasmas.  The main findings can be summarized as follows:
\subsection*{Case of laboratory plasmas}  
\begin{itemize}
\item The adiabatic dust-charge variation ($\nu_{\rm{ch}}\gg\omega_{\rm{pd}}$) introduces a new term, which modifies the expression of the phase velocity of DIA waves. In contrast to Maxwellian plasmas, the phase velocity in nonthermal nonextensive plasmas can exceed the ion-acoustic speed (supersonic) in the region of superextensive electrons.
\item At equilibrium, the dust-charge number can achieve a maximum value in collisionless isotropic nonthermal dusty plasmas with superextensive electrons. The effects of the ion pressure anisotropy and the ion-neutral collision are to significantly reduce the dust-charge number.
\item The DIA solitary waves get damped due to the effects of the adiabatic dust-charge variation and ion-dust and ion-neutral collisions. So, their evolution is governed by the damped KdV equation. The damped DIA solitary waves can appear as only compressive-type solitons (with positive potential). Their amplitudes decay with time, and so is the soliton energy, but they remain higher in the superextensive region ($0<q<1$) at a fixed time. 
\item The static magnetic field only contributes to the wave dispersion in decreasing the soliton width (but the amplitude remains unchanged) at increasing strength. 
\item The $q$-nonextensive parameter and the obliqueness angle of wave propagation $(\theta)$ have similar effects on the soliton amplitude and width, i.e., as one approaches from nonthermal superextensive region to the thermal one with $0.7\lesssim q\lesssim1$ or as one reduces the angle $\theta$ from $\theta=\pi/2$ to $\theta\approx 0$, both the amplitude and width of solitons decrease. Thus, the DIA solitons propagating close to the magnetic field can evolve with high energies in the presence of superthermal electrons in a fixed interval of time. 
\item The amplitude and width of damped DIA solitons always remain lower in isotropic plasmas than anisotropic ones, and in plasmas with strong collisional effects compared to collisionless plasmas. 
\item As the dusty plasma system changes from active (in the presence of source and sink) to quiescent medium or from collisionless (without collision enhancement current) to collisional (with collision enhancement current) plasmas, the amplitudes of DIA solitons get reduced, but their widths increase.  
\end{itemize}
\subsection*{Case of Saturn's E-ring space plasmas}
\begin{itemize}
\item A deviation from the limit of adiabatic dust-charge variation, i.e., either (i) $\nu_{\rm{ch}}<\omega_{\rm{pd}}$ or (ii) $\nu_{\rm{ch}}>\omega_{\rm{pd}}$, relevant for space plasmas, can lead to the evolution of either (i) DIA damped solitary waves or (ii) DIA shocks. While the former is governed by a damped KdV equation, the latter can be described by a KdV Burgers-type equation. In Case (i), the dust-charge variation and collisions appear as higher-order effects and cannot contribute to the linear wave phase velocity. However, the damping term in the KdV equation gets modified by these effects. On the other hand, Case (ii) corresponds to a collisionless plasma where the wave damping due to collisions can be neglected compared to the dissipation (Burgers term) 
due to the non-adiabatic dust-charge variation. 
\item In contrast to laboratory plasmas, depending on the plasma states we choose, i.e., nonthermal superextensive $(0.6<q<1)$ or thermal equilibrium state $(q\rightarrow1)$, the DIA waves can appear as either rarefactive (with negative potential) or compressive (with positive potential) types.
\item The qualitative features of the damped DIA solitons by the effects of the magnetic field $(\omega_i)$ and the angle of propagation $\theta$ remain similar to those for laboratory plasmas.
\item In contrast to laboratory plasmas with anisotropic pressure, the soliton amplitude gets maximized when the parallel component of ion pressure dominates over the perpendicular component.
\item Different sizes of charged dust grains with different radii influence the electron and ion currents and thus enhance the dust-charge number as the radius increases. Such an increase typically reduces the soliton amplitude but increases the width. 
\item A transition of DIA shocks from rarefactive to compressive types can occur as we approach from nonthermal states (with superextensive electrons) to thermal ones (with Boltzmann distributed electrons). 
\item It is noted that the effects of the magnetic field $(\omega_i)$ and the dust-charge fluctuation rate $(\nu_d)$ on the profiles of DIA shocks are similar, i.e., the shock amplitude (width) gets increased (decreased) with increasing values of any one of $\omega_i$ and $\nu_d$. In contrast, the effect of increasing the angle of propagation $\theta$ is to decrease the amplitude but to enhance the shock width.
\item The amplitude (width) of DIA monotonic shocks can be higher due to anisotropic pressure with  $P_\perp>P_\parallel$ $(P_\perp<P_\parallel)$ compared to the isotropic one with $P_\perp=P_\parallel$. 
\item Increasing sizes of the dust grains can also influence the shock profile in decreasing both amplitude and width. 
\end{itemize}
\par 
To conclude, the present results should help understand the propagation characteristics of DIA modes and the localization of DIA solitons and shocks in collisional nonthermal dusty magnetoplasmas that are relevant for laboratory and space plasmas. The present theory can be extended to study dust-acoustic solitary waves and shocks in nonthermal collisional dusty magnetoplasmas, including the effects of dust-charge fluctuations and dust-size distribution, but maybe a project of future work.  

\section*{ACKNOWLEDGMENTS}
N. P. Acharya would like to acknowledge the University Grants Commission, Bhaktapur, Nepal for the Ph.D. Fellowship with grant number PhD-78/79-S\&T-17. Authors S. Basnet and R. Khanal would like to acknowledge the Research Coordination and Development Council, Tribhuvan University, Kirtipur, Nepal, for the support via the National Priority Area Research Project: TU-NPAR-077/78-ERG-12.\\

\section*{DATA AVAILABILITY}
The data that support the findings of this study are available from the corresponding author upon reasonable request.\\

\section*{REFERENCES} 
\nocite{*}
\bibliography{Ref}
\end{document}